\documentclass[11pt]{article}
\pdfoutput=1
\usepackage{graphicx}
\usepackage{xcolor}
\graphicspath{{./figures/}}
\usepackage{appendix}
\usepackage{latexsym,amsmath,amsfonts,amssymb,booktabs}
\usepackage[font=small]{caption}
\usepackage{slashed,upgreek,amscd,cancel,tensor,color}
\usepackage{adjustbox}
\usepackage{soul}
\usepackage[normalem]{ulem}
\usepackage[numbers,compress,square]{natbib}
\usepackage{epsfig,latexsym}
\usepackage{amsmath}
\usepackage{enumerate}  
\usepackage{tensor}
\usepackage{braket}
\usepackage{comment}
\usepackage{makecell}
\usepackage[pdfencoding=auto]{hyperref} 
\usepackage{url}
\numberwithin{equation}{section}
\usepackage{doi}
\usepackage{subcaption}
\definecolor{MyBlue}{rgb}{0.15,0.15,0.70}
\definecolor{linkblue}{rgb}{0,0,0.8}
\definecolor{linkgreen}{rgb}{0,0.5,0}

\hypersetup{
colorlinks=true,
citecolor=linkgreen,
linkcolor=linkblue,
urlcolor=linkblue
}

\input epsf
\usepackage{amssymb}
\usepackage{amsmath}
\usepackage{amsfonts}
\usepackage{upgreek}
\usepackage{latexsym}
\usepackage{stfloats}
\usepackage{afterpage}

\newcommand{\hinvMpc}{\,h\, {\rm Mpc}^{-1}\,}

\newcommand*\diff{\mathop{}\!\mathrm{d}}

\newcommand{\mpl}{M_{\rm pl}}

\def\beq{\begin{equation}}
\def\eeq{\end{equation}}
\def\be{\begin{equation}}
\def\ee{\end{equation}}
\def\bea{\begin{eqnarray}}
\def\eea{\end{eqnarray}}

\newcommand{\ta}{\tilde{a}}
\newcommand{\mG}{\mathcal{G}}

\newcommand{\mU}{\mathcal{U}}
\newcommand{\mV}{\mathcal{V}}

\begin{document}
\renewcommand{\thefootnote}{\arabic{footnote}}
\setcounter{page}{1} \baselineskip=15.5pt \thispagestyle{empty}

\vspace*{-25mm}

\vspace{0.5cm}

\begin{center}
{\Large \bf Preference for evolving dark energy \\ in light of the galaxy bispectrum
}\\[0.7cm]

{\large   Zhiyu Lu${}^{1,2,3}$,  Théo Simon${}^{4}$, and Pierre Zhang${}^{5,6,7}$\\[0.7cm]}

\end{center}

\begin{center}

\vspace{.0cm}

\begin{small}

{ \textit{  $^{1}$ Department of Astronomy, School of Physical Sciences, University of Science and Technology of China\\Hefei, Anhui 230026, China}}
\vspace{.05in}

{ \textit{  $^{2}$ CAS Key Laboratory for Research in Galaxies
and Cosmology, School of Astronomy and Space Science, University of Science and Technology of China, Hefei, Anhui 230026, China}}
\vspace{.05in}

{ \textit{  $^{3}$ Deep Space Exploration Laboratory, Hefei 230088, China}}
\vspace{.05in}

{ \textit{  $^{4}$ Laboratoire Univers \& Particules de Montpellier,
CNRS \& Université de Montpellier, 34095 Montpellier, France}}
\vspace{.05in}

{ \textit{  $^{5}$ Institute for Particle Physics and Astrophysics, ETH Z\"urich, 8093 Z\"urich, Switzerland}}
\vspace{.05in}

{ \textit{  $^{6}$ Institut f\"ur Theoretische Physik, ETH Z\"urich, 8093 Z\"urich, Switzerland}}
\vspace{.05in}

{ \textit{  $^{7}$ Dipartimento di Fisica “Aldo Pontremoli”, Universit\`a degli Studi di Milano, 20133 Milan, Italy}}
\vspace{.05in}

\end{small}
\end{center}

\vspace{0.5cm}

\begin{abstract}
We analyse pre-DESI clustering data using a dark energy equation of state $w(z)$ parametrised by $(w_0, w_a)$, finding a $2.8-3.9\sigma$ preference for evolving dark energy over the cosmological constant $\Lambda$ when combined with cosmic microwave background data from \textit{Planck} and supernova data from Pantheon+, Union3, or DESY5. 
Our constraints, consistent with DESI Y1 results, are derived from the power spectrum and bispectrum of SDSS/BOSS galaxies using the Effective Field Theory of Large Scale Structure (EFTofLSS) at one loop. 
The evidence, estimated as the Gaussian-equivalent significance from the $\Delta \chi^2$-statistics to $(w_0, w_a) = (-1, 0)$ for two degrees of freedom, remains robust across analysis variations but disappears without the one-loop bispectrum. 
When combining DESI baryon acoustic oscillations with BOSS full-shape data, while marginalising over the sound horizon in the latter to partially mitigate potential correlations, the significance increases to $3.7-4.4\sigma$, depending on the supernova dataset. 
Using instead a data-driven reconstruction of $w(z)$, we show that deviations from $\Lambda$ can be observed at multiple redshifts, though at a low level of significance: only when imposing the specific parametrisation implied by $(w_0, w_a)$ the above evidence emerges.  
In addition, our findings are interpreted within the Effective Field Theory of Dark Energy (EFTofDE), from which we explicitly track the non-standard time evolution in EFTofLSS predictions.
For perturbatively stable theories in the $w < -1$ regime, the preference for $w_0w_a$CDM persists notably in the clustering limit $(c_s^2 \rightarrow 0)$ when higher-derivative corrections are present. 
\end{abstract}

\newpage

\tableofcontents

\vspace{.5cm}


\section{Introduction}\label{sec:introduction}

What drives the accelerated expansion of the Universe if not a cosmological constant $\Lambda$?
Understanding the nature of dark energy remains one of the most pressing questions in cosmology, and is central to major ongoing redshift survey programmes such as DESI~\cite{DESI:2024mwx} and Euclid~\cite{Euclid:2024yrr}. 
By mapping the large-scale structure (LSS) of the Universe over increasingly large volumes, these surveys will measure the cosmic expansion with percent precision, providing stringent tests for potential deviations from $\Lambda$. 
Notably, the recently released DESI Year 1 (Y1) data revealed a mild ($2-4\sigma$) preference for evolving dark energy based on the baryon acoustic oscillations (BAO) and the full-shape modelling of the power spectrum~\cite{DESI:2024hhd}. 
Despite these promising prospects, an inherent limitation remains: 
the observable Universe at low redshifts, where effects of dark energy are detectable, provides only a finite data volume.
Maximising the information extraction from LSS is therefore crucial.

\begin{figure}[!h]
    \centering
    \small
    \includegraphics[width=0.48\textwidth]{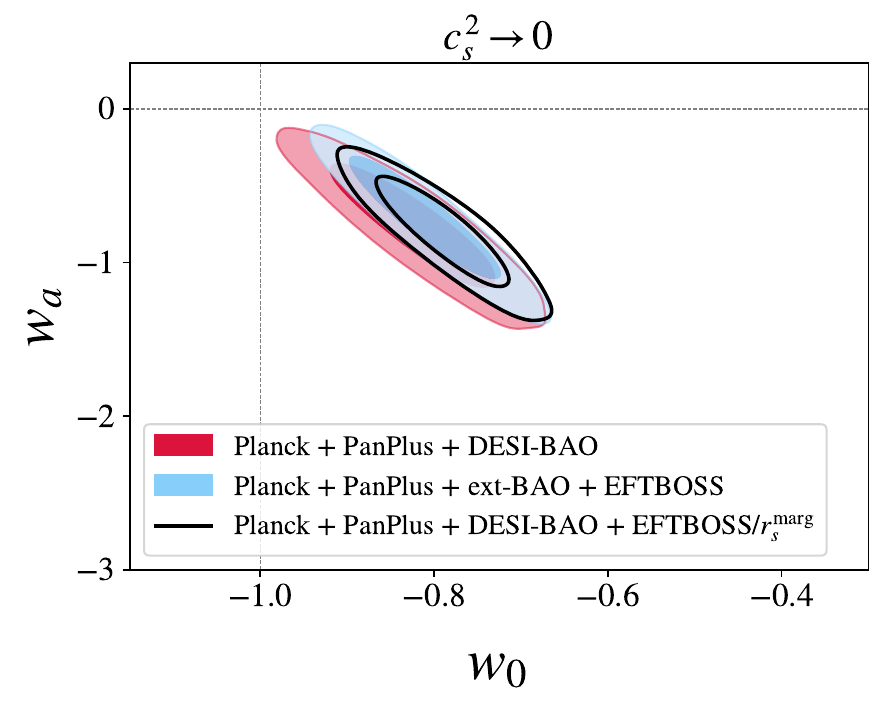}
    \includegraphics[width=0.48\textwidth]{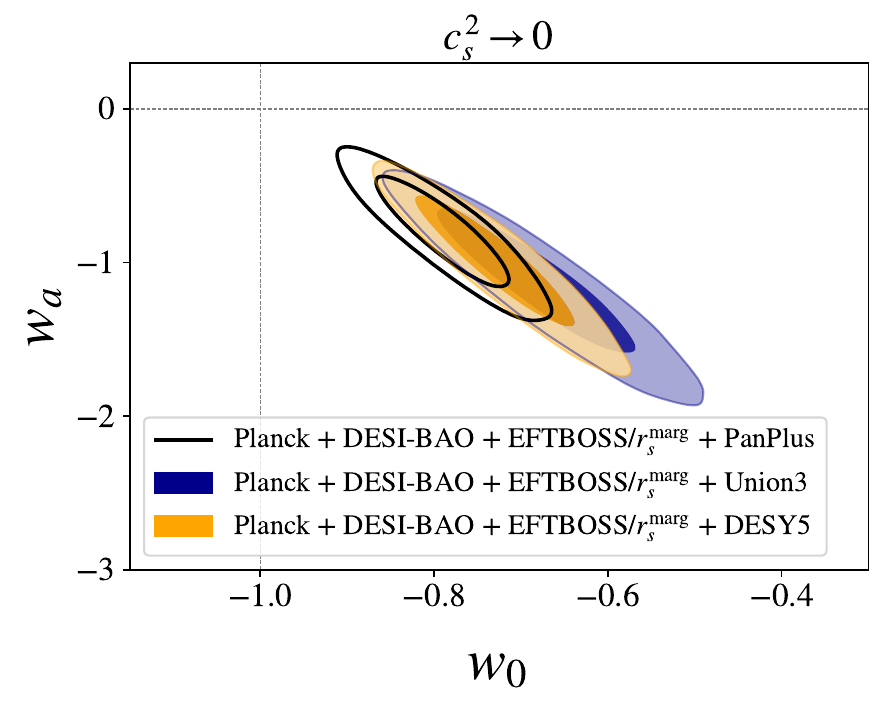}  \\ \vspace{0.5em}
    \begin{tabular}{|lcccc|}
    \hline
        Dataset & $w_0$ & $w_a$ & $p>\Lambda$ & $\Delta \chi^2$\\ \hline
        \textit{Planck} + PanPlus + DESI-BAO & $-0.821\pm 0.063$ & $-0.77\pm 0.27$ & $2.5\sigma$ & $-8.9$ \\ 
        \textit{Planck} + PanPlus + ext-BAO + EFTBOSS & $-0.809^{+0.053}_{-0.061}$ & $ -0.72^{+0.28}_{-0.25}$ & $2.8 \sigma$ & $-10.3$ \\ 
        \textit{Planck} + PanPlus + DESI-BAO + EFTBOSS/$r_s^{\rm marg}$ & $-0.789\pm 0.050$ & $-0.80^{+0.25}_{-0.22}$ & $3.7\sigma$ & $-16.9$ \\ 
        \textit{Planck} + Union3 + DESI-BAO + EFTBOSS/$r_s^{\rm marg}$ & $-0.677\pm 0.075$ & $-1.14^{+0.33}_{-0.29}$ & $3.8\sigma$ & $-17.7$ \\ 
        \textit{Planck} + DESY5 + DESI-BAO + EFTBOSS/$r_s^{\rm marg}$ & $-0.726\pm 0.060$ & $-1.00^{+0.30}_{-0.25}$ & $4.4\sigma$ & $-22.9$ \\ \hline
    \end{tabular} \\ \vspace{0.5em}
    \caption{\textbf{Constraints on $(w_0, w_a)$ across dataset combinations} --- 2D posterior distributions in the $w_0-w_a$ plane obtained by fitting cosmic microwave background data from \textit{Planck} alongside various galaxy clustering datasets (\textit{upper left}) and supernova datasets (\textit{upper right}), within the $w_0w_a$CDM model.
    Corresponding $68\%$ credible intervals, as well as the preference over $\Lambda$ ($p>\Lambda$) and the associated $\Delta \chi^2$, are also shown (\textit{lower panel}).\protect\footnotemark~
    For the comparison of galaxy clustering data, we consider SDSS/BOSS~\cite{BOSS:2016wmc} full-shape (FS) power spectrum and bispectrum (dubbed EFTBOSS), DESI Y1 BAO~\cite{DESI:2024mwx}, or their combination, using Pantheon+ supernovae.
    For the comparison of supernova data, we consider Pantheon+~\cite{Brout:2022vxf}, Union3~\cite{Rubin:2023ovl}, or DESY5~\cite{DES:2024tys}, using EFTBOSS + DESI-BAO as the galaxy clustering dataset. 
    When combining EFTBOSS with DESI-BAO, the sound horizon information is marginalised over in EFTBOSS (dubbed $r_s^{\rm marg}$), {mitigating part of the correlations between the two datasets}. 
    When analysed separately, we complete BOSS data with additional pre-DESI BAO measurements~\cite{Beutler:2011hx,Ross:2014qpa,eBOSS:2019ytm} (dubbed ext-BAO). 
    Here we show the results corresponding to the $c_s^2 \rightarrow 0$ (\textit{i.e.}, clustering quintessence) limit in the EFTofDE, using the EFTofLSS at one loop. 
    See main text for details. All the cosmological results are shown in sec.~\ref{sec:results}.
    }\label{fig:main}   
\end{figure}

In this work, we revisit the analysis of pre-DESI data with three main objectives. 
First, we demonstrate the constraining power of the Effective Field Theory of LSS (EFTofLSS)~\cite{Baumann:2010tm,Carrasco:2012cv} on dark energy properties, extending beyond the now-conventional one-loop power spectrum (see \textit{e.g.}, refs.~\cite{Senatore:2014via,Senatore:2014eva,Senatore:2014vja,Perko:2016puo,DAmico:2019fhj,Ivanov:2019pdj,Chen:2021wdi,Simon:2022csv,Maus:2024sbb,DESI:2024hhd} for an overview of progress in the field). 
The bispectrum of galaxies at one loop, as developed in refs.~\cite{DAmico:2022ukl,Anastasiou:2022udy,DAmico:2022osl}, enables unprecedented precision on the dark energy equation of state $w$, as recently highlighted in ref.~\cite{Spaar:2023his}. 
Here, we extend this investigation by allowing for time variations in $w$. 
Second, we provide an independent cross-check of the preference for evolving dark energy over $\Lambda$ observed in DESI data by analysing earlier galaxy surveys, SDSS/BOSS~\cite{BOSS:2016wmc,eBOSS:2019ytm}. 
Additionally, we assess the significance for evolving dark energy when combining DESI BAO with BOSS full-shape data, using the sound horizon information from the former while marginalizing over it in the latter. 
Third, we explore how constraints on $w$ depend on theoretical assumptions about dark energy, based on the Effective Field Theory of Dark Energy (EFTofDE)~\cite{Gubitosi:2012hu,Gleyzes:2013ooa}. 
In conjunction with the EFTofLSS, our symmetry-based approach provides a unified theoretical framework for consistently treating dark energy fluctuations in LSS.
A key novelty in this aspect is an explicit derivation of the exact time dependence in the EFTofLSS in presence of dark energy for the bispectrum, considering two phenomenologically distinct limits: the clustering ($c_s^2 \rightarrow 0$) and smooth ($c_s^2 \rightarrow 1$) regime. 
Our main results are summarised in fig.~\ref{fig:main}. 

This work is organised as follow. 
In sec.~\ref{sec:theory}, we review the EFTofDE and the EFTofLSS, highlighting the observational consequences for LSS in the presence of dark energy. 
Our inference setup is presented in sec.~\ref{sec:setup}. 
Cosmological results are presented in sec.~\ref{sec:results}, and their significance is further discussed in sec.~\ref{sec:discussion}. 
We finally conclude in sec.~\ref{sec:conclusion}. 
Supplementary materials are provided in appendices.  App.~\ref{app:eftoflss} provides details on the EFTofLSS in presence of dark energy. 
{Our modelling of observational effects and their impact on our cosmological constraints are discussed in length in app.~\ref{app:sys}. 
In particular, we derive there a new treatment of the impact of fibre collisions from a forward-model perspective at the field level, effectively correcting all $N$-point functions. }
Additional analysis products are given in app.~\ref{app:supp_material}, where we also {provide two additional discussions: \textit{(i)} on the relevance of the one-loop bispectrum to access informative smaller scales, supported by a Fisher matrix study, and \textit{(ii)}} {where we provide an order-of-magnitude estimate suggesting that residual correlations between BOSS full shape, wherein the sound horizon information is marginalised, and DESI BAO, are small (while emphasizing that this does not replace a full joint-covariance analysis).} 
Finally, a methodology to reconstruct $w(z)$ from the data or within a specific model is detailed in app.~\ref{app:recon}. 

\footnotetext{{Throughout this paper, when quoting the preference for evolving dark energy with an equation of state parametrised by $(w_0, w_a)$, we compute the Gaussian-equivalent significance in units of $\sigma$ from the $\Delta \chi^2$-statistic relative to the $\Lambda$CDM point, $(w_0, w_a) = (-1, 0)$, counting two degrees of freedom and assuming that Wilks' theorem applies. 
Similar to computing a likelihood-ratio test (as discussed in app.~\ref{app:recon}), this however differs from a full Bayesian model comparison, which is left to future work. 
}}

\paragraph{Conventions} In this paper, we adopt the following conventions. 
To denote quantities that evolve across the cosmic history, we use interchangeably the physical time $t$, the scale factor $a$ (normalised today at $a_0 \equiv 1$), or the redshift $z$. 
Dotted quantities are derivatives with respect to $t$, \textit{e.g.}, $\dot{x} \equiv dx/dt$.  
$H = \dot{a}/a$ is the Hubble parameter, and $\mathcal{H} \equiv aH$ is the comoving Hubble parameter. 
Quantities in bold like $\pmb{x}$ or written with one index like $x_i$ refer to vectors, and with more indices like $x_{ij}$ represent tensors. 
Repeated indices are understood to be summed over following Einstein conventions. 
We use interchangeably lower or upper cases for the indices, \textit{e.g.}, $x_i \equiv x^i$. 
$\partial_i$ refers to a derivatives with respect to space dimension $i$, and $\partial^2$ is the Laplacian operator.


\section{Dark energy meet galaxies}\label{sec:theory}

In this work, we aim to probe deviations from an expanding Universe accelerated by a cosmological constant $\Lambda$. 
We assume the existence of a physical clock, the so-called dark energy, whose evolution we track by profiling its equation of state $w(t)$. 
To do so, we will consider various phenomenological parametrisations for $w(t)$ in sec.~\ref{sec:wform}. 
We work under the lamppost of effective-field theories (EFT), guiding our explorations in what are the physically-allowed observational consequences (especially at the perturbative level) by ensuring that basic principles, such as spacetime symmetries, are respected. 
In sec.~\ref{sec:EFTofDE}, we review the EFTofDE. 
We then isolate in sec.~\ref{sec:limits} three phenomenologically-distinct classes of stable theories of dark energy. 
Finally in sec.~\ref{sec:EFTofLSS}, we derive their observational imprints in gravitational clustering under the framework of the EFTofLSS. 

\subsection{EFTofDE}\label{sec:EFTofDE}
Departure from $\Lambda$ can be probed under the assumption that (spontaneously-broken) time translations are nonlinearly realised around a Friedmann-Lema\^itre spacetime. 
The EFTofDE~\cite{Creminelli:2006xe,Cheung:2007st,Gubitosi:2012hu,Gleyzes:2013ooa} allows us to systematically organise the expansions of the fluctuations that remain invariant under the preserved (time-dependent) spatial diffeomorphisms, providing us a consistent and model-independent parametrisation of deviations from $\Lambda$. 
We focus on an EFT with one scalar degree of freedom, corresponding to the Goldstone boson associated to spontaneously-broken time translations. 
Working in the unitary gauge, we consider the following gravitational action~\cite{Creminelli:2006xe,Creminelli:2008wc}, 
\begin{equation} \label{eq:action}
S_G = \int d^4 x \, \sqrt{-g} \bigg[ \frac{\mpl^2}{2} R - \Lambda(t) - c(t) g^{00} + \frac{m_2^4(t)}{2} ( \delta g^{00} )^2 
- \frac{m_3^3}{2} \delta g^{00} \delta K \bigg] \, ,
\end{equation}
where the metric $g_{\mu\nu}$, and the operators built from it, are in the unitary gauge as well. 
Here $\delta g^{00} = 1 + g^{00}$ are the perturbations in the $00$-component of the metric $g^{00}$, $\delta K_\mu^\nu = K_\mu^\nu - H \delta_\mu^\nu$ are the perturbations of the extrinsic curvature, and $\delta K = K - 3H$ is the trace of the latter.  
In full generality, other operators can show up at quadratic order in perturbations and leading order in spatial derivatives, such as combinations involving $\delta K_\mu^\nu \delta K_\nu^\mu$ or $\delta K^2$. 
Moreover, additional ones appear at cubic and quartic order, and one could also consider a time-dependent Planck mass, $\mpl \rightarrow M_* f(t)$~\cite{Gleyzes:2013ooa}. 
For simplicity, in this work we consider eq.~\eqref{eq:action}.  
Moreover, we take the action for matter $S_M$ to be fully diffeormorphism invariant such that no couplings between the fluctuations of dark energy and matter are considered. 

\paragraph{Background equations}
The matter sector contributes to the background equations through the matter energy density $\bar \rho_m(t)$, while the matter pressure is null, $\bar p_m(t) = 0$. 
By analogy, it is useful to perform a change of variables of the zeroth-order functions $c(t)$ and $\Lambda (t)$ in eq.~\eqref{eq:action} to the energy density $\bar \rho_d(t)$ and pressure $\bar \rho_d(t)$ of a dark component, 
\begin{equation}
2c(t) = \bar \rho_d(t) + \bar p_d(t) \ , \quad 2\Lambda (t) = \bar \rho_d(t) - \bar p_d(t) \ . 
\end{equation}
Solving the zeroth-order Einstein equations and re-arranging them, leads to the usual Friedmann equations, 
\begin{align}
H^2 & = \frac{1}{3\mpl^2}(\bar \rho_m + \bar \rho_d) \ , \label{eq:H2}\\
\dot H & = - \frac{1}{2\mpl^2} (\bar \rho_m + \bar \rho_d + \bar p_d) \ , \label{eq:dotH}
\end{align}
where the time dependence is dropped from the notation to remove clutter. 
Defining the usual fractional energy densities as
\begin{equation}
\Omega_i = \frac{\bar \rho_i}{3\mpl^2 H^2} \ , \quad i = m, d \ ,
\end{equation}
satisfying $\Omega_m + \Omega_d = 1$, 
the equation of state of dark energy is then given by
\begin{equation}\label{eq:w}
w \equiv \frac{\bar p_d}{\bar \rho_d} = - \frac{1}{1-\Omega_m} \left( 1 + \frac{2}{3}\frac{\dot H}{H^2} \right) \ .
\end{equation}
Conservation of the matter stress-energy tensor implies the usual continuity equation for matter, 
\begin{equation}\label{eq:0continuity}
\dot{\bar \rho}_m + 3 H \bar \rho_m = 0 \ .
\end{equation}
Dotting~\eqref{eq:H2} and using~\eqref{eq:dotH}~and~\eqref{eq:0continuity}, lead to an analogous equation for the evolution of the dark component, 
\begin{equation}\label{eq:de_cont}
\dot{\bar \rho}_d + 3 (1+w)H \bar \rho_d = 0 \ . 
\end{equation}
Given an equation of state $w(t)$, we can solve $\bar \rho_d$ in~\eqref{eq:de_cont} using the first Friedmann equation~\eqref{eq:H2}. 
It is convenient to rewrite the Hubble and the matter density parameters with respect to their values at present time $H_0$ and $\Omega_{m,0}$, as
\begin{equation}\label{eq:H}
H = H_0 \sqrt{\Omega_{m,0} a^{-3} + (1 - \Omega_{m,0}) a^{-3 (1 + \tilde w)} }\ , \quad \Omega_m = \frac{\Omega_{m,0} a^{-3}}{(H/H_0)^2} \ , 
\end{equation}
where
\begin{equation}
\tilde w = \frac{1}{\ln a} \int_1^a \frac{w(a')}{a'} da' \ .
\end{equation}

\paragraph{Perturbation equations}
From the action~\eqref{eq:action} in unitary gauge, we can reintroduce the Goldstone mode $\pi$ via the Stückelberg trick, restoring manifest general covariance. 
This amounts to perform a time coordinate change $t \rightarrow t + \pi$ and track the resulting transformations in the quantities appearing in~\eqref{eq:action}. 
To describe gravitational perturbations, we consider the perturbed FLRW metric in Newtonian gauge, 
\begin{equation}
ds^2 = - (1+ 2\Phi)dt^2 + a(t)^2 (1-2\Psi)\delta_{ij}dx^i dx^j \ , 
\end{equation}
where we ignore tensor fluctuations. 
Expanding the gravitational action~\eqref{eq:action} together with the one of matter up to quadratic order in perturbations leads to the quadratic action governing the linear propagation of scalar fluctuations (see \emph{e.g.},~\cite{Lewandowski:2016yce} for details). 
Variation of the quadratic action with respect to $\pi$ then leads to an equation of motion for the propagating mode $\pi$~\cite{Gleyzes:2013ooa,Lewandowski:2016yce}. 
The dispersion relation $\omega^2 = c_s^2 k^2$ can be read off from the kinematic part, defining a speed of sound for $\pi$,
\begin{equation}\label{eq:soundspeed}
c_s^2 = \frac{\nu}{\alpha} \ , 
\end{equation}
where
\begin{align}
\nu & = - \alpha_B^2 + \alpha_B\left( \alpha_w -1 + \frac{3}{2}\Omega_m \right)  + \alpha_w - \frac{\dot \alpha_B}{H} \ , \\
\alpha & = \alpha_w + \alpha_K + 3 \alpha_B^2 \ . \label{eq:alpha}
\end{align}
Using the background equations and the definition of $w$, \textit{i.e.}, eq.~\eqref{eq:w}, we have defined dimensionless time-dependent parameters,
\begin{equation}
\alpha_w \equiv \frac{c}{\mpl^2 H^2} = \frac{3}{2}(1+w)(1-\Omega_m) \ , 
\end{equation}
together with\footnote{Notice that our definitions differ slightly from the ones of \textit{e.g.},~\cite{Bellini:2014fua,Gleyzes:2014rba}. }
\begin{equation}\label{eq:alphaB}
\alpha_K = \frac{2 m_2^4}{\mpl^2 H^2} \ , \qquad \alpha_B = -\frac{m_3^3}{2 \mpl^2 H} \ .
\end{equation}
Varying the quadratic action with respect to metric perturbations yields the linear perturbed Einstein equations~\cite{Gleyzes:2013ooa,Lewandowski:2016yce}. 
There are two constraint equations, the first relating $\Phi - \Psi$ to the anisotropic stress tensors,\footnote{At linear order, the anisotropic stress tensors are zero thus setting $\Psi = \Phi$. 
In general in this work, we consider $\Psi = \Phi$ (up to small corrections from neutrinos) as we neglect relativistic corrections, sub-leading at LSS survey scales.} and the second being a generalised Poisson equation relating $\partial^2 \Psi$ with the fluctuations in matter or dark energy. 
General formulae can be found in~\cite{Gleyzes:2013ooa}. 
To close the linear system of equations, conservation of the stress-energy tensor further leads to the continuity and Euler equations for the matter fluctuations. 
Since we will be interested in describing the galaxy distribution at the shortest possible distance, it is useful to consider the full nonlinear equations for the matter fluctuations~\cite{Baumann:2010tm,Cusin:2017wjg}, 
\begin{align}
\dot \delta + \frac{1}{a}\theta & = - \frac{1}{a}\partial_i (\delta v^i) \ , \label{eq:continuity} \\
\dot \theta + H \theta + \frac{1}{a}\partial^2 \Phi & = -\frac{1}{a}\partial_i(v^j\partial_j v^i) \ , \label{eq:Euler} 	
\end{align}
where $\delta \equiv \rho_m / \bar \rho_m -1$ is the matter overdensity field, $v^i$ is the matter velocity field, and $\theta \equiv \partial_i v^i$  is the velocity divergence.\footnote{Here we have left out the vorticity component of the velocity, $w_i \sim \epsilon_{ijk}\partial^j v^k$, which starts contributing at forth order in fluctuations (more on that latter). 
This does not affect the present discussion, as we focus only on modifications due to the presence of dark energy up to third order for reasons that we explain later. }

\paragraph{Quasi-static limit}
Scales observed in galaxy surveys are much shorter than the Hubble radius such that relativistic corrections can be neglected. 
Moreover, for sufficiently large speed of sound, $c_s^2 \rightarrow 1$, they are well below the sound horizon. 
When so, gravitational and field fluctuations on these scales can be assumed to evolve in the quasi-static approximation where time derivatives are much smaller than spatial derivatives. 
In this limit, the field equations take a particularly simple form~\cite{Cusin:2017mzw}, 
\begin{align}\label{eq:modPoisson}
\partial^2 \Psi & = \frac{\bar \rho_m a^2}{2\mpl^2} \left(1 + \frac{\alpha_B^2}{\nu} \right) \delta \ , \\ 
\partial^2(\Phi + \Psi) &= \frac{\bar \rho_m a^2}{M_{\rm pl}^2} \left(1 + \frac{\alpha_B^2}{\nu} \right) \delta \ , \label{eq:modWeyl} \\
H \partial^2 \pi & = \frac{\bar \rho_m a^2}{2\mpl^2}  \frac{\alpha_B}{\nu}  \delta \ . \label{eq:QSLpi}
\end{align} 
Here the propagation of $\pi$ is fully determined by the matter fluctuations $\delta$. 
The sole difference with general relativity lies in a modification of the Poisson equation and the Weyl potential equation. 
If $\alpha_B = 0$, there is no dark energy fluctuations participating to clustering, and we recover standard smooth quintessence. 
In principle, one can also consider nonlinear corrections stemming from higher-order operators that have been neglected in~\eqref{eq:action}. 
Formulae for the generalised Poisson equation, that we provide in app.~\ref{app:time_MG}, and their analog equations for $\Phi + \Psi$ and $\pi$ in the EFTofDE within the quasi-static approximation have been derived up to third order in fluctuations in ref.~\cite{Cusin:2017mzw}. 

\paragraph{Clustering limit}
Another phenomenologically interesting limit is obtained for vanishing sound speed, $c_s^2 \rightarrow 0$. 
When so, fluctuations in dark energy fall in potential wells, thus participating to gravitational clustering. 
Without lack of generality, we set $m_3^3 = 0$ in the following discussion, as the resulting equations relevant for LSS remains unchanged otherwise~\cite{Lewandowski:2016yce}. 
In the limit $c_s^2 \rightarrow 0$, eq.~\eqref{eq:soundspeed} yields $m_2^4 \approx \bar \rho_d (1+w) / (4c_s^2)$, and the linear equation for $\pi$ reads~\cite{Creminelli:2008wc,Lewandowski:2016yce} 
\begin{equation}\label{eq:pi}
\frac{1}{a^3 m_2^4}\frac{d}{dt} \big[a^3 m_2^4 (\dot \pi - \Phi) \big] = \frac{1}{a^2} \frac{c_s^2}{1-c_s^2}\partial^2 \pi \ .
\end{equation}
Counting powers of $c_s^2$, we see that the r.h.s. is negligible. 
This shows that, once the decaying mode has fade away, $\dot \pi - \Phi = 0$, implying $\partial_i \dot \pi - \partial_i \Phi = 0$.  
Since $\partial_i \Phi = - \frac{d}{dt}(av^i)$ from linearising the Euler equation~\eqref{eq:Euler}, we find that the two species are comoving, 
\begin{equation}\label{eq:comoving}
\partial_i \pi = -a v^i \ .
\end{equation}
The Poisson equation is then sourced by a single growing adiabatic mode~\cite{Creminelli:2008wc,Sefusatti:2011cm}, 
\begin{equation}
\partial^2 \Phi = \frac{\bar \rho_m a^2}{2\mpl^2} \delta_A \ , \qquad \delta_A =\delta + \delta_{d} \ ,
\end{equation}
where we have defined a quintessence overdensity field as
\begin{equation}\label{eq:delta_d}
\delta_d = \frac{4 m_2^4}{\bar \rho_m}(\dot \pi - \Phi) \approx \frac{1+w}{c_s^2} \frac{\bar \rho_d}{\bar \rho_{m}} (\dot \pi - \Phi) \ .
\end{equation} 
Using~\eqref{eq:continuity} and~\eqref{eq:pi}, together with $\partial^2 \pi = - a^2 \theta$ that follows from~\eqref{eq:comoving}, dotting $\delta_A$ leads to the linear continuity equation for the adiabatic mode, 
\begin{equation}
\dot \delta_A + \frac{1}{a}C(a) \theta = 0 \ , 
\end{equation}
where we have defined
\begin{equation}\label{eq:C}
C(a) = 1 + (1+w) \frac{\bar \rho_d}{\bar \rho_{m}} \ . 
\end{equation}
The Euler equation for the adiabatic mode simply follows from relating $\theta_A = \theta$ (as the two species are comoving) to the gravitational potential using~\eqref{eq:Euler}. 
Derivation of the equations of motion at nonlinear order, relying on an argument that shows that the two species remain comoving, can be found in~\cite{Lewandowski:2016yce}. 
The nonlinear continuity and Euler equations for the adiabatic mode read~\cite{Lewandowski:2016yce}
\begin{align}
\dot \delta_A + \frac{1}{a}C(a)\theta & = - \frac{1}{a}\partial_i (\delta_A v^i) \ , \label{eq:CQcontinuity} \\
\dot \theta + H \theta + \frac{1}{a}\partial^2 \Phi & = -\frac{1}{a}\partial_i(v^j\partial_j v^i) \ . \label{eq:CQEuler} 	
\end{align}

\paragraph{Stability}
To ensure the absence of gradient instabilities, the sound speed $c_s^2$ defined in~\eqref{eq:soundspeed} has to be positive. 
To avoid ghost instabilities, we can further impose that the kinematic energy of $\pi$ (proportional to $\alpha$) has to be positive as well. 
These requirements yield
\begin{equation}\label{eq:stable}
\alpha > 0 \ , \qquad \nu \geq 0 \ . 
\end{equation}
Allowing $\alpha_B$ and $\alpha_K$ to take large values $\sim \mathcal{O}(1)$ (see refs.~\cite{Nicolis:2008in,Deffayet:2010qz}), one can show that the stability requirements~\eqref{eq:stable} can be generically satisfied even for $w<-1$. 
This is also the case when generalising the action~\eqref{eq:action} with a time-dependent Planck mass or in the presence of the quadratic operator leading to tensor modes propagating at non-luminal speed~\cite{Gleyzes:2013ooa}.

For $\alpha_w \sim \alpha_B \ll \alpha_K \sim \mathcal{O}(1)$, \textit{i.e.}, $m_3^3 \sim m_2^4$ and $w \sim -1$, the propagation of $\pi$ can still be stabilised in a region $w<-1$, not too far from $w = -1$: the kinetic and the gradient terms remain positive provided that $\alpha \approx \alpha_K > 0 \ , \ \nu \approx \alpha_w - \alpha_B (p + 1+ \frac{3}{2} \Omega_m) > 0$. 
When so, the speed of sound is small, $c_s^2 \ll 1$. 
Note that this limit is technically natural: $c_s^2 \equiv 0$ is protected by the presence of a shift symmetry~\cite{Arkani-Hamed:2003pdi,Creminelli:2006xe}, which in particular enforces $\alpha_B \equiv 0$. 
Higher-derivative operators not shown in~\eqref{eq:action} can also lead to stable propagations even when $c_s^2 < 0$ for $w < -1$~\cite{Creminelli:2006xe,Creminelli:2008wc}: leading to a modified dispersion relation of the type $w^2 \sim k^4$, they act as a cutoff --- confining the gradient instability to (quantum-safe) large scales. 
However, this is effective only if $c_s^2 \rightarrow 0$ as shown to be necessary for the rate of the instability to not grow too fast within a Hubble time~\cite{Creminelli:2008wc}. 

When all operators (but the zeroth-order ones) are small, we are in the regime of $k$-essence with unit sound speed, where nothing in the EFT can prevent ghost instabilities when the null energy condition is violated. 
To summarise, based on stability requirements, we consider three classes of theories: \textit{i)} $k$-essence where $w<-1$ is not allowed, \textit{ii)} EFTofDE in the quasi-static limit, stable for all $w$ with large $c_s^2 \rightarrow 1$, yielding a modified Poisson equation~\eqref{eq:modPoisson}, and \textit{iii)} clustering quintessence with $c_s^2 \rightarrow 0$ where $w\lesssim -1$ is allowed. 
We now review them in regards of their distinct phenomenology, whose description we complete in sec.~\ref{sec:EFTofLSS} for their imprints on the LSS.

\subsection{Three shades of quintessence}\label{sec:limits}

From considerations on the stability of fluctuations in dark energy, we have distinguished three broad classes of theories. 
In this section, we describe their distinct observational signatures. 
Their main features, summarised in tab.~\ref{tab:quintessence}, are detailed below. 

\begin{table}[!h]
    \centering
    \small
    \begin{tabular}{|c|c|c|c|}
    \hline
        Dark energy theories & $c_s^2$ & $w<-1$ & Signatures in LSS perturbations  \\ \hline
        Standard $k$-essence & 1 & Prohibited & non-EdS time evolution \\ \hline
        \makecell{Smooth quintessence \\ 
        in modified gravity} & 1 & Stable & \makecell{Modified time evolution \\ from generalised Poisson equation}  \\ \hline
        Clustering quintessence & 0 & Stable & \makecell{Modified time evolution \\ from dark energy clustering}  \\ \hline    
        \end{tabular}
\caption{\small Classification of dark energy theories within the EFTofDE according to their distinct observational consequences in LSS (galaxy clustering at the perturbative level). 
Stability requirements in the propagations of dark energy fluctuations impose restrictions when probing $w(t)$. 
See main text for more details.  }\label{tab:quintessence} 
\end{table}

\paragraph{Smooth quintessence}
This is the standard $k$-essence with unit sound speed, $c_s^2 = 1$. 
At the background level, the cosmological constant is replaced by a dark energy component with a general equation of state $w(t)$. 
At the perturbation level, the structure of the time dependence in the EFTofLSS is modified when $w \neq -1$: this will be detailed in the next section. 
EFTofDE operators are negligible, \textit{e.g.}, $\alpha_K, \alpha_B = 0$. 
There is thus nothing in the EFT that prevents from ghost instabilities when the null energy condition is violated. 
When considering such theory in sec.~\ref{sec:stability}, we safely impose a physical cut on the equation of state, $-1 < w$. 
See ref.~\cite{DAmico:2018mnx} for a class of $k$-essence models appearing natural and stable under quantum corrections. 

\paragraph{Smooth quintessence in modified gravity}
This is the regime of the EFTofDE that is stable in regions where $w<-1$, \textit{i.e.}, where the stability requirements~\eqref{eq:stable} are met, preventing ghost and gradient instabilities.  
We consider the limit $c_s^2 \rightarrow 1$ where the quasi-static approximation is valid at scales observed in galaxy surveys. 
This leads to a modified Poisson equation~\eqref{eq:modPoisson}, generalisable to higher order in fluctuations~\cite{Cusin:2017mzw}, which in turn modifies the time evolution of EFTofLSS operators~\cite{Cusin:2017wjg} (see app.~\ref{app:time_MG}). 
When considering smooth quintessence, additional degrees of freedom, \textit{e.g.}, $\alpha_B$, need to be marginalised to probe the $w<-1$ region safely. 
See refs.~\cite{Nicolis:2008in,Deffayet:2010qz} for the cubic Galileon --- a class of models where $\alpha_B$ can naturally takes large ($\sim \mathcal{O}(1)$) values, thereby preventing gradient instabilities. 

\paragraph{Clustering quintessence}
Instabilities can be avoided when $w<-1$ in the limit of a vanishing sound speed ($c_s^2 \rightarrow 0$) without extra degrees of freedom (relevant at cosmological scales), provided that $w$ stays relatively close to $w=-1$~\cite{Creminelli:2008wc}. 
We explicitly check that those theoretical bounds are satisfied in sec.~\ref{sec:stability}.
In the limit $c_s^2 \rightarrow 0$, the dark energy and matter fluctuations form a single growing adiabatic mode relevant to gravitational clustering. 
The equations of motion for the matter fluctuations, eqs.~\eqref{eq:continuity}~and~\eqref{eq:Euler}, are replaced by analogous ones for the adiabatic mode sourcing the gravitational potential, eqs.~\eqref{eq:CQcontinuity}~and~\eqref{eq:CQEuler}. \\

In this work, we analyse the equation of state $w(t)$ both in the presence of smooth ($c_s^2 \rightarrow 1$) and clustering ($c_s^2 \rightarrow 0$) quintessence. 
The main results presented in sec.~\ref{sec:results} are obtained imposing no prior on $w$ and without marginalising over additional EFTofDE parameters. 
In sec.~\ref{sec:stability}, we revisit the constraints on dark energy in light of the stability of the fluctuations.

\subsection{EFTofLSS}\label{sec:EFTofLSS}

To make contact with observations, \textit{i.e.}, maps of distant massive objects such as galaxies, we make use of the EFTofLSS~\cite{Baumann:2010tm,Carrasco:2012cv,Lewandowski:2016yce,Cusin:2017wjg,DAmico:2022ukl}. 
At sufficiently long distances, the equivalence principle allows for a perturbative description of gravitational clustering: all dust-like fields at long distances are governed by the same symmetries. 
A long-wavelength field $\phi_\ell$ at position $\pmb{x}$ in the Universe is invariant under translations and Galilean boosts --- \textit{i.e.}, the Newtonian limit of diffeomorphism~\cite{Peloso:2013zw,Kehagias:2013yd,Creminelli:2013mca},
\begin{equation}
\pmb{x} \rightarrow \pmb{x} + \pmb{n} \ , \qquad \pmb{v} \rightarrow  \pmb{v}  + \pmb{\chi} \ , 
\end{equation}
where $\pmb{v} = \dot{\pmb{x}}$, and $\pmb{n}$ and $\pmb{\chi}$ are constant in space. 
Correlation functions of $\phi_\ell$ are further preserving rotational invariance. 
In the following, we describe the steps to compute the power spectrum and bispectrum of galaxies in the EFTofLSS at one loop in the presence of smooth or clustering quintessence. 
Our goal is to highlight the main difference with the original computations of~\cite{DAmico:2022ukl}, which are modifications of the time dependence of the EFTofLSS operators.  

\paragraph{Dark matter - quintessence effective fluid}
The first step is to smooth all fields $\phi \rightarrow \phi_\ell^{(\Lambda_s)}$ over a length scale $\Lambda_s$ such that an EFT can be written for the resulting long-wavelength fluctuations on the scales where their variance are smaller than unity. 
For example, one can apply a sharp cutoff in Fourier space, namely imposing $\phi_\ell(\pmb{k}) = \phi(\pmb{k})$ for $|\pmb{k}| < \Lambda_s^{-1}$, $\phi_\ell(\pmb{k})=0$ otherwise. 
Defining so a long-wavelength density $\delta_\ell$ and momentum $\pmb{p_\ell} = \bar \rho_m (1+\delta_\ell) \pmb{v_\ell}$ of the adiabatic mode, as well as a long-wavelength gravitational potential $\Phi_\ell$, we obtain a smoothed version of the system of equations (continuity, Euler, and Poisson) governing the propagation of the long-wavelength fields~\cite{Baumann:2010tm,Cusin:2017wjg},\footnote{Here we have again left out the curl component of the velocity from the presentation, although when computing the one-loop contributions to the bispectrum we are actually also solving for it~\cite{DAmico:2022ukl}.}
\begin{align}
\dot \delta_\ell + \frac{1}{a}C(a) \theta_\ell & = - \frac{1}{a}\partial_i (\delta_\ell v_\ell^i) \ , \label{eq:EFTcontinuity} \\
\dot \theta_\ell + H \theta_\ell + \frac{1}{a}\partial^2 \Phi_\ell & = -\frac{1}{a}\partial_i(v_\ell^j\partial_j v_\ell^i) - \frac{1}{a}\partial_i \left( \frac{1}{\rho_m}\partial_j \tau^{ij} \right) \ , \label{eq:EFTEuler}  \\
\partial^2 \Phi_\ell & = \frac{3}{2} \Omega_m \mathcal{H}^2 \delta_\ell \ . \label{eq:EFTPoisson}
\end{align}
For clarity, the effects of modified gravity within the EFTofLSS are discussed in app.~\ref{app:time_MG}, while the main text retains the standard Poisson equation. 
In smooth quintessence ($c_s^2 \rightarrow 1$), the adiabatic mode are simply the matter fluctuations such that $C\equiv 1$, while in clustering quintessence ($c_s^2 \rightarrow 0$), the adiabatic mode also includes dark energy fluctuations~\eqref{eq:delta_d} such that $C$ is given by~\eqref{eq:C}. 
Upon smoothing, we consider, on the r.h.s. of~\eqref{eq:EFTEuler}, the divergence of a stress tensor $\partial_j \tau^{ij}$ enclosing the effects of the short-wavelength fluctuations sourcing the long-wavelength fields~\cite{Carrasco:2012cv}. 
In the following, we drop the lower-case index $\ell$ to avoid clutter.

Leaving aside counterterms for now, \textit{i.e.}, solutions generated from the sourcing of the Euler equation by the EFT expansion of $\tau^{ij}$, the perturbation theory contributions for $\psi_\alpha \equiv (\delta_\ell, \theta_\ell)^T$ are obtained by solving recursively the system of equations~\eqref{eq:EFTcontinuity}~to~\eqref{eq:EFTPoisson}. 
The $n$-th order solutions for the fields $\psi_\alpha$ read
\begin{equation}
\psi_\alpha^{(n)}(\pmb{k}, a) = \int \frac{d^3q_1}{(2\pi)^3} \dots \frac{d^3q_n}{(2\pi)^3} \ (2\pi)^3 \delta_D(\pmb{k} - \pmb{q_1} - \dots - \pmb{q_n}) \, K_\alpha (\pmb{q_1}, \dots, \pmb{q_n}, a) \, \delta_{\pmb{q_1}}^{(1)}(a) \dots \delta_{\pmb{q_n}}^{(1)}(a) \  ,
\end{equation}
where $\delta_D$ is the Dirac delta-distribution and $\delta_{\pmb{q}}^{(1)}(a)$ is the linearly-evolved adiabatic mode up to scale factor $a$ evaluated at Fourier mode $\pmb{q}$. 
Explicit expressions for the time-dependent Fourier kernels $K_\alpha^{(n)}$ up to third order in clustering quintessence can be found in~\cite{DAmico:2020tty}, and read schematically as
\begin{equation}\label{eq:dm_kernel}
K_\alpha^{(n)}(\pmb{q_1}, \dots, \pmb{q_n}, a) = \sum_\sigma \Pi_\sigma(\pmb{q_1}, \dots, \pmb{q_n}) \mathcal{G}_\sigma^{(n),\alpha}(a) \ ,
\end{equation}
where we see that they can be decomposed into a sum of momentum-dependent functions $\Pi_\sigma$ (whose form, universal, is dictated by the equivalence principle~\cite{DAmico:2021rdb}) multiplied by time-dependent functions $\mathcal{G}_\sigma^{(n),\alpha}$ that are found by solving for the Green's functions associated to the system of equations~\eqref{eq:EFTcontinuity} to~\eqref{eq:EFTPoisson}~\cite{Carrasco:2012cv,Lewandowski:2016yce}. 
Smooth quintessence is recovered taking the limit $C(a) \equiv 1$ in $\mathcal{G}_\sigma^{\alpha}$ and taking further $w = -1$ lands us in $\Lambda$CDM. 
Explicit expressions are given in app.~\ref{app:time_quint}. 

\paragraph{Exact time dependence}
An exact treatment of time dependence in the EFTofLSS contrasts with the standard approach based on the Einstein-de Sitter (EdS) approximation.
In the latter, the time evolution of perturbations is approximated by simple powers of the growth factor $D$, as it would be in a matter-dominated Universe ($\Omega_m \equiv 1$), \textit{e.g.}, $\delta_{\pmb{k}}^{(n)}(a) \simeq D(a)^n \delta_{\pmb{k}}^{(n)}(a_i)$, where $D$ is normalised at some initial time $a_i$ (deep inside matter domination).  
This approximation has been shown to be highly accurate within $\Lambda$CDM and in smooth quintessence for the data volume of LSS Stage-4 surveys~\cite{Zhang:2021uyp}. 
In the case of clustering quintessence, exact time dependence was discussed in ref.~\cite{DAmico:2020tty}. 
These previous analyses, based on the power spectrum, assumed a constant dark energy equation of state. 

In this work, we explore general time-varying $w(t)$ in light of the one-loop bispectrum of galaxies. 
Unlike the power spectrum, where deviations from the EdS approximation arise only at the loop level, the bispectrum is already affected at tree level (through $\delta^{(2)}$). 
In the present cosmological analysis, we incorporate the exact time dependence in both the one-loop power spectrum and the tree-level contribution of the galaxy bispectrum. As we find that exact time dependence has a negligible impact on cosmological constraints (see sec.~\ref{sec:results}), we omit it in the loop contributions of the bispectrum.

\paragraph{Galaxies}
Positions of biased tracers (that we refer as galaxies) of the underlying dark matter - quintessence fluid, are what we are offered in LSS surveys --- these are ultimately what we want to describe. 
At sufficiently long distances, we can again make use of the equivalence principle to write the most general expressions for the galaxy density in terms of all possible scalars constructed from the tidal field $\partial_i \partial_j \Phi$ and spatial derivatives~\cite{McDonald:2006mx,McDonald:2009dh,Senatore:2014eva}. 
This expansion is integrated over the past lightcone within which galaxies form, and is evaluated along the fluid elements throughout their history flowing into the final observed positions~\cite{Senatore:2014eva,Mirbabayi:2014zca}. 
Leaving aside stochastic terms for now, the bias expansion of the galaxy density at final position $\pmb{x}$ and time $t$ is given by
\begin{equation}\label{eq:delta_g_sym}
\delta_g(\pmb{x}, t) = \sum_i \int^t dt' \, H(t')\, c_i(t, t') \, \mathcal{O}_i \left[\frac{\partial_i \partial_j \Phi(\pmb{x_{\rm fl}}, t')}{H(t')^2}, \frac{\partial_i}{k_{\rm M}} \right] \ ,
\end{equation}
where $\mathcal{O}_i$ are all possible Galilean-invariant scalars built from powers of $\partial_i \partial_j \Phi$ and spatial derivatives $\partial_i$ (normalised by $k_{\rm M}$, the spatial extension of the observed objects) that are evaluated at the fluid element $\pmb{x_{\rm fl}}$ defined from displacing recursively the position as  
\begin{equation}\label{eq:x_fl}
\pmb{x_{\rm fl}}(\pmb{x}, a, a') = \pmb{x} - \int_{a'}^{a}\frac{d\tilde a}{a \mathcal{H}(\tilde a)} \,\pmb{v}(\pmb{x_{\rm fl}}(\pmb{x}, a, \tilde a), \tilde a) \ . 
\end{equation} 
Note that for clustering quintessence, the bias expansion is constructed from the adiabatic mode. 
After Taylor expanding around $\pmb{x}$ to the relevant order in fields and accounting for the degeneracies, the galaxy density schematically reads as
\begin{equation}\label{eq:delta_g}
\delta_g(\pmb{x}, t) = \sum_i b_i(t) \mathcal{D}_i(t) \mathbb{C}_i(\pmb{x}, t) \ ,
\end{equation}
where $b_i$ are time-dependent free coefficients to adjust to the data, $\mathcal{D}_i$ are calculable generalised growth functions, and $\mathbb{C}_i$ are operators carrying the position dependence constructed from powers of $\delta^{(1)}$ and derivatives. 
The time dependence of the operator $\mathbb{C}_i$ at order $n$ in perturbations is $D(t)^n$, such that in the EdS time approximation, $\mathcal{D}_i(t) = 1$. 
At nonlinear order ($n \geq 2$), $\mathcal{D}_i$ can take more complicated forms~\cite{Donath:2020abv}. 
The bias expansion up to third order has been derived within the EdS approximation in~\cite{Mirbabayi:2014zca,Angulo:2015eqa,Fujita:2016dne}, later with exact time dependence in~\cite{Donath:2020abv} for $\Lambda$CDM and smooth quintessence, and finally in~\cite{DAmico:2020tty} for clustering quintessence (see also~\cite{Fujita:2020xtd,DAmico:2021rdb}). 
Explicit expression for the galaxy density field can be found in app.~\ref{app:eftoflss}. 

\paragraph{Redshift-space distortions}
Because galaxies and dark matter are comoving at large scales, their velocities are the same (up to higher-derivative terms). 
As shown in app.~\ref{app:eftoflss}, it is convenient to write $\theta_g \equiv \theta$ in the form of~\eqref{eq:delta_g} with specific values of the biases. 
For general cosmologies, those are calculable time-dependent functions identified with the ones appearing in the dark matter velocity expansion~\eqref{eq:dm_kernel}. 
This is important as galaxies are observed in redshift space where their positions are distorted by peculiar velocities along the line-of-sight $\hat z$. 
Upon this change of variables, 
\begin{equation}
\pmb{x} \rightarrow \pmb{x} + \frac{\pmb{v} \cdot \hat z}{\mathcal{H}} \hat z \ , 
\end{equation}
new terms are generated in the expansion of the galaxy density involving products of density and momentum operators~\cite{Matsubara:2007wj,DAmico:2022ukl}. 
At lowest order in the Taylor expansion around $\pmb{x}$, we have
\begin{align}
\delta_{g,r}(\pmb{x}) & = \delta_g(\pmb{x}) + \partial_i \left( (1+\delta_g(\pmb{x}))  \frac{\pmb{v} \cdot \hat z}{\mathcal{H}} \hat z \right) + \dots \nonumber \\
& = \delta_g(\pmb{x}) + \frac{\hat z^i \hat z^j}{\mathcal{H} \bar \rho_m} \partial_i p^j + \dots \ , \label{eq:delta_g_r_exp}
\end{align}
where we remind that the momentum reads
\begin{equation}\label{eq:momentum}
\pmb{p} = \bar \rho_m (1+\delta) \, \pmb{v} \ .
\end{equation}
The final structure is similar to~\eqref{eq:delta_g} but now with terms that are explicitly depending on the line-of-sight direction, 
\begin{equation}\label{eq:delta_g_r}
\delta_{g,r}(\pmb{k}, t) = \sum_j \mu^{\nu_j} b^\mathcal{G}_j(t) \mathcal{D}_j(t) \mathbb{C}_j(\pmb{k}, t) \ ,
\end{equation}
where $\mu = \hat k \cdot \hat z$, $\nu$'s are some even integer powers, and we use a different index $j$ compared to $i$ in~\eqref{eq:delta_g} to stress that the terms have been shuffled and redefined. 
In particular, $b^\mathcal{G}_j$ symbolically represents that sometimes it is not a free bias coefficient but a calculable time function when $\mathbb{C}_j$ stems from a product of momentum operators only. 
This means that thanks to redshift-space distortions, the (partial) degeneracies of the time-dependent growth functions with the galaxy biases $b_i$ are further broken, thus allowing in principle for a measurements of their dependence --- in particular on $w(t)$. 

\begin{figure}[!h]
    \centering
    \includegraphics[width=0.68\textwidth]{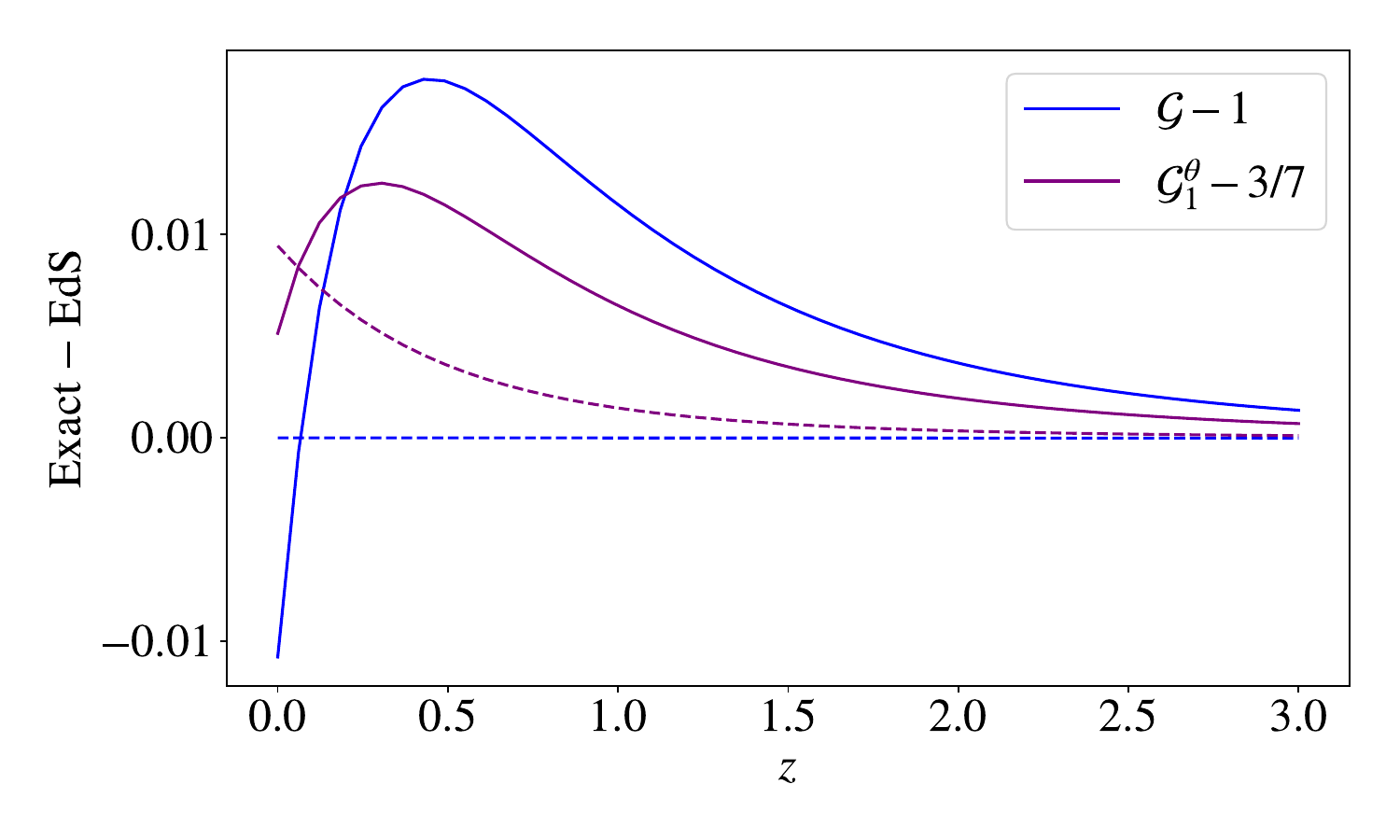}
    \caption{ Relative difference of the nonlinear time functions $\mG$ and $\mG_1^\theta$ between their exact values and their EdS approximations (using the best fit values found in this work in $w_0w_a$CDM). 
    The continuous and dashed lines are for $c_s^2 \rightarrow 0$ and $c_s^2 \rightarrow 1$, respectively. }\label{fig:G}
\end{figure}

\paragraph{Modified time dependence}
To illustrate a nontrivial modification of the time dependence in the presence of dark energy, let us inspect a few contributions to $\delta_g$ or $\theta_g$ from~\eqref{eq:delta_g_expand}~and~\eqref{eq:theta_g_expand}. 
For example, the operator $\mathbb{C}_{\delta,1}^{(2)}$ comes multiplied by a time function $\mathcal{G}$ defined as
\begin{equation}\label{eq:G}
\mathcal{G}(a) = \frac{1}{D(a)} \int^{a}  \frac{da' }{a'} \frac{f(a')}{C(a')} \ ,
\end{equation}
where $f \equiv \frac{d\ln D}{d\ln a}$ is the growth rate. 
It is relevant only for the clustering case, as when $C(a) \equiv 1$, we have $\mathcal{G}(a) \equiv 1$. 
Another example comes from the contribution $\frac{7}{2}(1-\mathcal{G}_1^\theta)\mathbb{C}_{\delta,2}^{(2)}$ to $\theta_g$, relevant for both smooth and clustering quintessence. 
Because $\mathbb{C}_{\delta,1}^{(2)}$ and $\mathbb{C}_{\delta,2}^{(2)}$ are second order, we expect $\mathcal{G}$ and $\mathcal{G}_1^\theta$ to enter in the loop of the power spectrum (through the $22$-diagram) and linearly in the tree-level bispectrum in redshift space. 
In fig.~\ref{fig:G}, we plot $\mathcal{G}$ and $\mathcal{G}_1^\theta$ as function of the redshift for $w(z)$ given by the best fit found with the $(w_0, w_a)$-parametrisation in this work. 
For relative data uncertainties of $\sim 10\%$, we expect this rescaling of $\sim 1\%$ (with respect to the EdS case)\footnote{For clustering quintessence, when referring to the EdS approximation we further assume $C=1$ beyond linear order~\cite{DAmico:2022ukl}, yielding $\mathcal{G}=1$. } on the leading redshift-space distortions in the bispectrum to be small. 

\paragraph{Counterterms}
So far, we have only focus on the perturbation theory contributions. 
There are various counterterms to consider. 

\begin{itemize}
\item Counterterms that arise from the expansion of $\tau^{ij}$ in the r.h.s. of the smoothed Euler equation~\eqref{eq:EFTEuler}. 
This has nothing but the same structure written for the galaxy density~\eqref{eq:delta_g_sym} but equipped with $\delta_{ij}$, where we further allow for Galilean tensors under rotations of the $i$ and $j$ indices. 
The scale controlling this expansion, $k_{\rm NL}^{-1}$, corresponds roughly to when the variance of the unsmoothed field becomes nonlinear, \textit{i.e.}, $\sim \mathcal{O}(1)$. 
There are two types of contributions that arise from the expansion of $\tau^{ij}$: either terms constructed from the tidal tensor $\partial_i \partial_j \Phi$, or stochastic terms. 
We discuss the latter below and we focus now on the former, that we call response terms. 
They start contributing at third order in perturbations, since it is $\partial_i\partial_j \tau^{ij}$ that sources the Euler equation~\eqref{eq:EFTEuler}. 
The leading response $\sim \partial^2 \delta^{(1)} / k_{\rm NL}^{2}$ is degenerate with the leading spatial derivative correction $\sim  \partial^2 \delta^{(1)} / k_{\rm M}^{2}$, where $k_{\rm M}^{-1}$ represents the spatial extension of the galaxy, roughly the size of the host halo. 
We comment on the next-to-leading response terms below. 

In presence of modified gravity, we have to further consider the Galilean-invariant tensors $\partial_i \partial_j \Psi$ and $\partial_i \partial_j \pi$ when expanding $\tau^{ij}$ to account all possible feedbacks of short-wavelength fluctuations on the long ones~\cite{Cusin:2017wjg}. 
However, in the limiting cases that we study, those are constrained such that they do not lead to new terms in the EFT expansions other than the ones generated by $\partial_i \partial_j \Phi$ (see~\eqref{eq:QSLpi}~and~\eqref{eq:comoving}). 

\item The expansion of the density field in redshift space~\eqref{eq:delta_g_r_exp} involves (products of) the momentum. 
Because the momentum is a composite operator~\eqref{eq:momentum}, it has to be renormalised by adding suitable counterterms to remove its UV-sensitivity. 
Those are precisely provided by the expansion of $\tau^{ij}$ through the Euler equation~\eqref{eq:EFTEuler}. 
Importantly, some of the next-to-leading response terms entering the momentum renormalisation are not present in the bias expansion~\eqref{eq:delta_g} up to forth order~\cite{DAmico:2020tty}. 
This is because the spatial Green's function from the Poisson equation is nonlocal, \textit{i.e.}, $\Phi \propto \partial^{-2}\delta$, and is not cancelled when considering the traceless part of $\partial_i \partial_j \Phi$ contributing to the vector part of the momentum $p_V^i = \epsilon^{ijk}\partial_j p^k$ (related to the vorticity).\footnote{Here $\epsilon^{ijk}$ is the three-dimensional totally antisymmetric Levi-Civita symbol.}
As a consequence, appear in the density in redshift space at forth order non-locally-contributing counterterms~\cite{DAmico:2020tty}. 
Products involving momentum operators in the redshift-space expansion are renormalised by a new scale $k_{\rm R}^{-1}$~\cite{Lewandowski:2015ziq,DAmico:2021ymi}. 
All response terms are found to be necessary and sufficient to renormalise the one-loop contributions in the power spectrum and bispectrum involving $\delta_g^{(3)}$ or $\delta_g^{(4)}$ with a self-loop~\cite{DAmico:2020tty}. 

\item Stochastic contributions arising in the expansions of both the galaxy density field and the stress tensor. 
Those are quantities $\epsilon$ and $\epsilon_{ij}$, whose correlation functions can be written as an expansion in powers of $\partial_i$ of terms invariant by rotations~\cite{Carrasco:2013mua,DAmico:2020tty}.  
In the expansion of the galaxy density, their size is controlled by $1/\bar n_g$, where $\bar n_g$ is the average number density of observed galaxies, while in  the stress tensor, their size is controlled by $1/\bar n_{\rm NL} \sim k_{\rm NL}^3$, the occupation number density in regions of size $\sim k_{\rm NL}^{-1}$. 
Analogously as the response terms, there are non-locally contributing stochastic terms at next-to-leading order~\cite{DAmico:2020tty}. 
\end{itemize}

\begin{figure}[!ht]
    \centering
    \begin{minipage}{0.49\linewidth}
    \centering
    \includegraphics[width=.99\linewidth]{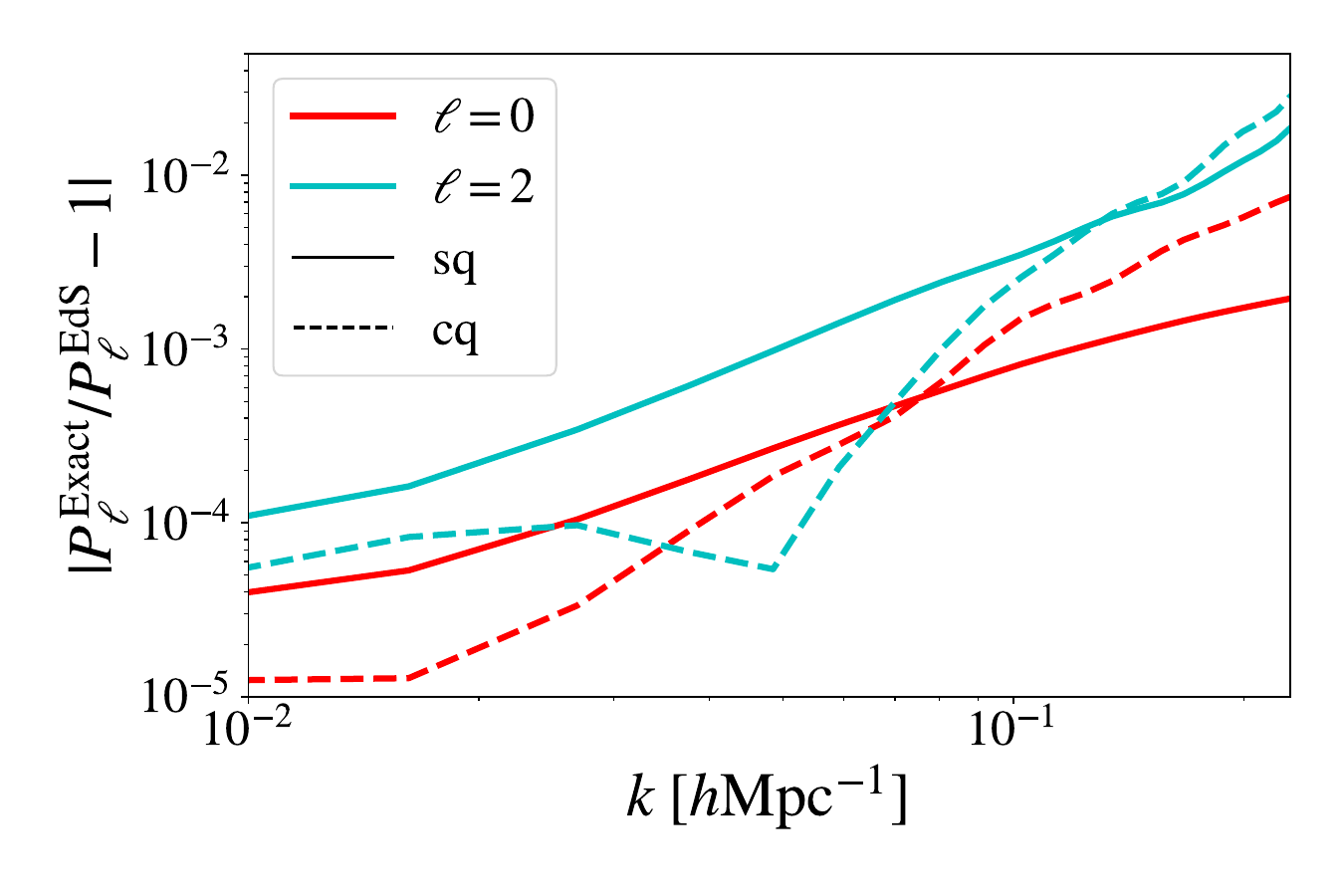}
    \includegraphics[width=.99\linewidth]{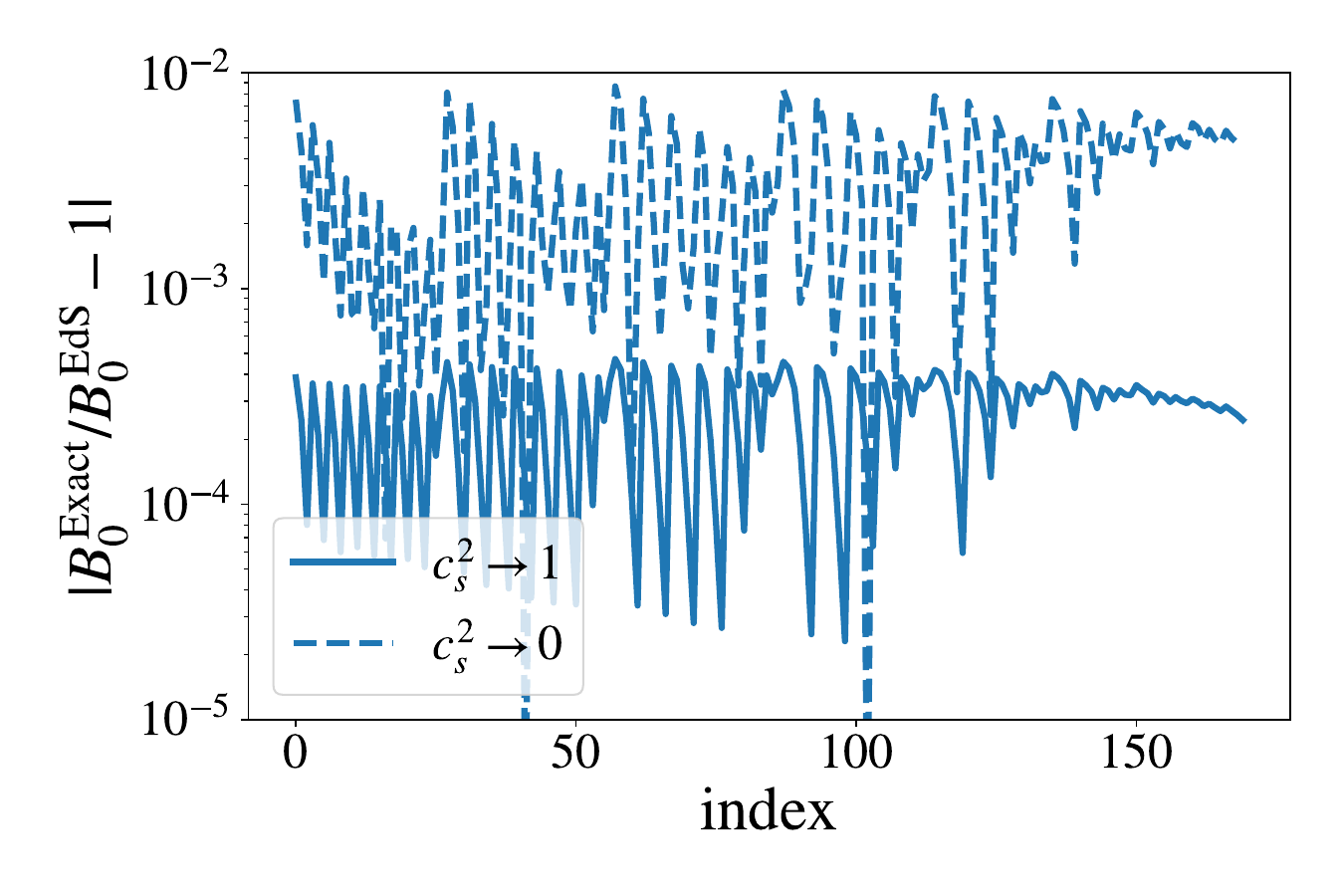}
    \end{minipage}
    \begin{minipage}{0.49\linewidth}
    \centering
    \includegraphics[width=.99\linewidth]{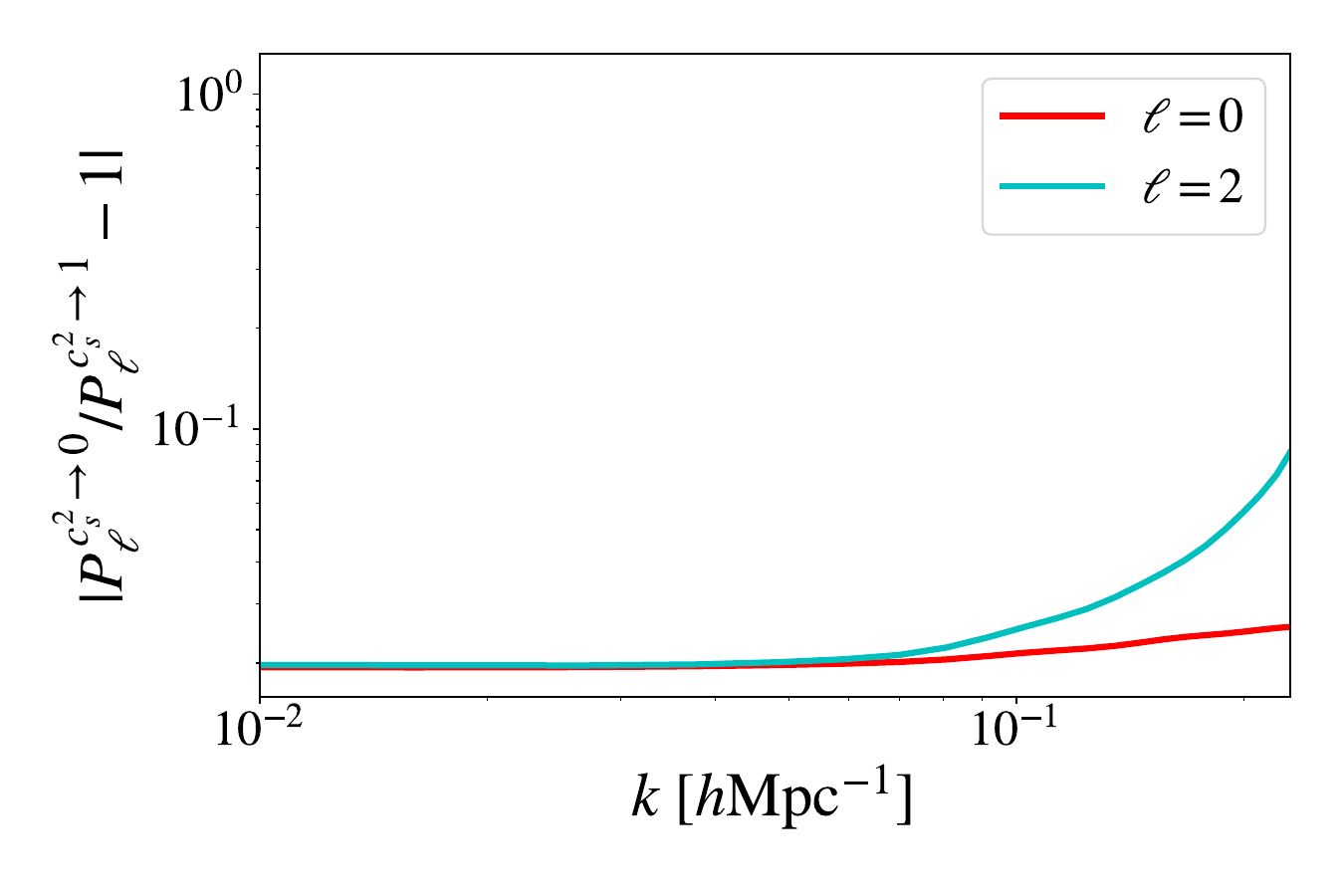}
    \includegraphics[width=.99\linewidth]{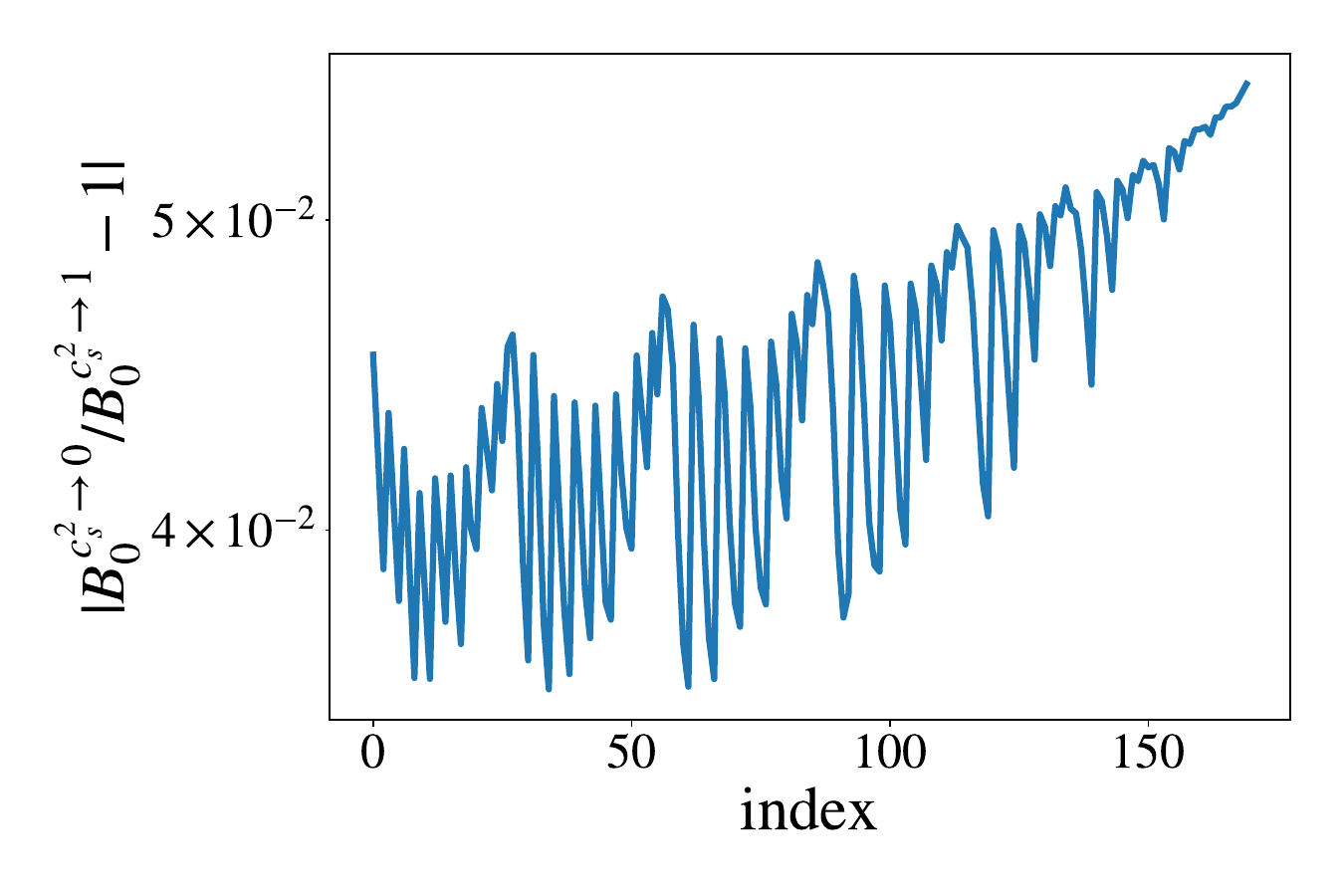}
    \end{minipage}
    \caption{Relative difference on the power spectrum multipoles (\textit{upper panels}) and bispectrum monopole (\textit{bottom panels}) at fixed cosmology ($\sim w_0=-0.8,w_a=-0.6$) {and fixed EFT parameters, closed to the best fit obtained in the analysis,} for various setups considered in this work. 
    The $x$-axes in the bottom panel correspond to the triangle bin indices. 
    \textit{Left panels}: Exact-time dependence vs. EdS approximation for $c_s^2 \rightarrow 1$ (solid line) or $c_s^2 \rightarrow 0$ (dashed line). \textit{Right panels}: $c_s^2 \rightarrow 1$ vs. $c_s^2 \rightarrow 0$. 
    \label{fig:bestfit_ratio}
    }
\end{figure}

\paragraph{Observables}
We consider the two- and three-point function in Fourier space, defined as
\begin{align}
\braket{\delta_{g,r}(\pmb{k}) \delta_{g,r}(\pmb{k}')} & = (2\pi)^3\delta_D(\pmb{k} + \pmb{k}') P(k, \mu)  \ ,  \\
\braket{\delta_{g,r}(\pmb{k_1}) \delta_{g,r}(\pmb{k_2})\delta_{g,r}(\pmb{k_3}) } & = (2\pi)^3\delta_D(\pmb{k_1} + \pmb{k_2} + \pmb{k_3}) B(k_1, k_2, k_3, \mu, \phi)   \ .
\end{align}
The Dirac delta-distributions reflect translation invariance. 
Rotation invariance, partially broken in redshift space, implies that the power spectrum $P$ is a function of the norm $k$ and the cosine $\mu = \hat k \cdot \hat z$.  
As for the bispectrum $B$, we can express it with the three sides of the triangle formed by $(\pmb{k_1}, \pmb{k_2}, \pmb{k_3})$, the cosine $\mu \equiv \mu_1 = \hat k_1 \cdot \hat z$, and the azimuthal angle $\phi$~\cite{Scoccimarro:1999ed}. 
At the one-loop precision, the power spectrum $P$ and bispectrum $B$ of galaxies in redshift space read~\cite{DAmico:2020tty}
\begin{align}
P & = P_{11} + (P_{13} +  P_{13}^{ct} ) + ( P_{22}   + P_{22}^{\epsilon} )  \ , \\
B & = B_{211} + ( B_{321}^{(II)} + B_{321}^{(II),ct}  )  + ( B_{411} + B_{411}^{g,r,ct} )  + ( B_{222} + B_{222}^{g,r,\epsilon} ) + ( B_{321}^{(I)} +  B_{321}^{(I),\epsilon} ) \ ,
\end{align}
where in each loop diagram the UV-sensitivity is absorbed by the appropriate counterterms mentioned above. 
{In app.~\ref{app:loop}, we show the size of each of the loop contributions to the bispectrum, illustrating their relevance to reach smaller scales in a controlled way. }
For our data analysis, we consider the multipoles ($\ell = 0, 2$) of the power spectrum and the monopole of the bispectrum, 
\begin{align}
P_\ell(k) = & \frac{2\ell+1}{2} \int_{-1}^{+1} d\mu \ P(k, \mu) \mathcal{L}_\ell(\mu) \ ,  \\
B_0(k_1, k_2, k_3) & = \frac{1}{4\pi} \int_{-1}^{+1} d\mu  \int_0^{2\pi} d\phi \  B(k_1, k_2, k_3, \mu, \phi) \ , 
\end{align}
where $\mathcal{L}_\ell$ is the Legendre multipole of order $\ell$. 
In fig.~\ref{fig:bestfit_ratio}, we show the relative difference in the power spectrum and bispectrum when accounting for the modified time dependence in the presence of dark energy. 
Compared to current data uncertainties, the EdS approximation appears likely to be under control for the best fit of $w(z)$ obtained in this work --- we nevertheless explicitly check its impact on the constraints in sec.~\ref{sec:results}.
In contrast, the difference between the $c_s^2 \rightarrow 0$ and $c_s^2 \rightarrow 1$ limits is significant and can impact the constraints, as we find in sec.~\ref{sec:results}.

\paragraph{IR-resummation}
The variance of long-wavelength displacements $\pmb{s_\ell}$ is close to unity for scales around the BAO peak, making the expansion of the density field converge slowly in this parameter~\cite{Porto:2013qua,Senatore:2014via}. 
To cure this limitation, one can displace the position of the galaxies with the linear displacement, which captures most of the large-scale bulk flow~\cite{Senatore:2017pbn}. 
The linear displacement is
\begin{equation}\label{eq:lin_disp}
s_i^{(1)}(\pmb{x}, a) = \int^{\ln a}\frac{d\ln a'}{\mathcal{H}(a)} \,v_i^{(1)}(\pmb{x}, a') = - \mathcal{G}(a) \frac{\partial_i}{\partial^2}  \delta^{(1)} \ , 
\end{equation}
where in the second equality we have used $v_i^{(1)} = \frac{\partial_i}{\partial^2} \theta^{(1)} = -\frac{\mathcal{H} f}{C}\frac{\partial_i}{\partial^2} \delta^{(1)}$ following the linearised continuity equation~\eqref{eq:EFTcontinuity}, together with the definition~\eqref{eq:G}. 
The power spectrum and bispectrum are therefore IR-resummed as usual following~\cite{Lewandowski:2015ziq,DAmico:2020kxu,DAmico:2022osl} but with the replacement $P_{\rm lin} \rightarrow \mathcal{G}^2 P_{\rm lin}$ in the exponent of the damping~\cite{DAmico:2020tty}. 
This is relevant only in clustering quintessence as we remind that $\mathcal{G} = 1$ if $C=1$. 
We find that setting $\mathcal{G}$ in eq.~\eqref{eq:lin_disp} to its values associated with the best fit found in this work in $w_0w_a$CDM leads to an extra relative damping of $\sim 0.1\%$. 
Thus, in practice, the modification of the BAO features from the presence of clustering dark energy is negligible. 


\section{Inference setup}\label{sec:setup}

We aim to find an appropriate way to bridge candidate models of dark energy that can produce departure from $w=-1$, with what the data can tell us. 
Because there is no consensus on what dark energy is, we make use of several parametric forms for $w(t)$ in our search. 
The various forms for the equation of state of dark energy $w(t)$  considered in this work are presented in sec.~\ref{sec:wform}. 
In sec.~\ref{sec:dataset}, we then describe the datasets we use to constrain them. 
In sec.~\ref{sec:eft_lkl} we give some extra details on the likelihood and prior used to analyse the power spectrum and bispectrum of galaxies.

\subsection{Probing evolution in the dark}\label{sec:wform}
To probe potential evolution in dark energy equation of state, we consider the following parametrisations for $w(a)$: 
\begin{itemize}
\item $w(a) = w_0$: How on average is the cosmic acceleration not driven by $\Lambda$, \textit{i.e.}, is $w\neq -1$?
From a UV perspective, this raises the following question: if we give up $\Lambda$, how flat must the scalar field potential remain throughout cosmic history?
This naturally leads to the question of the extent to which cosmological data constrain $w$ to be close to (or far from) $-1$.
Pragmatically, we want to get a sense of the overall data sensitivity on $w$. 
$w_0$CDM was explored in ref.~\cite{Spaar:2023his} in light of the galaxy bispectrum, yielding $w_0 = -0.975 \pm 0.019$ at $68\%$CL using a similar dataset as the baseline considered in this work (see below). Given the high sensitivity to $w$, this motivates us to explore more general parametric forms. 
\item $w(a) = w_0 + w_a(1-a)$: Commonly referred to as the Chevalier-Polarsky-Linder (CPL) parametrisation~\cite{Chevallier:2000qy,Linder:2002et}, it can be thought as the first (linear) correction in a Taylor expansion of $w(a)$ around the scale factor today $a_0 \equiv 1$. 
$w_0w_a$CDM is explored in sec.~\ref{sec:results} in light of various dataset combinations and analysis setups. 
To help visualise and understand our results, we also use an equivalent, more interpretable, parametric form, $w_\star = w_0 + w_a (1-a_\star)$, where $a_\star$ is the pivot epoch at which $w$ is best constrained throughout cosmic history (see sec.~\ref{sec:histories}).
\item Beyond-CPL parametrisation: Is the data sensitive to more subtle time variations? 
We investigate this question in details in sec.~\ref{sec:histories} and app.~\ref{app:recon}, using a piece-wise $w(z)$ in redshift bins, arbitrarily generalisable by increasing the number of bins upon saturating the $\chi_{\rm min}^2$.
If no particular model of dark energy is assumed, a general flexible parametric form allows to assess the consistency with $\Lambda$ in a data-driven way. 
\end{itemize}

\subsection{Datasets and inference framework}\label{sec:dataset}

In our cosmological inference, we consider different combinations of the following datasets:
\begin{itemize}
    \item \textbf{Planck:} The temperature, polarization and lensing power spectrum data from \textit{Planck} PR3.
    We use the \texttt{Commander} likelihood for low-$l$ TT, the \texttt{Simall} likelihood for EE, and the \texttt{Plik} likelihood for high-$l$ TT, EE, TE \cite{Planck:2019nip}, as well as the reconstructed gravitational lensing potential power spectrum from \cite{Planck:2018lbu}.\footnote{We checked that the difference in the lensing likelihood compared to that used in the joint analysis of DESI Y1~\cite{DESI:2024mwx} results in a negligible impact on the posteriors.}

    \item \textbf{EFTBOSS:} The full-shape analysis of the power spectrum and bispectrum of BOSS Luminous Red Galaxies (LRG) from the EFTofLSS at one loop. 
    The SDSS-III BOSS DR12 galaxy sample data are described in ref.~\cite{BOSS:2016hvq,BOSS:2016wmc}. 
    The power spectrum and bispectrum measurements, obtained in ref.~\cite{DAmico:2022osl}, {using the standard FKP-like estimators~\cite{Feldman:1993ky,Yamamoto:2005dz,Hand:2017irw,Bianchi:2015oia,1312.4611,Scoccimarro:2015bla,BOSS:2015npt,Gil-Marin:2016wya} by \texttt{Rustico} code\footnote{\url{https://github.com/hectorgil/Rustico}}}~\cite{BOSS:2015npt,Gil-Marin:2016wya}, are from BOSS catalogs DR12 (v5) combined CMASS-LOWZ\footnote{Publicly available at \url{https://data.sdss.org/sas/dr12/boss/lss/}.}~\cite{BOSS:2015ewx}, and are divided in two redshift bins (and four sky-cuts): low-$z$, $0.2<z<0.43 \  (z_{\rm eff}=0.32)$, and high-$z$, $0.43<z<0.7  \ (z_{\rm eff}=0.57)$, with north and south galactic skies for each. The covariance, including the correlation between power spectrum and bispectrum, is measured through 2048 Patchy mocks~\cite{Kitaura:2015uqa}, while the window function is measured from \texttt{fkpwin}\footnote{\url{https://github.com/pierrexyz/fkpwin}}~\cite{Beutler:2018vpe}. 
    More details are provided in sec.~\ref{sec:eft_lkl}, {and app.~\ref{app:sys} provides an extensive assessment of our modelling of observational systematics in the bispectrum. }
    
    \item \textbf{ext-BAO:} The angle-averaged distance ratio $D_V/r_s$ extracted from baryon acoustic oscillations (BAO) at $z=0.106$ from 6dFGS \cite{Beutler:2011hx} and at $z=0.15$ from SDSS DR7 \cite{Ross:2014qpa}, as well as the joint constraints from eBOSS DR14 Ly-$\alpha$ absorption auto-correlation at $z = 2.34$ and cross-correlation with quasars at $z = 2.35$ \cite{eBOSS:2019ytm,Cuceu:2019for}.
    
    \item \textbf{DESI-BAO:} The DESI Y1 BAO data~\cite{DESI:2024uvr}, which include the distance ratios $D_M/r_s$ and $D_H/r_s$ at $z=0.51$ (LRG1), 0.706 (LRG2), 0.930 (LRG3+ELG1), 1.317 (ELG2), and 2.33 (Ly-$\alpha$ QSO), as well as $D_V/r_s$ at $z=0.295$ (BGS), and 1.491 (QSO). 
    
    \item \textbf{PanPlus:} Pantheon+ catalog of uncalibrated luminosity distance of SNeIa in the range $0.01 < z < 2.26$ \cite{Brout:2022vxf}.

    \item \textbf{Union3:} Union3 catalog of uncalibrated luminosity distance of SNeIa in the range $0.05 < z < 2.26$ \cite{Rubin:2023ovl}.

    \item \textbf{DESY5:}  DES year 5 catalog of uncalibrated luminosity distance of SNeIa in the range $0.10 < z < 1.13$, combined with an external sample of SNeIa at low redshifts $0.024<z<0.10$~\cite{DES:2024tys}.\footnote{For Union3 and DESY5, we have adapted the public \texttt{Cobaya}~\cite{Torrado:2020dgo} likelihoods to \texttt{Montepython-v3}.
    }

\end{itemize}

When inferring cosmological parameters from various combinations of the datasets above, we simply add their likelihoods together, marginalising over the nuisance parameters upon sampling. 
In particular, and in line with the literature, we neglect the small correlations between \textit{Planck} lensing and the integrated Sachs-Wolfe effects with galaxy clustering data. 

\paragraph{DESI-BAO + EFTBOSS/$r_s^{\rm marg}$}
DESI Y1 and SDSS/BOSS share a substantial overlap in their observed sky coverage. 
Specifically, $27\%$ of DESI BGS and LRG1 galaxies, along with $9\%$ of DESI LRG2 galaxies, were already observed by BOSS~\cite{DESI:2024aax}. 
To mitigate potential correlations when combining these two datasets, the information from the sound horizon ($r_s$) is taken from DESI-BAO while being marginalised in EFTBOSS full-shape analysis, following the methodology of refs.~\cite{Farren:2021grl,Smith:2022iax}. 
Explicitly, when jointly fitting these datasets, we further marginalise over a nuisance parameter, $\alpha_{r_s}$. 
This parameter modifies the EFT predictions for the BOSS power spectrum and bispectrum by scaling the BAO wiggles through the (input) linear matter power spectrum as
\begin{equation}
P_{\rm lin}(k) \rightarrow P_{\rm nw}(k) + P_{\rm w}(\alpha_{r_s} k) \ ,
\end{equation}
thus effectively marginalising over the sound horizon information. 
Here, $P_{\rm nw}$ and $P_{\rm w}$ represent the no-wiggle and wiggle components of the linear power spectrum, split following the method of ref.~\cite{1003.3999}. 
In app.~\ref{app:bossxdesi}, we provide estimates of the residual correlation between DESI-BAO and EFTBOSS/$r_s^{\rm marg}$ {(since the our sound-horizon marginalisation does not remove all overlapping information)} and their impact on the final constraints, {which we estimate to be small within our approximate treatment.} 
{We caution, however, that our results obtained by combining DESI and BOSS would benefit a confirmation from a full likelihood analysis with a joint covariance matrix that properly accounts for their correlations.}

\paragraph{Inference framework}
To sample posteriors from the likelihoods of the datasets above, we run Markov chain Monte Carlo (MCMC) chains through the Metropolis-Hasting algorithm implemented in \texttt{Montepython-v3}\footnote{\url{https://github.com/brinckmann/montepython_public}} \cite{Brinckmann:2018cvx,Audren:2012wb}, interfaced with the Boltzmann code \texttt{class}\footnote{\url{http://class-code.net/}} \cite{Lesgourgues:2011re, Blas:2011rf} for background and linear cosmological quantities, and \texttt{PyBird} for computing the nonlinear galaxy power spectrum and bispectrum\footnote{\url{http://github.com/pierrexyz/pybird/}} \cite{DAmico:2020kxu}. 
When marginalising over EFTofDE parameters described in sec.~\ref{sec:EFTofDE}, we use a modified version of \texttt{hi\_class}\footnote{\url{https://miguelzuma.github.io/hi_class_public/}}~\cite{Zumalacarregui:2016pph,Bellini:2019syt} to compute the linear cosmological observables in the EFTofDE, while using \texttt{PyBird}\footnote{Adapted from \url{https://github.com/billwright93/pybird}~\cite{Piga:2022mge}} with modified nonlinear time functions (see sec.~\ref{sec:EFTofLSS}).

For all analyses performed in this work, we impose large flat priors on the $\Lambda$CDM cosmological parameters $\{ \omega_b,\omega_{\rm cdm}, h, \ln (10^{10}A_s), n_s, \tau_{reio} \}$, as well as on the dark energy equation-of-state parameters described in sec.~\ref{sec:wform}.
Following \textit{Planck} convention for neutrinos, we consider two massless neutrinos and one massive neutrino, with $m_\nu = 0.06 e$V.
We consider that our chains have converged when the Gelman-Rubin criterion $R-1<0.05$. 
We use simulated annealing from \texttt{Procoli}\footnote{\url{https://github.com/tkarwal/procoli}} \cite{Karwal:2024qpt} to find the maximum a posteriori (that we also refer as the best
fit). 
Finaly, marginalised posteriors and credible intervals are obtained using~\texttt{Getdist}\footnote{\url{https://getdist.readthedocs.io/en/latest/}} \cite{Lewis:2019xzd}.

\subsection{EFT likelihood} \label{sec:eft_lkl}
We analyse the full-shape power spectrum ($P$) and bispectrum ($B$) of BOSS galaxies in redshift space based on the one-loop predictions from the EFTofLSS presented in sec.~\ref{sec:EFTofLSS}.  
Additional modelling required to make contact with observations --- such as Alcock-Paczynski distortions, window functions, and binning --- have been thoroughly tested and detailed in refs.~\cite{DAmico:2019fhj,DAmico:2022osl}.

\paragraph{Likelihood, data specification, and covariance}
The likelihood used in this work follows the methodology outlined in ref.~\cite{DAmico:2022osl}. 
The BOSS survey data are divided into two redshift bins:
low-$z$ ($0.2<z<0.43 \  (z_{\rm eff}=0.32)$) and high-$z$ ($0.43<z<0.7  \ (z_{\rm eff}=0.57)$),
each further split into north and south galactic cuts, resulting in four independent sky regions~\cite{BOSS:2016hvq,BOSS:2015npt}.
For each region, we analyse the monopole and quadrupole of the power spectrum, and the monopole of the bispectrum, {measured in ref.~\cite{DAmico:2022osl} using standard FKP-like estimators~\cite{Feldman:1993ky,Yamamoto:2005dz,1312.4611,Scoccimarro:2015bla,BOSS:2015npt,Gil-Marin:2016wya}, making use of the \texttt{Rustico} code\footnote{\url{https://github.com/hectorgil/Rustico}} implementing the estimator developped in ref.~\cite{BOSS:2015npt,Gil-Marin:2016wya}}. 
The chosen scale ranges are $k_{\rm min}=0.01 \hinvMpc$ and $k_{\rm max} = 0.20/0.23 \hinvMpc$ for low-$z$/high-$z$ skies, based on prior estimates of the theoretical error relative to BOSS uncertainties (see below).
{For each sky region, we construct a data vector $\mathcal{D}_\alpha$
containing the power spectrum multipoles and bispectrum measurements. The
power spectrum part contains a total of 36 (40) $k$-bins across $P_0$ and
$P_2$ for the low-$z$ (high-$z$) sample, while the bispectrum part contains
115 (150) triangle bins. The resulting data-vector dimensions are therefore
151 and 190 per sky region for the low- and high-redshift samples,
respectively. The combined BOSS likelihood therefore contains
$682$ data points across the four sky cuts.
The covariance matrix $C_{\alpha\beta}$ is estimated from 2048
Patchy mocks~\cite{Kitaura:2015uqa}, with the inverse covariance corrected using the Hartlap factor~\cite{Hartlap:2006kj}, {corresponding to values of $0.93$ (low-$z$) $0.91$ (high-$z$), respectively}.
Given the large number of mocks relative to the data-vector dimensions, the residual effect of finite-mock covariance noise on the parameter uncertainties is expected to be subdominant. A detailed validation using an analytical covariance matrix is left for future work.}
At each likelihood evaluation, the theory model vector $m_\alpha(\theta)$ is computed according to the predictions of sec.~\ref{sec:EFTofLSS}, where $\theta$ contains the cosmological and EFTofLSS parameters. 
{$m_\alpha(\theta)$ is further dressed with additional modelling for observational effects such as the Alcock-Paszinski effect, window functions, binning, and fibre collisions, as described in ref.~\cite{DAmico:2022osl}, with their accuracy re-assessed for the current analysis setup in app.~\ref{app:sys}. }
The likelihood function $\mathcal{L}$ for each sky is given by:  
\begin{equation}\label{eq:lkl}
-2 \ln \mathcal{L}(\mathcal{D}|\theta) = \sum_{\alpha, \beta} (\mathcal{D}_\alpha - m_\alpha(\theta)) C^{-1}_{\alpha \beta} (\mathcal{D}_\beta - m_\beta(\theta)) \ .
\end{equation}  

\paragraph{Prior and marginalisation}
Details on priors and analytical marginalisation can be found in ref.~\cite{DAmico:2022osl} {and tab.~\ref{tab:prior}}.  
For the predictions to remain within the validity of perturbation theory, EFT parameters are expected to be $\mathcal{O}(b_1)$~\cite{DAmico:2019fhj,DAmico:2022osl}. 
Thus, they are marginalized over using Gaussian priors centred at zero with a width of $\sim2$, except for $b_1$, which is constrained to be positive via an equivalent log-normal prior.  
To account for variations across different sky regions, separate sets of EFT parameters are assigned for each region, incorporating correlations due to redshift evolution and north-south observational differences. 
Specifically, EFT parameters are expected to vary by at most $10\%$ between north and south regions within the same redshift bin, and by $20\%$ between the low-$z$ and high-$z$ bins within the same hemisphere. This is enforced through a multivariate Gaussian prior. 
{This correlation is motivated from the fact that the typical redshift evolution of galaxy biases follows roughly $b(a) \sim D(a)^\gamma \sim a^\gamma$ where $\gamma$ is a small, order-one power. 
For $-2 \lesssim \gamma \lesssim 2$, the corresponding change in the bias values between $z_{\rm low} = 0.43$ to $z_{\rm high} = 0.57$ is at most $20\%$. 
In fact, our bound on the difference between the EFT parameter values of low-$z$ and high-$z$ is consistent with the assumption that their values are slowly varying with the width of the {redshift} bin, allowing to assume time-independent biases within a given redshift bin. 
If the biases vary too rapidly with redshift, so that the difference between their low-$z$ and high-$z$ values exceeds 
$20\%$, then the constant-bias approximation within a redshift bin would break down, introducing a non-negligible error~\cite{Zhang:2021uyp}.
As for the difference between the north and south atmosphere, since the target galaxies are the same, we do not expect that difference in their selection function can lead to a difference of more than $10\%$ in the values of the EFT parameters. 
}

Since our primary focus is on cosmological parameters, we analytically marginalise over the EFT parameters that enter linearly in the predictions (and quadratically in the likelihood) using Gaussian integral properties. 
This significantly reduces computational cost, as only three EFT parameters --- $b_1, b_2,$ and $b_5$ (following the notation of ref.~\cite{DAmico:2022ukl}) --- need to be explicitly sampled, out of a total of $41$ per sky. 
For the characteristic scales governing the EFTofLSS expansions described in sec.~\ref{sec:EFTofLSS}, we set $k_{\rm NL} = k_{\rm M} = 0.7 \ \hinvMpc$ and $k_{\rm R} = 0.25 \ \hinvMpc$, while adopting a mean galaxy number density of $\bar n_g = 4 \cdot 10^{-4} (\textrm{Mpc}/h)^{3}$. 
In sec.~\ref{sec:checks}, we verify that our prior choice has a negligible impact on the cosmological constraints by increasing the prior width by a factor of 2 and by checking the distance of the posterior mean with the maximum a posteriori estimate. 
{A similar test on our prior correlating BOSS galaxy biases is presented in app.~\ref{app:supp_material}. }

\begin{table}[] \centering
\begin{tabular}{|cc|}\hline
Parameter & Prior               \\ \hline
$\omega_{\rm b},\omega_{\rm cdm}, h, \ln (10^{10}A_s), n_s, \tau_{\rm reio},w_0,w_a $   & Flat  \\ \hline
$b_1$     & Lognormal(0.8, 0.8944) \\
$b_i$  ,  $ i = 2 , \dots , 11$    & $\mathcal{N}(0, 2)$ \\
$c_1^h$, $c_1^\pi$, $c_1^{\pi v}$	& $\mathcal{N}(0, 4)$ \\
$c^{\pi v}_3$ , $c^h_2$ , $c^h_3$, $c^h_4$, $c^h_5$, $c^{\pi}_5$, $c^{\pi v}_2$, $c^{\pi v}_4$, $c^{\pi v}_5$	& $\mathcal{N}(0, 2)$ \\
$c^{\rm St}_2$	& $\mathcal{N}(0, 4)$ \\
$(2/3) f c^{\rm St}_3$, $ c^{\rm St}_4$, $c^{\rm St}_5$,  $2 c^{\rm St}_9$, $- c^{\rm St}_7 - c^{\rm St}_9$, $ - c^{\rm St}_8 - 2 c^{\rm St}_9$, $2 c^{\rm St}_{12}$, $c^{\rm St}_{11}$, $c^{\rm St}_{10}$, $c^{\rm St}_{13}$    &  $\mathcal{N}(0, 2)$   \\
$c^{\rm St}_1$	& $\mathcal{N}(0, 0.2)$ \\
$c^{(222)}_1$	& $\mathcal{N}(0, 0.4)$ \\
$2c^{\rm St}_6$	& $\mathcal{N}(0, 1)$ \\
\hline
\end{tabular}
\caption{\label{tab:prior} {Priors on cosmological and EFT parameters chosen for all analyses performed in this work.
Lognormal($\mu,\sigma$) and $\mathcal{N}(\mu,\sigma)$ denote a log-normal and Gaussian prior of mean $\mu$ and standard deviation $\sigma$, respectively. 
We further impose a correlation between the parameters assigned to the various skies of BOSS, consistent with conservative expectations on the differences implied by the redshift evolution of the EFT parameters and the BOSS selection functions, see main text for details. 
See ref.~\cite{DAmico:2022gki} for details on the notation used for the EFT parameters. 
}
}
\end{table}

\paragraph{Validations}
Extensive tests of the power spectrum and bispectrum have been performed using high-fidelity simulations with their $k$-reach further assessed using perturbative arguments~\cite{Nishimichi:2020tvu,DAmico:2019fhj, Chen:2020zjt,DAmico:2021ymi,Zhang:2021uyp,DAmico:2022osl}. 
In particular, the BOSS LRG power spectrum likelihood used in this work has been validated against the following $N$-body simulations with various galaxy-to-halo connection models: the BOSS lettered challenge~\cite{DAmico:2019fhj, Colas:2019ret}, the PT blind challenge~\cite{Nishimichi:2020tvu}, with additional tests in extended cosmologies or alternative setups in refs.~\cite{DAmico:2020kxu, DAmico:2020tty, DAmico:2021ymi, Zhang:2021yna, Zhang:2021uyp, Simon:2022lde}.  
The pipeline has also been tested using eBOSS quasars and ELGs against EZmocks~\cite{Simon:2022csv, Zhao:2023ebp}. 
Finally, the $P+B$ likelihood for BOSS has been validated against the Nseries and Patchy mocks in ref.~\cite{DAmico:2022osl}, which incorporate realistic observational effects such as lightcone evolution and sky footprints.  
{In app.~\ref{app:sys}, we validate our whole analysis pipeline against the Nseries $N$-body simulations. }

\section{Cosmological results}\label{sec:results}

In this section, we present our cosmological results for the dark energy equation of state $w(t)$ parametrised by $(w_0, w_a)$, using the theoretical developments presented in sec.~\ref{sec:theory} together with the inference setup laid in sec.~\ref{sec:setup}.
Our main results are presented in sec.~\ref{sec:mainresults} with several combinations of data, before investigating in sec.~\ref{sec:checks} where does the preference for evolving dark energy comes from in our analysis.  
We complete our analysis of $w_0w_a$CDM by releasing the curvature in sec.~\ref{sec:omega_k} or EFTofDE parameters in sec.~\ref{sec:mg}.

\subsection{$w_0w_a$CDM}\label{sec:mainresults}

\begin{table}[!h]
    \centering
    \small
    \begin{tabular}{|lcccc|}
        \hline
        Smooth quintessence ($c_s^2\rightarrow 1$) & $w_0$ & $w_a$ & $p > \Lambda$ & $\Delta \chi^2$\\
        \hline
        \textit{Planck} + PanPlus + ext-BAO & $-0.856\pm 0.085$ & $-0.60^{+0.43}_{-0.37}$ & $1.1 \sigma$ & $-2.5$\\
        \textit{Planck} + PanPlus + DESI-BAO & $-0.821\pm 0.063$ & $-0.77\pm 0.27$ & $2.5 \sigma$ & $-8.9$\\
        \textit{Planck} + ext-BAO + EFTBOSS & $-0.724^{+0.087}_{-0.11}$ & $-0.91^{+0.42}_{-0.33}$ & $0.0 \sigma$ & $0.0$ \\
        \textit{Planck} + PanPlus + ext-BAO + EFTBOSS & $-0.844\pm 0.055$ & $-0.53^{+0.26}_{-0.23}$ & $2.6 \sigma$ & $-9.2$ \\
        \textit{Planck} + Union3 + ext-BAO + EFTBOSS & $-0.735^{+0.073}_{-0.085}$ & $-0.87^{+0.35}_{-0.29}$ & $3.4 \sigma$ & $-14.5$ \\
        \textit{Planck} + DESY5 + ext-BAO + EFTBOSS & $-0.776\pm 0.063$ & $-0.76^{+0.29}_{-0.27}$ & $3.7 \sigma$ & $-16.8$ \\ 
        \hline
        \hline
         Clustering quintessence ($c_s^2\rightarrow 0$) & $w_0$ & $w_a$ & $p >\Lambda$ & $\Delta \chi^2$\\
        \hline
        \textit{Planck} + PanPlus + ext-BAO & $-0.856\pm 0.085$ & $-0.60^{+0.43}_{-0.37}$ & $1.1 \sigma$ & $-2.5$\\
        \textit{Planck} + PanPlus + DESI-BAO & $-0.821\pm 0.063$ & $-0.77\pm 0.27$ & $2.5 \sigma$ & $-8.9$\\
        \textit{Planck} + ext-BAO + EFTBOSS & $-0.679^{+0.091}_{-0.11}$ & $-1.15^{+0.42}_{-0.35}$ & $0.6 \sigma$ & $-1.3$\\
        \textit{Planck} + PanPlus + ext-BAO + EFTBOSS & $-0.809^{+0.053}_{-0.061}$ & $ -0.72^{+0.28}_{-0.25}$ & $2.8 \sigma$ & $-10.3$\\
        \textit{Planck} + Union3 + ext-BAO + EFTBOSS & $-0.692\pm 0.078$ & $-1.10\pm 0.33$ & $3.6 \sigma$ & $-16.0$\\
        \textit{Planck} + DESY5 + ext-BAO + EFTBOSS & $-0.739\pm 0.061$ & $-0.97^{+0.30}_{-0.27}$ & $3.9 \sigma$ & $-18.2$\\
        \textit{Planck} + PanPlus + DESI-BAO + EFTBOSS/$r_s^{\rm marg}$ & $-0.789\pm 0.050$ & $-0.80^{+0.25}_{-0.22}$ & $3.7\sigma$ & $-16.9$\\ 
        \textit{Planck} + Union3 + DESI-BAO + EFTBOSS/$r_s^{\rm marg}$ & $-0.677\pm 0.075$ & $-1.14^{+0.33}_{-0.29}$ & $3.8\sigma$ & $-17.7$\\ 
        \textit{Planck} + DESY5 + DESI-BAO + EFTBOSS/$r_s^{\rm marg}$ & $-0.726\pm 0.060$ & $-1.00^{+0.30}_{-0.25}$ & $4.4\sigma$ & $-22.9$\\ 
        \hline
    \end{tabular}

    \caption{ $68\%$ credible intervals of $w_0$ and $w_a$, as well as the preference over $\Lambda$ ($p>\Lambda$) and the associated $\Delta \chi^2$, obtained within $w_0w_a$CDM for both smooth and clustering quintessence across various datasets, varying galaxy clustering (ext-BAO, ext-BAO + EFTBOSS, DESI-BAO, or DESI-BAO + EFTBOSS/$r_s^{\rm marg}$) and supernova (PanPlus, Union3 or DESY5) data.}
    \label{tab:main_table}
\end{table}

\begin{figure}[!h]
    \centering
    \includegraphics[width=0.7\textwidth]{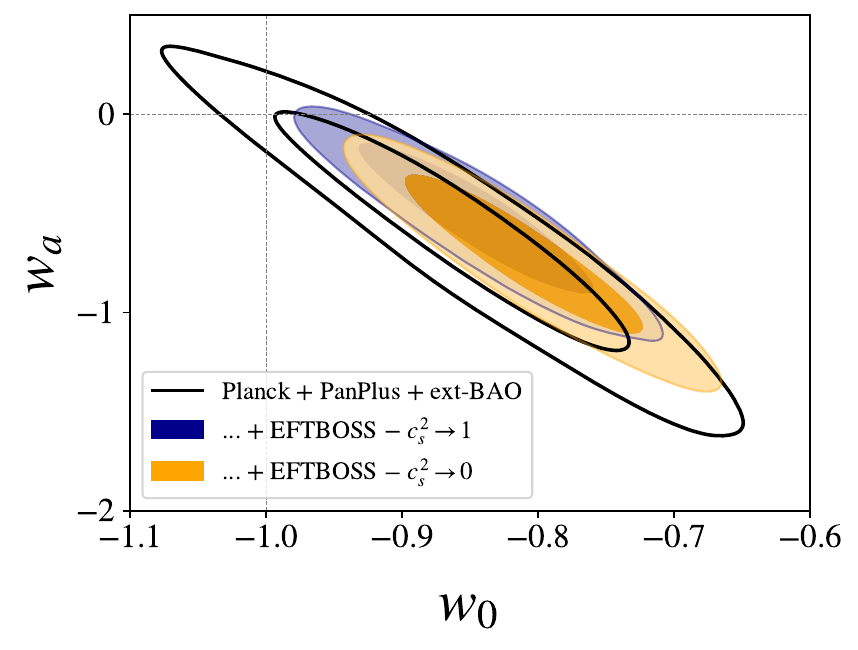}
    \caption{
    \textbf{Baseline constraints on $(w_0, w_a)$ from pre-DESI data} ---
    2D posterior distributions of $(w_0, w_a)$ from \textit{Planck} + PanPlus + ext-BAO + EFTBOSS within $w_0w_a$CDM for both smooth and clustering quintessence. 
    Constraints obtained without EFTBOSS are shown for comparison. 
    }
    \label{fig:main_plot_w0wa}
\end{figure}

\paragraph{Baseline `pre-DESI' dataset}
In fig.~\ref{fig:main_plot_w0wa}, we show the 2D posterior distributions in the $w_0 - w_a$ plane for the $w_0w_a$CDM model, while the 1D and 2D posterior distributions for the whole parameter space are shown in fig.~\ref{fig:w0wa_all} of app.~\ref{app:supp_material}.
In these figures, we consider the \textit{Planck} + PanPlus + ext-BAO + EFTBOSS dataset, for the two phenomenologically-distinct limits of the dark energy fluctuations identified in sec.~\ref{sec:limits}, \textit{i.e.}, the smooth ($c_s^2 \rightarrow 1$) and clustering ($c_s^2 \rightarrow 0$) quintessence. 
The $68\%$-credible intervals and the best fit values are given in tab.~\ref{tab:bestfit_params} of app.~\ref{app:supp_material}, together with the $\Delta \chi^2$ and the $\sigma$-deviation from $\Lambda$CDM. \\

\begin{figure}[!h]
    \centering
    \includegraphics[width=0.49\textwidth]{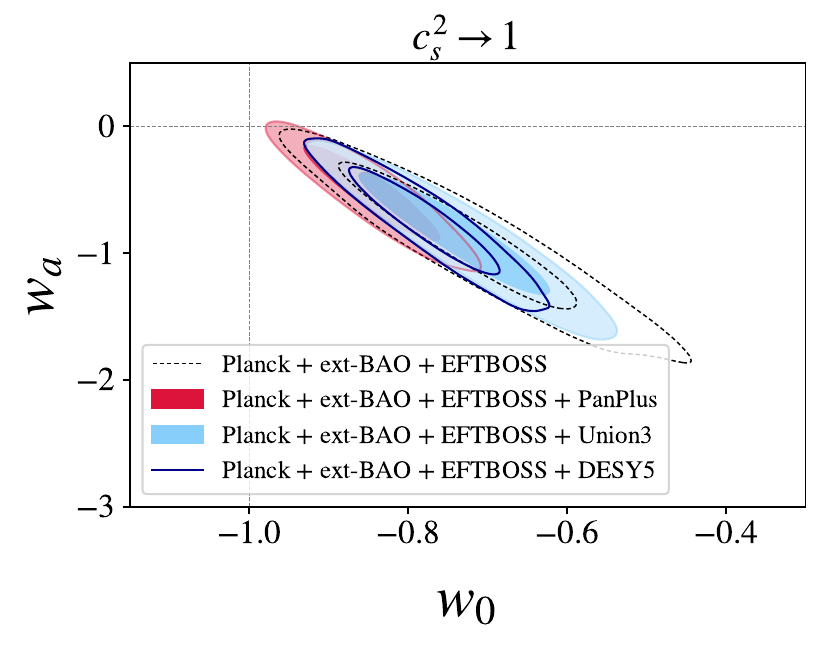}
    \includegraphics[width=0.49\textwidth]{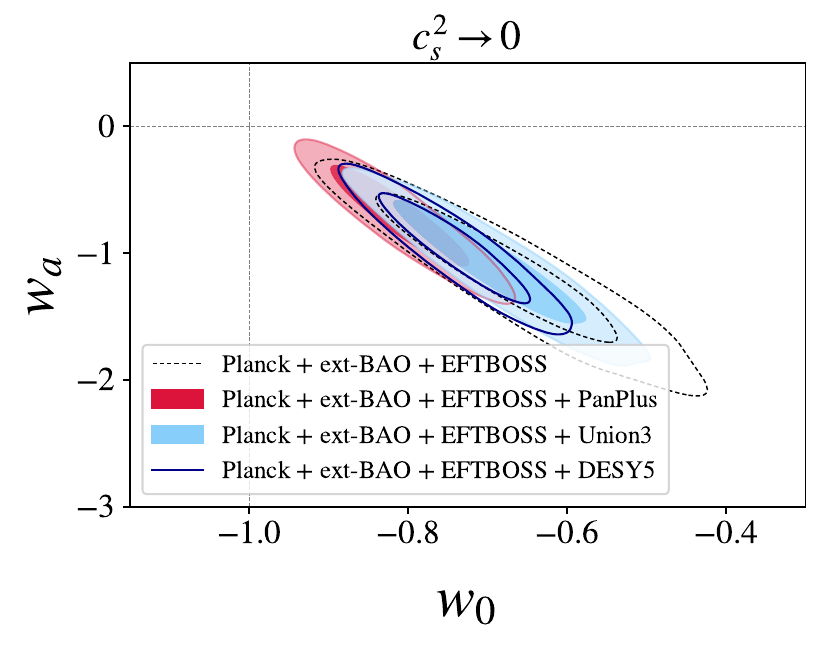}
    \caption{\textbf{Impact of supernova data on $(w_0, w_a)$ constraints} ---
    2D posteriors of $(w_0, w_a)$ from \textit{Planck} + ext-BAO + EFTBOSS combined with various supernova data (\textit{i.e.}, no supernovae data, PanPlus, Union3, or DESY5) within $w_0 w_a$CDM for smooth (left) and clustering (right) quintessence. See also fig.~\ref{fig:main}. }
    \label{fig:w0wa_DESI-PanPlus}
\end{figure}

Within $w_0 w_a$CDM, we obtain, at $68\%$CL, $w_0 = -0.844\pm 0.055$ and $w_a = -0.53^{+0.26}_{-0.23}$ in smooth quintessence, and $w_0 = -0.809^{+0.053}_{-0.061}$ and $w_a = -0.72^{+0.28}_{-0.25}$ in clustering quintessence. 
Compared to the results without EFTBOSS, yielding $w_0 = -0.856\pm 0.085$ and $w_a = -0.60^{+0.43}_{-0.37}$, the uncertainties are improved by $\sim 35\%$. 
In light of these results, we can make several observations:
\begin{itemize}
    \item For the dataset combination considered, \textit{Planck} + PanPlus + ext-BAO + EFTBOSS, we find a preference for evolving dark energy over $\Lambda$ at $2.6 \sigma$ and $2.8 \sigma$ whether $c_s^2 \rightarrow 1$ or  $c_s^2 \rightarrow 0$ is considered, corresponding to $\Delta \chi^2_{\rm min } = -9.2$ and $\Delta \chi^2_{\rm min } = -10.3$, respectively. 
    \item The preferences arise only with the inclusion of the bispectrum in EFTBOSS (carefully studied in sec.~\ref{sec:checks}). 
    Without this contribution, the significance decreases to $1.4 \sigma$ for $c_s^2 \rightarrow 1$ and $1.5 \sigma$ for  $c_s^2 \rightarrow 0$.
    \item Our findings, relying on pre-DESI clustering data, are consistent with DESI Y1 results, that yield a $2.5 \sigma$ preference over $\Lambda$CDM when combined with \textit{Planck} and PanPlus~\cite{DESI:2024hhd}, in the same preferred region as displayed in fig.~\ref{fig:w0wa_DESI-PanPlus}.
    While posteriors in the $w_0-w_a$ plane are consistent at $\sim 0.5\sigma$, our 2D constraints are approximately $35\%$ $ (25 \%)$ tighter for $c_s^2 \to 1$ $ (c_s^2 \to 0)$ compared to those from DESI BAO \cite{DESI:2024mwx}, and comparable to those from DESI FS + BAO, where FS stands for the full-shape power spectrum~\cite{DESI:2024hhd}.\footnote{Here and throughout the paper, comparisons of credible region areas in the $w_0-w_a$ plane are based on the ratio of the dark energy Figure of Merit (FoM)~\cite{Albrecht:2006um,DESI:2024hhd}. Especially, we use the fact that the $\rm{FoM} \propto \mid {\rm det} \, C \mid^{-1/2}$, where $C$ corresponds to the $2 \times 2$ covariance matrix of $(w_0,w_a)$.} 
    \item Constraints on $(w_0, w_a)$ depend not only on the background but also on the propagation of dark energy perturbations --- controlled by $c_s^2$ --- affecting clustering differently, as anticipated in sec.~\ref{sec:limits}.
    Our results are further discussed in light of the stability of dark energy fluctuations in sec.~\ref{sec:stability}. 
\end{itemize}

\begin{figure}[!h]
    \centering
    \includegraphics[width=0.7\textwidth]{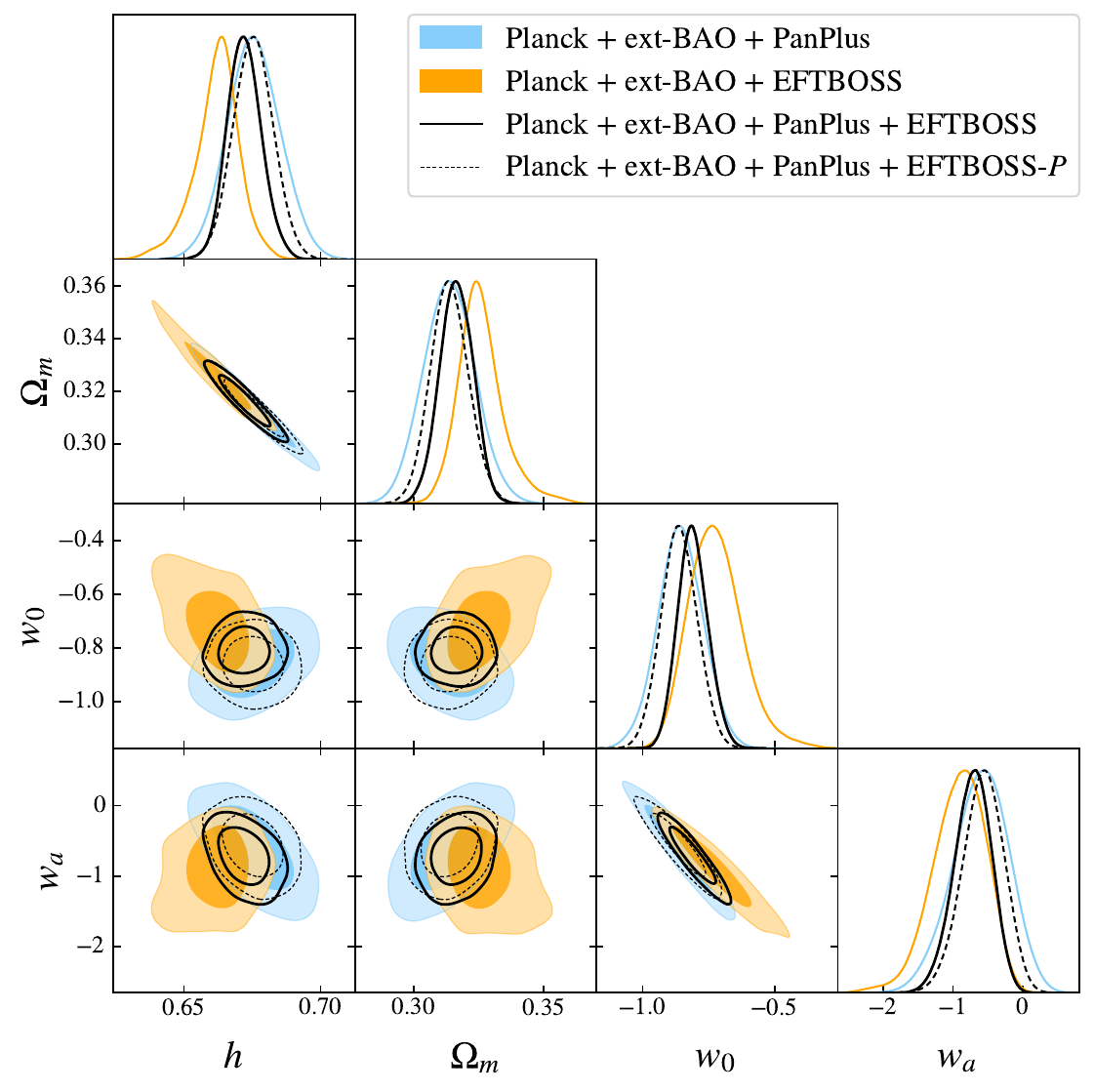}
    \caption{
     \textbf{Breaking degeneracies between PanPlus and EFTBOSS} --- 2D posterior distributions of $(w_0, w_a)$ from \textit{Planck} + ext-BAO combined with PanPlus, EFTBOSS, or PanPlus + EFTBOSS, within $w_0w_a$CDM for clustering quintessence. For comparison, we show \textit{Planck} + ext-BAO + PanPlus + EFTBOSS with the contribution of the power spectrum only for the latter.
    }
    \label{fig:combination}
\end{figure}

\paragraph{Breaking degeneracies} In tab.~\ref{tab:main_table}, we report the preference for $(w_0,w_a)$ over $\Lambda$ (and the associated $\Delta \chi^2$) for several dataset. Interestingly, the combinations of \textit{Planck} + ext-BAO with either PanPlus or EFTBOSS do not show any preference for evolving dark energy. Yet, the preference arises when PanPlus and EFTBOSS are combined, implying that some degeneracies are broken. In fig.~\ref{fig:combination}, we show how the degeneracies in the $h - w_0/w_a$ and $\Omega_m - w_0/w_a$ planes are orthogonal between PanPlus and EFTBOSS. The combination of both thus allows to break those degeneracies efficiently, leading to a strong constraint in the $w_0 - w_a$ plane, which is $1.9$ $(2.2)$ times better than the dataset including PanPlus (EFTBOSS) only. In addition, the inclusion of EFTBOSS pulls the contours towards small $h$ and large $\Omega_m$, leading to an increase in $w_0$ and a decrease in $w_a$ due to their geometrical degeneracies in the angular diameter distance. Anticipating the results of sec.~\ref{sec:checks}, we can see that the analysis with the power spectrum only is less able to break those degeneracies and prefers higher values of $h$ and smaller values of $\Omega_m$.

\paragraph{Impact of DESI BAO data}
In the top left panel of fig.~\ref{fig:main}, we display, for clustering quintessence, the 2D posterior distributions in the $w_0 - w_a$ plane from \textit{Planck} + PanPlus + ext-BAO + EFTBOSS and  \textit{Planck} + PanPlus + DESI-BAO + EFTBOSS/$r_s^{\rm marg}$, where the sound horizon information is marginalised in the latter (see sec.~\ref{sec:dataset}). The corresponding $68 \%$ credible intervals of $w_0$ and $w_a$ along with the preference over $\Lambda$ (and the associated $\Delta \chi^2$) are reported in tab.~\ref{tab:main_table}. 
Posteriors of the scaling parameter $\alpha_{r_s}$ are shown in app.~\ref{app:supp_material}. 
The preference for evolving dark energy increases from $ 2.8 \sigma$ to $3.7 \sigma$ (for $c_s^2 \rightarrow 0$) when replacing ext-BAO by DESI-BAO. 
This improvement is accompanied by uncertainty reductions of $\sim 45\%$ and $\sim 15\%$ on the 2D posterior of $(w_0, w_a)$ compared to \textit{Planck} + PanPlus + DESI-BAO and \textit{Planck} + PanPlus + ext-BAO + EFTBOSS, respectively.
Interestingly, our joint \textit{Planck} + PanPlus + DESI-BAO + EFTBOSS analysis yields a significantly stronger preference than the $2.5\sigma$ significance reported in ref.~\cite{DESI:2024hhd} from \textit{Planck} + PanPlus + DESI-BAO + DESI FS, where DESI FS relies solely on the full-shape power spectrum. 
These results highlight the crucial role of the bispectrum in probing evolving dark energy, further demonstrating that it carries substantial information beyond the sound horizon. 

\paragraph{Impact of supernova data}
We compare in fig.~\ref{fig:w0wa_DESI-PanPlus} the 2D posteriors in the $w_0-w_a$ plane from \textit{Planck} +  ext-BAO + EFTBOSS without supernova data, or in combination with PanPlus, Union3, or DESY5. 
Corresponding $68 \%$ credible intervals of $w_0$ and $w_a$ with preference over $\Lambda$ are shown in tab.~\ref{tab:main_table}. 
With no supernovae, the significance reduces to $<0.6\sigma$ for both smooth and clustering quintessence.\footnote{As checked on fake synthetic data, we find prior volume projection effects at $>1 \sigma$ when not including any supernova data, in accordance with the observed shift between the posterior mean and maximum a posteriori.  } 
As already mentioned above, supernovae are thus crucial in driving the preference for evolving dark energy, in line with DESI Y1 results~\cite{DESI:2024mwx,DESI:2024hhd}. 
In addition, the significance over $\Lambda$ increases from $2.6 \sigma $ to $3.4 \sigma$ ($3.7 \sigma)$ when replacing PanPlus by Union3 (DESY5) in smooth quintessence, and from $2.8 \sigma $ to $3.7 \sigma$ ($3.9 \sigma)$ in clustering quintessence, following the same trend as found with DESI Y1 data.\footnote{Ref.~\cite{DESI:2024hhd} finds an increase from $2.5 \sigma$ to $3.4 \sigma$ ($3.8 \sigma$) when replacing PanPlus by Union3 (DESY5) data.}

Finally, in the right panel of fig.~\ref{fig:main}, we perform the same exercise replacing ext-BAO by DESI-BAO while properly marginalising over the sound horizon information in EFTBOSS. 
The corresponding $68 \%$ credible intervals of $w_0$ and $w_a$ with the preference over $\Lambda$ are shown in tab.~\ref{tab:main_table}.
For clustering quintessence, the preference over $\Lambda$ increases from $3.7\sigma$ to $3.8 \sigma$ ($4.4 \sigma$) when replacing PanPlus with Union3 (DESY5). 
Notably, in contrast to the DESI Y1 results~\cite{DESI:2024hhd}, the preference obtained from the three supernova datasets remains relatively stable ($\lesssim 1 \sigma$) when they are combined with \textit{Planck} +  DESI-BAO + EFTBOSS/$r_s^{\rm marg}$.

\subsection{Consistency checks}\label{sec:checks}

To assess the robustness of our results, we systematically examine and compare different analysis configurations against our baseline setup.
In particular, we remind that our baseline setup consists in 
\begin{enumerate}
    \item the power spectrum $P$ and bispectrum $B$ at one loop;
    \item the bispectrum cutoff scales $k_{\rm max}^{\rm lowz} = 0.20 \, h/{\rm Mpc}$ and $k_{\rm max}^{\rm highz} = 0.23 \, h/{\rm Mpc}$;
    \item the exact-time dependence of nonlinear EFTofLSS operators;
    \item the EFT priors presented in sec.~\ref{sec:eft_lkl};
    \item the monopole $P_0$ and the quadrupole $P_2$ of the power spectrum.
\end{enumerate}
In the following, we vary each of these configurations, using the \textit{Planck} + PanPlus + ext-BAO + EFTBOSS dataset.

\begin{figure}
    \centering
    \includegraphics[width=0.49\textwidth]{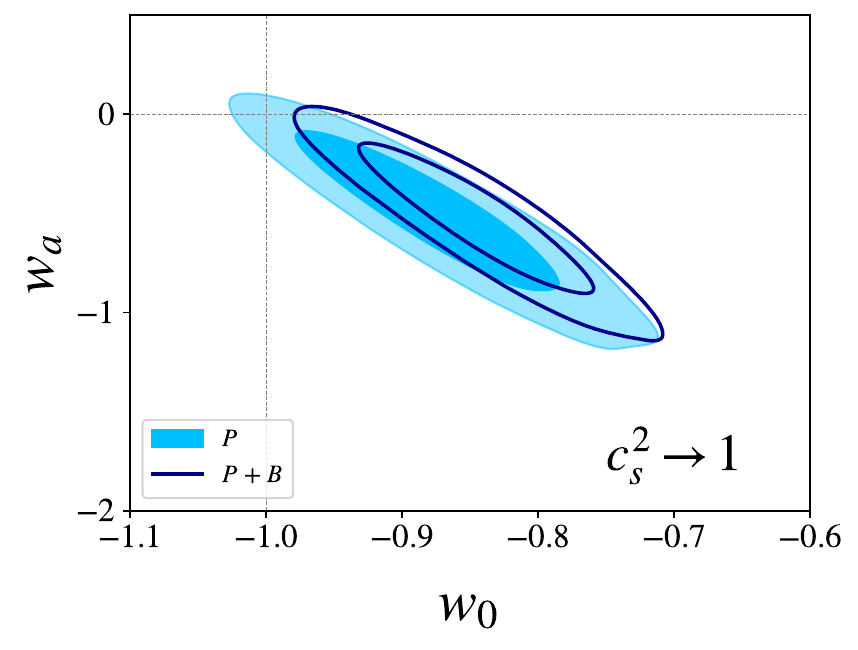}
    \includegraphics[width=0.49\textwidth]{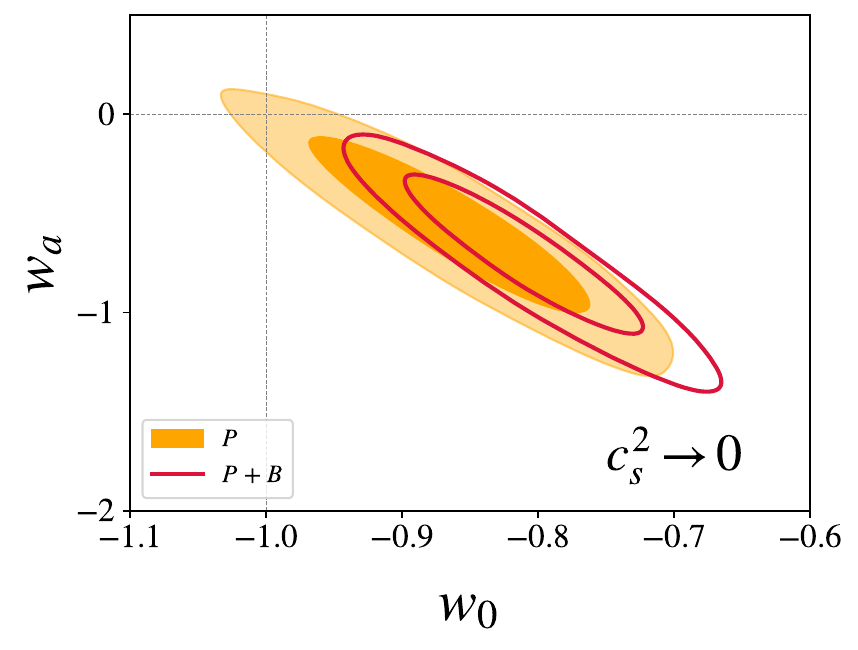} 
    \includegraphics[width=0.49\textwidth]{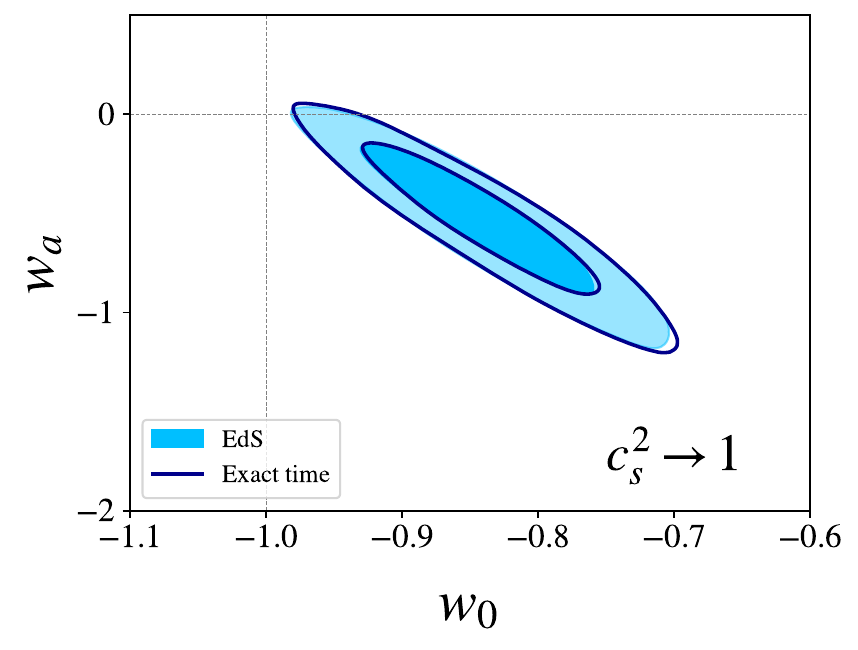}
    \includegraphics[width=0.49\textwidth]{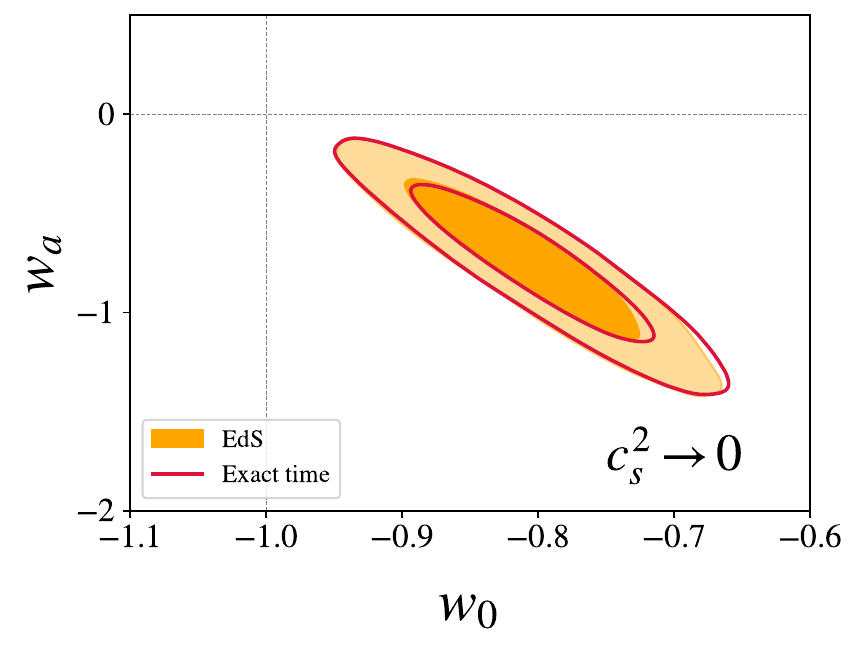}
    \includegraphics[width=0.49\textwidth]{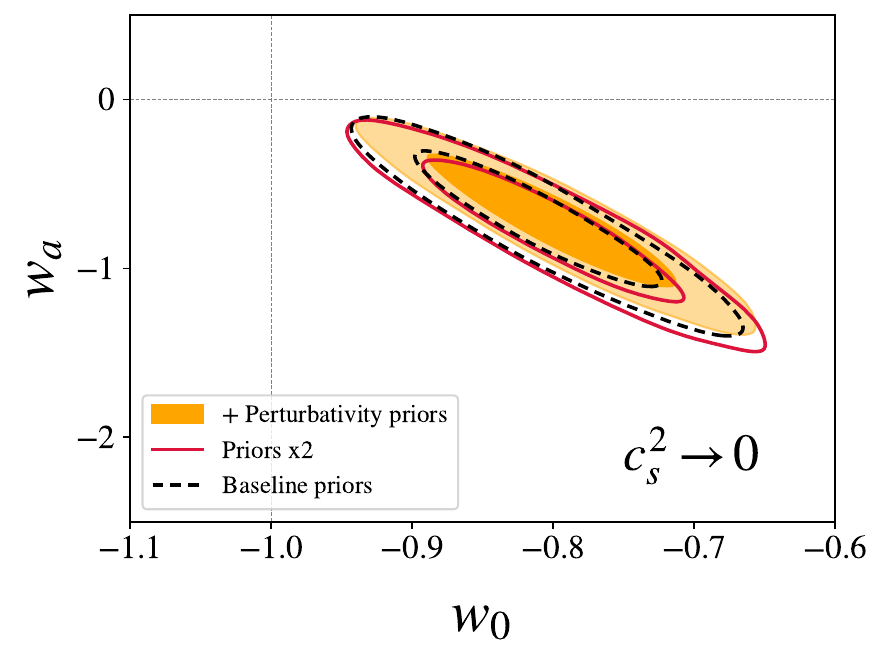}
    \includegraphics[width=0.49\textwidth]{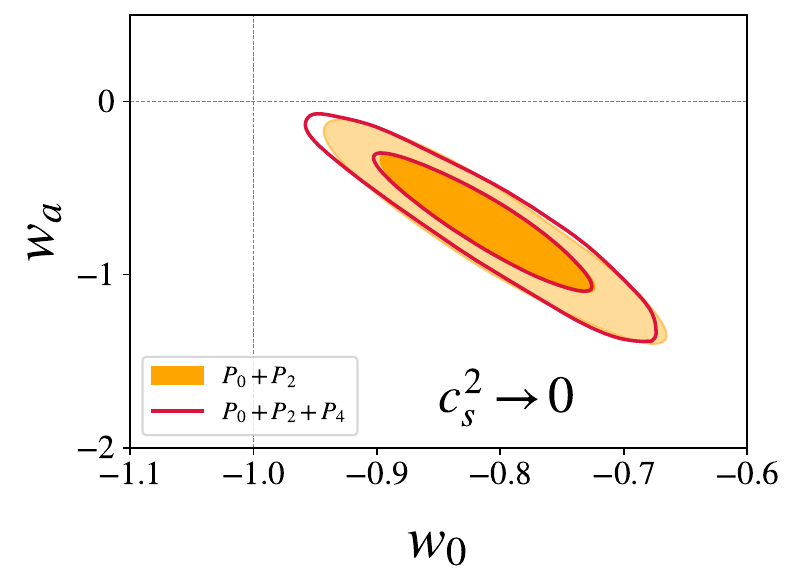}  
    \caption{
    \textbf{Constraints on $(w_0, w_a)$  across analysis variations} --- This figure illustrates the impact of different analysis choices relative to our baseline setup:  the addition of the bispectrum (\textit{upper panels}), the EdS time approximation (\textit{middle panels}), the prior choice on EFT parameters (\textit{lower left panel}), and the addition of the power spectrum hexadecapole (\textit{lower right panel}). Comparison are shown for both smooth ($c_s^2 \rightarrow 1$) and clustering ($c_s^2 \rightarrow 0$) quintessence, except in the lower panels, where only clustering quintessence is shown. See main text for more details. }
    \label{fig:analysis_choices}
\end{figure}

\begin{figure}
    \centering
    \includegraphics[width=0.6\textwidth]{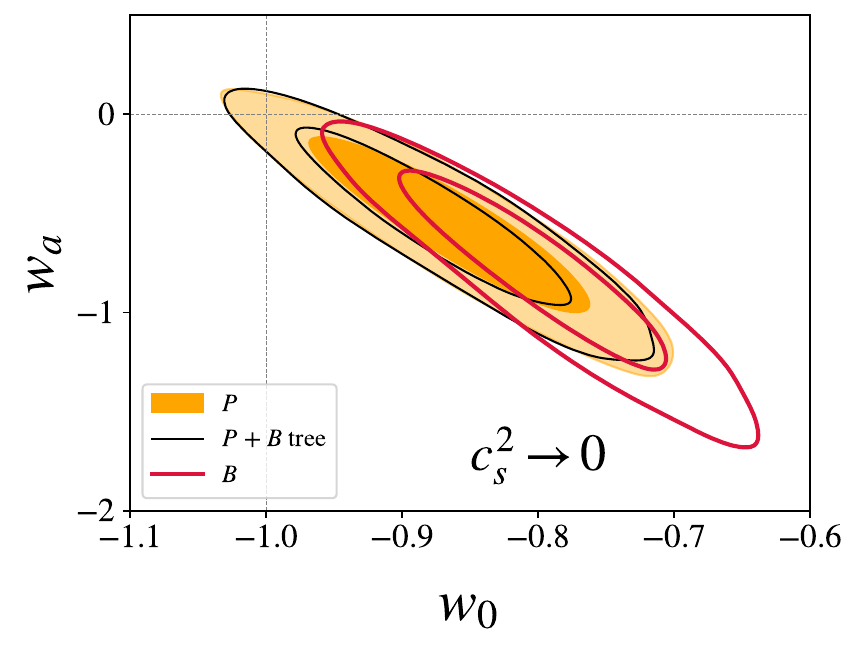}
    \includegraphics[width=1\textwidth]{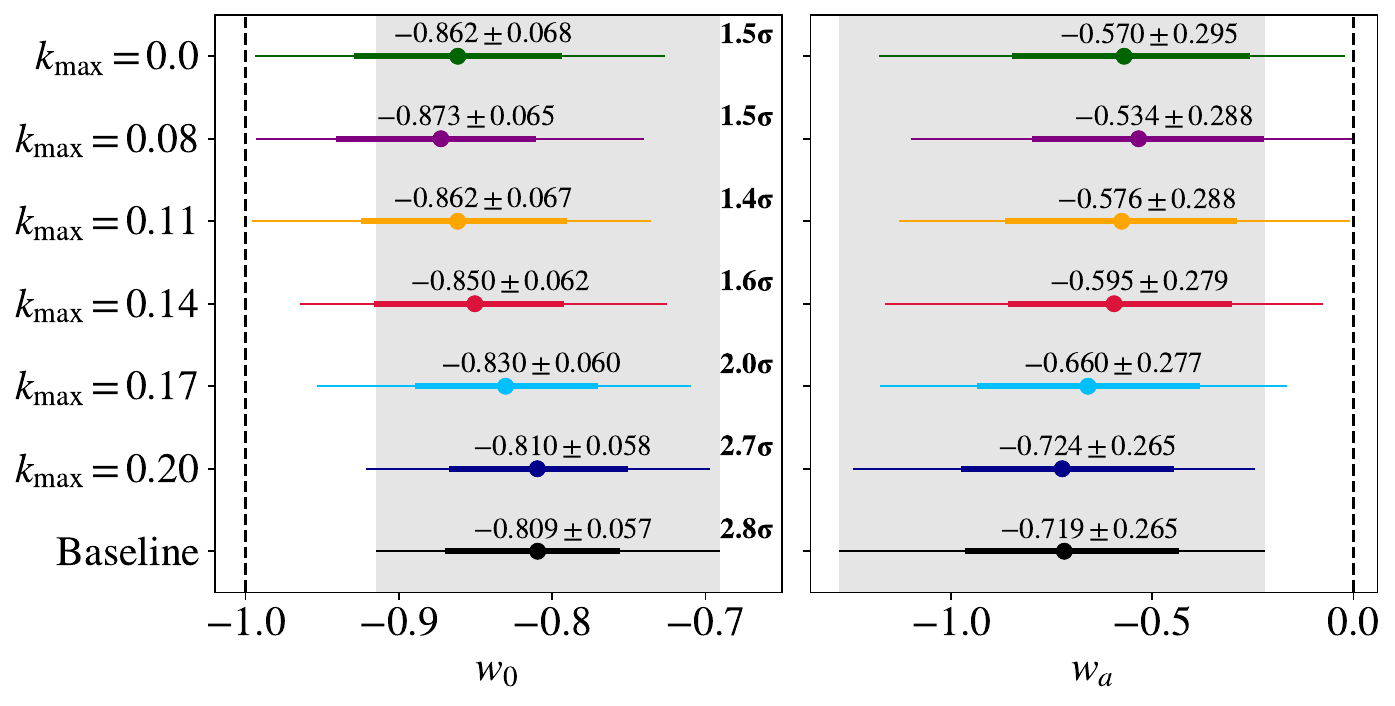} 
    \caption{
    \textbf{Impact of the bispectrum on $(w_0, w_a)$ constraints} --- \textit{Upper panel:} Comparison of the constraints in clustering quintessence derived from \textit{Planck} + PanPlus + ext-BAO + EFTBOSS, where EFTBOSS includes the one-loop power spectrum alone, the one-loop bispectrum alone, or the combined one-loop power spectrum and tree-level bispectrum. 
    Contrary to the one-loop bispectrum, the tree-level one is analysed only up to $k_{\rm max}=0.08 \ \hinvMpc$. 
    \textit{Bottom panel:} $68\%$ credible intervals and significance over $\Lambda$ for clustering quintessence as a function of the highest $k_{\rm max}$ (in $\hinvMpc$) used in the analysis of the bispectrum. 
    ``Baseline'' corresponds to the fiducial scale cuts used in our main analysis.
    See main text for details. 
    }
    \label{fig:w0wa_pk_bk}
\end{figure}

\paragraph{Power spectrum vs. bispectrum}
To assess the impact of the bispectrum on the constraints, we show in the top panels of fig.~\ref{fig:analysis_choices} the difference in the posteriors of $(w_0, w_a)$ upon its inclusion. 
Without the bispectrum, we obtain a $1.4\sigma$ ($1.5\sigma$) preference over $\Lambda$ for $c_s^2 \to 1$ ($c_s^2 \to 0$). 
Including this contribution increases the significance to $2.6\sigma$ ($2.8\sigma$), corresponding to a relative shift of $\sim 1\sigma$ in the 2D plane mostly along the principal axis of the ellipse, and an uncertainty reduction of $\sim 30\%$ on the 2D posterior (corresponding to $\sim 15\%$ on the 1D marginals). 
The improvement of $30\%$ of the FoM in the $w_0-w_a$ plane corresponds roughly to a doubling of the data volume. 
This goes in line with the findings of ref.~\cite{Spaar:2023his}, which find a similar $30\%$ improvement in the FoM of $w_0$ assuming a constant dark-energy equation of state, \textit{i.e.}, by restricting the posterior volume of $(w_0, w_a)$ to lie on the line parametrised by $w_a = 0$. 
Given the amount of information the bispectrum is adding, a $\sim 1\sigma$ shift in the mean is, a priori, within expectations. 
Interestingly, we also note that ref.~\cite{SolaPeracaula:2018wwm} already showed statistical hints of a preference for $w_0w_a$CDM in the pre-DESI era upon the inclusion of the bispectrum information of BOSS. 
{While the consistency tests presented in this work have not identified an obvious systematic origin (see in particular our test on simulations presented in app.~\ref{app:sims} validating the absence of a large bias from our pipeline), the possibility of a statistical fluctuation in
the data realisation remains to be duly quantified, \textit{e.g.},~by comparing the observed shift to an estimate from mocks of the typical scatter upon addition of the one-loop bispectrum. 
We leave this to future work. } 

To isolate the impact of the bispectrum, we compare in fig.~\ref{fig:w0wa_pk_bk} the constraints derived from the one-loop power spectrum alone, from the one-loop bispectrum alone, and from the combined one-loop power spectrum and tree-level bispectrum (restricting $k_{\rm max}=0.08 \ \hinvMpc$ in the latter). 
We observe that the bispectrum drives the constraints away from $\Lambda$, provided that it is analysed up to the $k$-reach accessible by one-loop predictions. 
A similar, albeit weaker, trend is seen when assuming a constant equation of state as discussed in ref.~\cite{Spaar:2023his}. 
{We provide additional support from the Fisher matrix on the constraining power from the scales accessible by the one-loop bispectrum in app.~\ref{app:loop}. }

\paragraph{Cutoff scale}
In fig.~\ref{fig:w0wa_pk_bk}, we show the $68\%$ credible intervals of $w_0$ and $w_a$, together with the significance over $\Lambda$, as a function of the highest Fourier mode $k_{\rm max}$ included in the analysis of the bispectrum.\footnote{The $k_{\rm max}$ of the power spectrum is kept unchanged in this exercice. }
In particular, we display the analysis without the bispectrum (\textit{i.e.}, $k_{\rm max} = 0 \ \hinvMpc$), the analysis with the tree-level bispectrum (\textit{i.e.}, $k_{\rm max} = 0.08 \ \hinvMpc$), the baseline analysis (\textit{i.e.}, $k_{\rm max}^{\rm lowz} = 0.20 \ \hinvMpc$ and $k_{\rm max}^{\rm highz} = 0.23 \ \hinvMpc$), as well as several intermediate cutoff scales, namely $k_{\rm max} \in [0.11, 0.14, 0.17, 0.29] \ \hinvMpc$.
Deviations from $\Lambda$, together with improvement in the constraints, start around $k_{\rm max} = 0.14 \ \hinvMpc$, progressively increasing until the highest $k_{\rm max}$ reachable at one loop which is used in our main analyses. 
This is to be contrasted with the marginal gain in precision from the inclusion of the tree-level bispectrum, reaching at most $k_{\rm max} \sim 0.08-0.10 \ \hinvMpc$.

\paragraph{EFTofLSS time dependence}
In the middle panels of fig.~\ref{fig:analysis_choices}, we show the difference in the 2D posteriors of $(w_0, w_a)$ treating the time dependence in the EFTofLSS exactly, or assuming the EdS approximation (see sec.~\ref{sec:EFTofLSS}).
As anticipated in figs.~\ref{fig:G} and~\ref{fig:bestfit_ratio}, the EdS approximation is valid for the current data precision \emph{in the region preferred by the data}. 
The exact-time dependence might reveal to be significant for upcoming surveys such as future DESI data releases or Euclid. 

\paragraph{EFT priors}
In the bottom left panel of fig.~\ref{fig:analysis_choices}, we illustrate the effect of broadening the prior width on the EFT parameters (as described in sec.~\ref{sec:eft_lkl}) by a factor $2$. 
As the constraints remain virtually unchanged, we conclude that our prior choice has a negligible impact on the results. 
Additionally, we impose a prior, dubbed ``perturbativity prior'' in the same plot, on the size of the one-loop contributions to the power spectrum and bispectrum based on a perturbative convergence criterion. 
Along with our fiducial prior choice, this ensures that our constraints are marginalised over EFT parameter values for which loop corrections remain within their expected range in the EFTofLSS~\cite{Braganca:2023pcp,DAmico:2022gki}. 
Again, the constraints remain practically unchanged, reinforcing the conclusion that the region favored by the data is consistently described within the EFTofLSS.  
Moreover, tab.~\ref{tab:bestfit_params} shows that the relative shifts between the maximum a posteriori estimates and the posterior means are $\lesssim 0.3\sigma$ for all cosmological parameters, and $\lesssim 0.2\sigma$ for $w_0$ and $w_a$, indicating that prior volume projection effects are not important. 
This stems from the efficient breaking of parameter degeneracies enabled by the dataset combination we use, consistent with the findings of ref.~\cite{DESI:2024hhd}. 
{In app.~\ref{app:supp_material}, we provide an additional check of the impact of our prior assumption on the correlation between the various BOSS skies, finding that the cosmological results remains largely consistent when removing this prior information. }

\paragraph{Power spectrum multipoles}
In the bottom right panel of fig.~\ref{fig:analysis_choices}, we show the impact of the addition of the power spectrum hexadecapole. 
Given its small signal-to-noise ratio in BOSS~\cite{Zhang:2021yna,Simon:2022lde},{this leads to $2\%$ improvement in terms of FoM in the $w_0-w_a$ plane.} \\

In summary, we conclude that the preference for evolving dark energy is entirely driven by the bispectrum over the scales accessible to the EFTofLSS at one loop and appears robust in regards of the consistency checks we have conducted.


\subsection{$w_0w_a$CDM + $\Omega_k$} \label{sec:omega_k}

\begin{figure}
    \centering
    \includegraphics[width=0.49\textwidth]{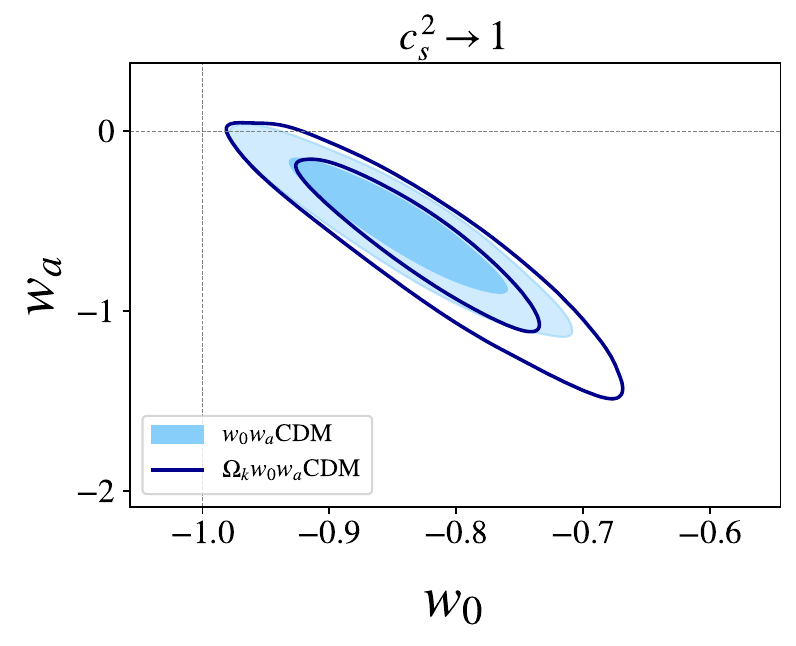}
    \includegraphics[width=0.49\textwidth]{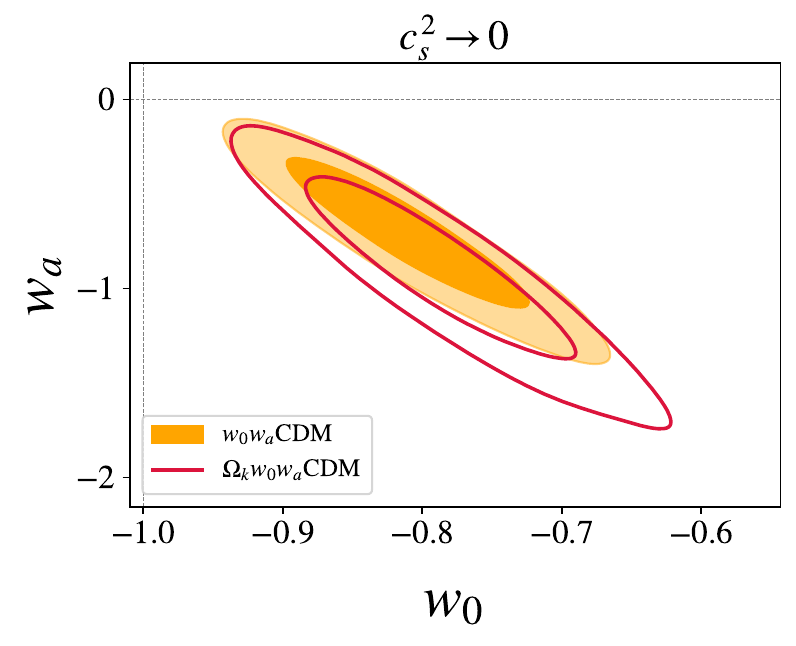} 
    \caption{
    \textbf{Impact of free curvature on $(w_0, w_a)$ constraints} --- 2D posterior distributions of $(w_0, w_a)$ from \textit{Planck} + PanPlus + ext-BAO + EFTBOSS within $w_0w_a$CDM with or without free $\Omega_k$ for both smooth (\textit{left}) and clustering (\textit{right}) quintessence. 
    }
    \label{fig:omega_k}
\end{figure}

Is the preference for evolving dark energy sensitive to the curvature of the Universe?
In fig.~\ref{fig:omega_k}, we compare the 2D posterior distributions in the $w_0 - w_a$ plane from the baseline dataset where $\Omega_k$ is fixed to zero or let free (within a large flat prior), for both smooth and clustering quintessence.
We respectively obtain $\Omega_k = -0.0014\pm 0.0027$ and $\Omega_k = -0.0027\pm 0.0025$ at $68\%$CL, both compatible with a flat Universe at $\lesssim 1 \sigma$.
As expected, the constraints on $w_0$ and $w_a$ are weakened because of the degeneracy with $\Omega_k$.\footnote{For smooth quintessence, we obtain $w_0 = -0.825^{+0.060}_{-0.067}$ and $w_a = -0.64^{+0.36}_{-0.28}$ at $68 \%$ CL, while for clustering quintessence, we obtain $w_0 = -0.787\pm 0.063$ and $w_a = -0.91^{+0.34}_{-0.30}$ at $68 \%$ CL.}
Yet, the preference over $\Lambda$, at $2.7 \sigma$ $(2.9\sigma)$ for $c_s^2 \to 1$ ($c_s^2 \to 0$), remains similar as in the flat case.

\subsection{$w_0w_a$CDM in modified gravity}\label{sec:mg}

\begin{figure}[!h]
    \centering
    \includegraphics[width=0.7\textwidth]{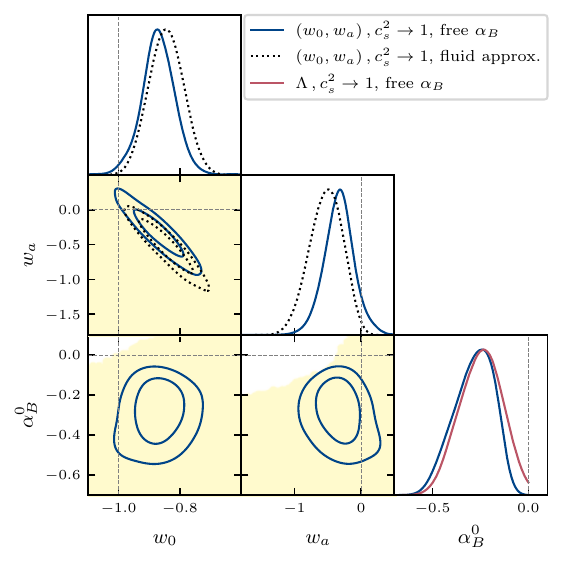}
    \caption{
    \textbf{$w_0w_a$CDM in modified gravity} ---
    1D and 2D posteriors of $(w_0, w_a, \alpha_B^0)$ from \textit{Planck} + PanPlus + ext-BAO + EFTBOSS within $\Lambda$CDM or $w_0w_a$CDM for $c_s^2 \rightarrow 1$, with free braiding $\alpha_B(t) = \alpha_B^0 \, \Omega_d(t) / \Omega_{d,0}$. 
    The beige area corresponds to the perturbatively stable region. 
    Constraints from smooth quintessence in the standard fluid approximation are shown for reference. 
    }
    \label{fig:alphaB}
\end{figure}

In the analyses presented previously, we provided results for smooth quintessence assuming the standard fluid approximation. 
However, the $c_s^2 \rightarrow 1$ limit allows stable dark energy fluctuations for $w < -1$ only if the EFTofDE operators are adjusted to avoid gradient instabilities~\eqref{eq:stable}, thereby leading to the modified Poisson equation~\eqref{eq:modPoisson}. 
The time dependence in the EFTofLSS is modified accordingly, as detailed in app.~\ref{app:time_MG}. 
In fig.~\ref{fig:alphaB}, we show the results obtained by fitting \textit{Planck} + PanPlus + ext-BAO + EFTBOSS within $w_0w_a$CDM for $c_s^2 \rightarrow 1$ with free $\alpha_B$, assumed to evolve as $\alpha_B(t) = \alpha_B^0 \Omega_d(t)/\Omega_{d,0}$.\footnote{As we follow the conventions of ref.~\cite{Gleyzes:2014rba}, $\alpha_B$ defined in eq.~\eqref{eq:alpha} is related to $\alpha_B'$ defined in \texttt{hi\_class} as $\alpha_B = -\alpha_B'/2$. Notice also the normalisation $\alpha_B \equiv 1$ at $a = 1$ when $\alpha_B^0 = 1$. }  
To ensure that the stability conditions are satisfied in the fit, $c_s^2$ is kept to unity by fixing $\alpha_K = \nu - \alpha_w - 3\alpha_B^2$ (see eq.~\eqref{eq:soundspeed}), while $\alpha_B$ is allowed to vary in the region that ensures $\nu > 0$.

Freeing $\alpha_B$ results in a visible volume increase in the posteriors of $(w_0, w_a)$, leading to $w_{0} = (-0.849)\, -0.868\pm 0.056$, $w_a = (-0.467) \, -0.33 \pm 0.24$, and $\alpha_B^0 = (-0.214) \, -0.283^{+0.12}_{-0.086}$ at $68\%$CL, where the best-fit values are in parenthesis.
Yet, the model is favoured at $\Delta\chi^2=-13.3$, yielding a preference over $\Lambda$ at the $2.9\sigma$ level, counting 3 additional degrees of freedom.
This preference is explained by a departure from $\alpha_B^0 = 0$ at the $\sim 2\sigma$ level, as shown in fig.~\ref{fig:alphaB}.

When assuming $\Lambda$ instead but free $\alpha_B$, we find $\alpha_B^0 = -0.25^{+0.12}_{-0.11}$ at $68\%$CL, yielding a departure at a similar level.
Bearing in mind the difference in the supernova and clustering datasets, we note that our results (both assuming $(w_0,w_a)$ or $\Lambda$ cosmologies) are in good agreement with those from the `$\alpha$-basis' analysis of CMB + DESY5 + DESI BAO+FS from~\cite{Ishak:2024jhs},\footnote{We note that ref.~\cite{Ishak:2024jhs} co-vary $\alpha_B$ with the running of the effective Planck mass.  An additional distinction lies in their treatment of the sound speed: instead of enforcing $c_s^2 = 1$, they fix $\alpha_K^0 = 10^{-2}$.} as well as with previous studies that found a slight preference for a non-zero $\alpha_B^0$ based \textit{Planck} PR3 data~\cite{Noller:2018wyv,Seraille:2024beb}.

To understand what drives a non-zero value for $\alpha_B$,
we isolate the modifications in the Poisson and Weyl potential equations~\eqref{eq:modPoisson}~and~\eqref{eq:modWeyl} by using instead phenomenological functions $\mu(a)$ and $\Sigma(a)$, 
\begin{equation}
\partial^2 \Psi =  \frac{\bar \rho_m a^2}{2\mpl^2} \mu(a) \delta \ , \qquad \partial^2 (\Phi + \Psi) \ =  \frac{\bar \rho_m a^2}{\mpl^2} \Sigma(a) \delta \ .
\end{equation}
In the EFTofDE limit we consider in this work, those correspond to $\mu = \Sigma = 1 + \alpha_B^2/\nu$. 
We compare this case to two other cases, where $\mu \equiv 1$ and $\Sigma = 1 + \alpha_B^2/\nu$, modifying only the Weyl potential equation, or $\mu = 1 + \alpha_B^2/\nu$ and $ \Sigma \equiv 1$, modifying only the Poisson equation. 
Assuming $\Lambda$, we find $\alpha_B^0 \neq 0$ at $\sim 2\sigma$ when $\Sigma$ is free to deviate from general relativity, whether $\mu$ is free or not, while $\alpha_B^0$ is consistent with $0$ within $1\sigma$ when $\Sigma = 1$.\footnote{On this test, we use \textit{Planck} PR3 likelihood only (including lensing) for simplicity, as we find that the constraints on $\mu$ and $\Sigma$ are practically unmodified when adding LSS data. We make use of \texttt{MGCLASS-II}~\cite{Sakr:2021ylx}. } 
This implies that the deviation we see in $\alpha_B$ is mainly driven by the modification in the Weyl potential equation, and thus originate from the lensing contributions and integrated Sachs-Wolfe (ISW) effects in the CMB data. 
We note that $\mu$ and $\Sigma$ are correlated with the $A_{\rm lens}$ anomaly of \textit{Planck} PR3 data~\cite{Calabrese:2008rt,Renzi:2017cbg,Mokeddem:2022bxa}, which is alleviated in the \textit{Planck} PR4 data~\cite{Rosenberg:2022sdy,Tristram:2023haj}, thus leading to a better consistency with general relativity \cite{Specogna:2024euz}. 
As suggested in ref.~\cite{Ishak:2024jhs}, the preference for a nonzero $\alpha_B^0$ could be reduced if we consider \textit{Planck} PR4 instead of \textit{Planck} PR3. 
See~ref.~\cite{Lu:2025sjg} for a recent analysis in this direction, in which $\alpha_B$ is found to be more consistent with $0$ due to the reduced lensing anomaly in \textit{Planck} PR4 and the inclusion of extra ISW information, while deviations from $\Lambda$ are still observed in the extended setup considered there.


\section{Discussions} \label{sec:discussion}
In this section, we first investigate how the preference we see for evolving dark energy arises across cosmic history (sec.~\ref{sec:histories}). 
We then assess the significance for $w<-1$ (sec.~\ref{sec:phantom}).  
Since this possibility is allowed by the data, we check that the stability conditions on the propagation of dark energy fluctuations laid in sec.~\ref{sec:theory} are met (sec.~\ref{sec:stability}).

\subsection{Dark energy histories}\label{sec:histories}
To understand the origin of the preference for $(w_0, w_a)$ over $\Lambda$, we reconstruct the evolution of dark energy $w(z)$ across cosmic history preferred by the data, delimiting the redshift range within which the data is sensitive to $w$. 
As a warm up, we identify the pivot epoch $z_* \sim 0.25$ for which $w(z)$ is best constrained. 
Next, we reconstruct $w(z)$, either assuming the CPL $(w_0, w_a)$-parametrisation or in a model-agnostic way.
The reconstruction methodology is relegated to app.~\ref{app:recon}. 
This comparison allows us to identify the deviations from $\Lambda$ across redshifts driving the preference for $(w_0, w_a)$. 

\begin{figure}[h]
    \centering
    \includegraphics[width=0.49\textwidth]{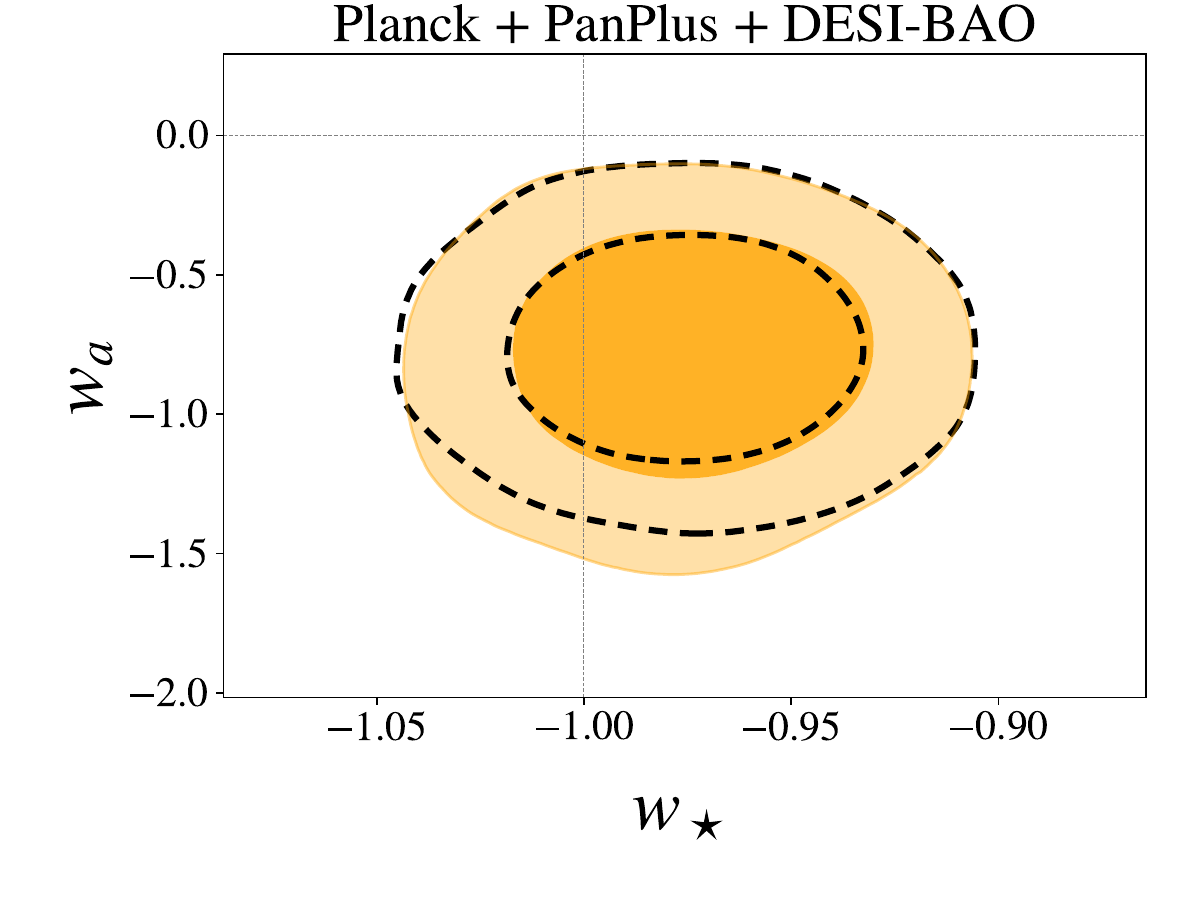} 
    \includegraphics[width=0.49\textwidth]{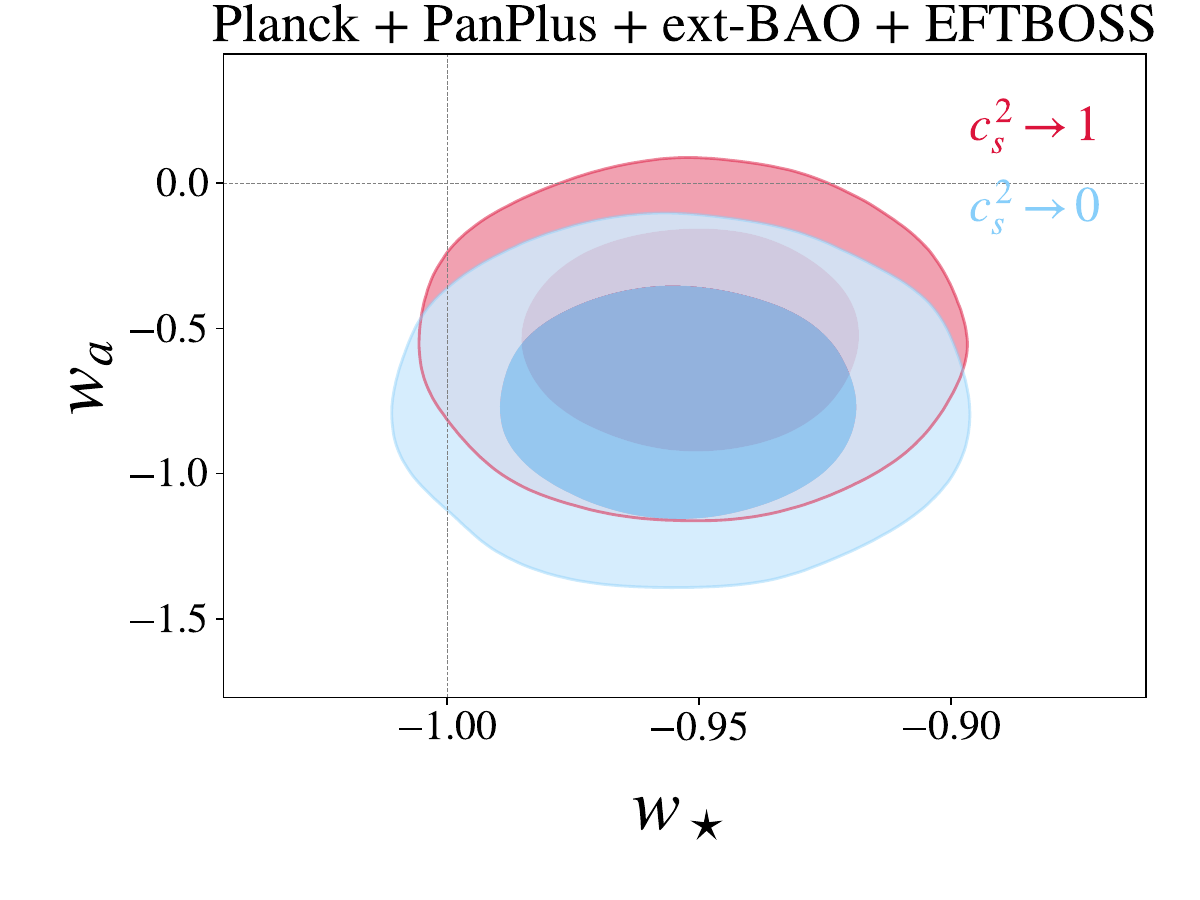} 
    \caption{
    \textbf{$w$ at the pivot $z_*$} --- Posterior distributions in the $w_\star-w_a$ plane from \textit{Planck} + PanPlus + DESI-BAO (\textit{left}) or from \textit{Planck} + PanPlus + ext-BAO  + EFTBOSS (\textit{right}) for smooth (red contour) or clustering (blue contour) quintessence. 
    While we obtain those results  by directly sampling  $w_*$ and $w_a$ in the MCMC, the dashed contour is obtained under the Gaussian approximation (\textit{i.e.}, minimising $\sigma_{w_*}$ in eq.~\eqref{eq:fisher}), displaying good agreement. 
    }
    \label{fig:wpwa}
\end{figure}

\paragraph{$w$ at the pivot} 
First, it is useful to look at the constraints on ($w_\star, w_a$)~\cite{Albrecht:2006um}, where $w_\star$ is defined at the pivot epoch $a_\star$, such that 
\begin{equation}
w_\star = w_0 + (1-a_\star) w_a ,
\end{equation}
yielding $w(a) = w_\star + (a_\star - a) w_a$. 
The pivot epoch $a_\star$ is chosen such that $w_\star$ has minimal variance. 
As such, $w_\star$ can be seen as the principal component of the ($w_0, w_a$)-contours, \textit{i.e.}, the major semi-axis of the ellipse~\cite{Cortes:2024lgw}, implying that $w_\star$ and $w_a$ are fully uncorrelated. 
While $w_\star$ is the best constrained value of $w$ across cosmic history, $w_a$ now becomes the local, linearised variation of $w(a)$ around $a_\star$. 
To find $a_\star$, we use the fact that the Fisher matrices $\mathcal{F}^\star$ and $\mathcal{F}^0$ of $\pmb{w^\star} \equiv (w_\star, w_a)$ and $\pmb{w^0} = (w_0, w_a)$, respectively, are related by 
\begin{equation}\label{eq:fish}
\mathcal{F}^\star_{\mu\nu} = J_{\mu\alpha} \mathcal{F}^0_{\alpha\beta} J_{\beta\nu} \ ,
\end{equation}
through the Jacobian 
\begin{equation}
J_{\mu\alpha} = \partial w^0_\mu / \partial w^\star_\alpha = \begin{pmatrix}1 & a_\star-1 \\ 0 & 1 \end{pmatrix} \ .
\end{equation} 
Writing
\begin{equation} \label{eq:fisher}
\left(\mathcal{F}^{\star}\right)^{-1}_{\mu\nu} = \begin{pmatrix} \sigma_{w_\star}^2 & \sigma_{w_\star} \sigma_{w_a} \rho \\
\sigma_{w_\star} \sigma_{w_a} \rho  & \sigma_{w_a}^2  \end{pmatrix} \ ,
\end{equation}
we can find the solution of $a_\star$ such that $\rho \equiv 0$. 
Equivalently, we can find $a_\star$ by minimising $\sigma_{w_\star}$ (and check that $\rho \simeq 0$). 
Regardless the method we use, we get $z_\star \simeq 0.25-0.27$ for \textit{Planck} + PanPlus + ext-BAO + EFTBOSS and \textit{Planck} + PanPlus + DESI.
We show the 2D posterior distributions in the $(w_\star,w_a)$ plane for both smooth and clustering quintessence in fig.~\ref{fig:wpwa}. 
We summarise our findings as follows:
\begin{itemize}
\item The best constrained value $w_*$ across cosmic history is above while being consistent with $-1$ at $\lesssim 1.7\sigma$ for all dataset combinations considered;
\item The  time derivative $w_a$ is nonzero at $1.5\sigma$ when analysing BOSS with $c_s^2 \rightarrow 1$ and $2.3\sigma$ when analysing either BOSS with $c_s^2 \rightarrow 0$ or DESI BAO. 
\end{itemize}
Therefore, while data mainly constrain $w$ around $z_* \sim 0.25$ at a value marginally consistent with $-1$, another culprit behind the deviation from $\Lambda$, that induces a non-negligible time variation (\textit{i.e.}, $w_a \neq 0$), is to be found elsewhere. 

\begin{figure}[!h]
    \centering
    \begin{minipage}{0.63\textwidth}
        \centering
         \includegraphics[width=1.\textwidth]{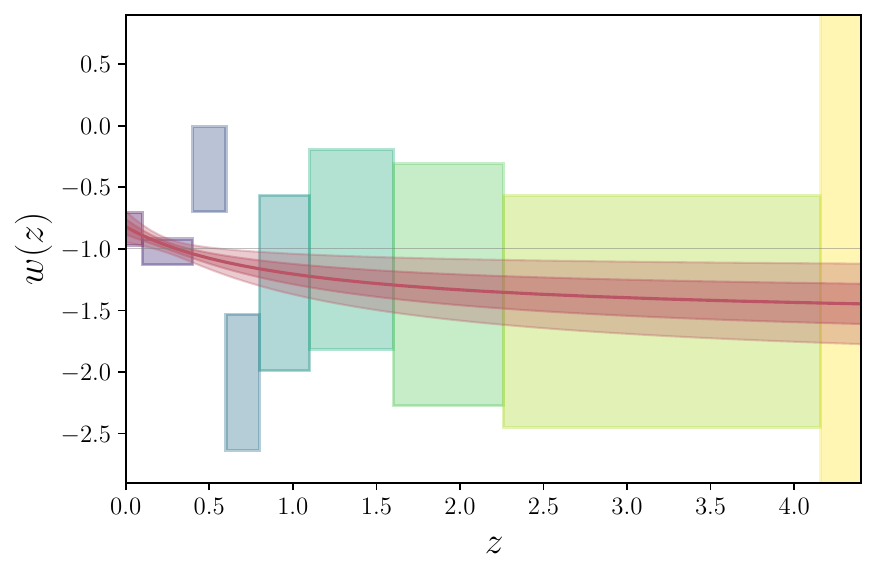}
    \end{minipage}
    \hfill
    \raisebox{0.15\height}{ 
        \begin{minipage}{0.34\textwidth}
            \centering
            \includegraphics[width=1.\textwidth]{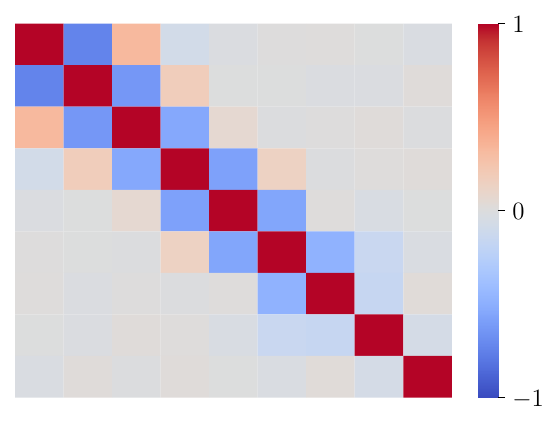}
        \end{minipage}
    }
    \caption{
\textbf{Reconstructed dark energy equation of state $w(z)$ across cosmic history} --- \textit{Left panel}: ``Model-independent'' piecewise parametrisation $\lbrace w_i^z \rbrace$, $i=1, \dots, 9$, where the coloured rectangles correspond to the $68\%$ credible intervals of $w(z)$ for the redshift ranges considered (with $w_9^z$ partially depicted), compared to the reconstructed $w(z)$ under the $(w_0, w_a)$ parametrisation, where the $68\%$ and $96\%$ credible regions are shown in shaded red. 
\textit{Right panel}: Correlation matrix of $\lbrace w_i^z \rbrace$. 
}
    \label{fig:reconst_wz}
\end{figure}

\paragraph{$w(z)$ histories}

\begin{figure}[!h]
    \centering
    \includegraphics[width=0.69\textwidth]{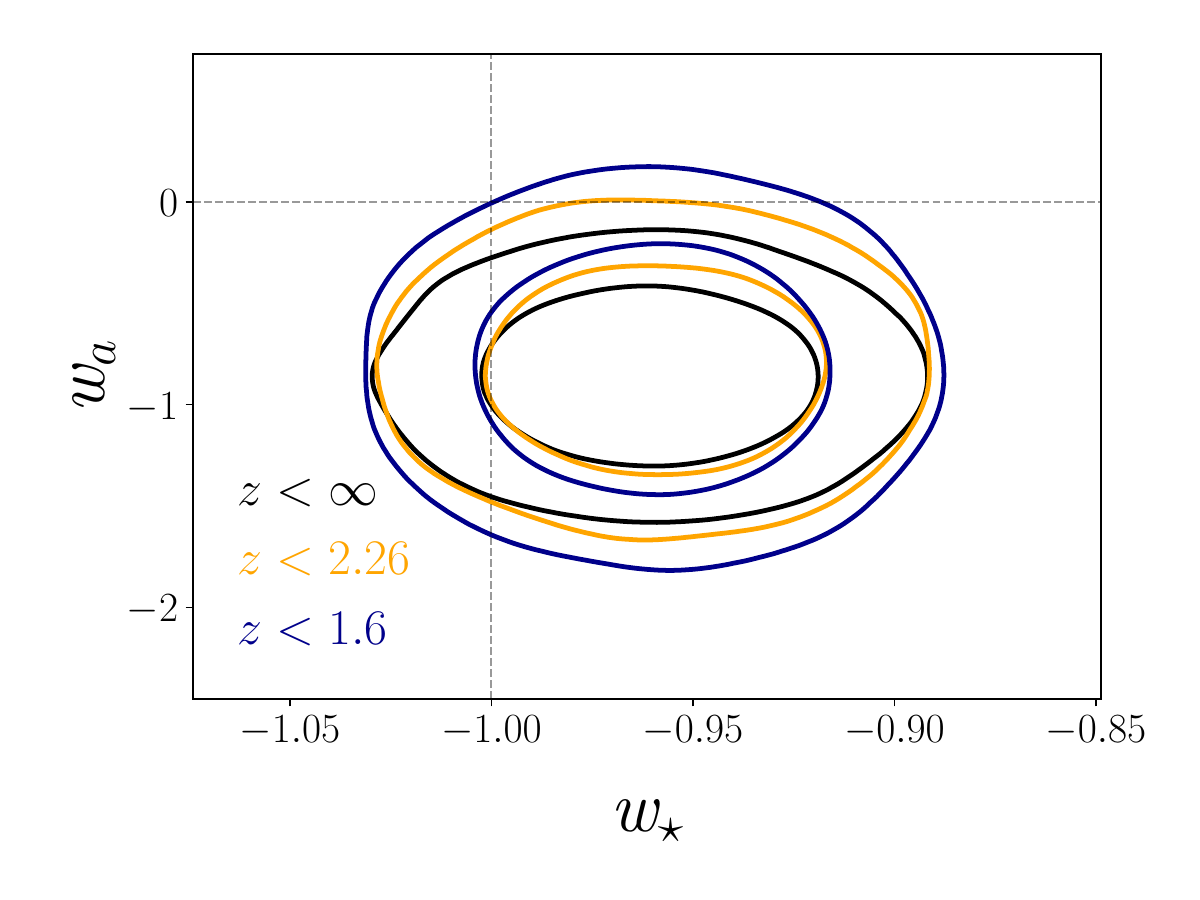}
    \caption{\textbf{Data redshift sensitivity window on $(w_*, w_a)$} --- Comparison of $(w_*, w_a)$ constraints when freeing the parametrisation of $w(z)$ for $z> z_t$, where $z_t = 1.6$ or $z_t = 2.26$, with floating parameters $w^{z}_i$ for $z_i>z_t$. 
    The constraint on $w_*$, coming from $z_* = 0.25$, remains practically untouched, while the one on $w_a$ is progressively relaxed as we scan over decreasing $z_t$. In this plot, we use the \textit{Planck} + PanPlus + DESI-BAO dataset and we reconstruct $(w_*, w_a)$ according to the method developed in app.~\ref{app:recon}.
    }
    \label{fig:w0wa_wzinf}
\end{figure}

In fig.~\ref{fig:reconst_wz}, we show $w(z)$ across cosmic history reconstructed from the posterior of $(w_0, w_a)$ obtained in sec.~\ref{sec:results}. 
This is compared with the reconstruction of $w(z)$ from a ``model-independent'' approach based on a general piecewise parametrisation $\{w_i^z\}$, $i=1, \dots, N$, with $N=9$. 
For simplicity, we use here \textit{Planck} + PanPlus + DESI-BAO. 
Given our findings in sec.~\ref{sec:mainresults}, we do not expect our conclusions to change significantly if alternative dataset combinations are used. 
Our choice of redshift bins $\{w^z_i\}$ roughly follows the division of galaxy clustering data from DESI, divided into $7$ redshift bins between $z \in [0.1, 4.16]$, that we further complement with a bin at $z \in [0, 0.1]$ and a bin at $z \in [4.16, \infty]$. 
Details on our reconstruction methodology are provided in app.~\ref{app:recon}, where we also explain how the reconstructed $w(z)$ which assumes the $(w_0, w_a)$ parametrisation can be obtained by adjusting $\{w_0, w_a\}$ to the posterior of $\{w_i^z\}$. 
From fig.~\ref{fig:reconst_wz}, we make several observations:
\begin{itemize}
\item The constraint on $w_*$ primarily originating from low redshifts around $z_* \sim 0.25$ can be traced to the error bars on $w_z^1$ and $w_z^2$ being significantly smaller than those for the other $w^z_i$. 
The mild preference for $w_* > -1$ seen in fig.~\ref{fig:wpwa} is driven by $w_z^1$, \textit{i.e.}, low-redshift supernovae. 
For $z \gg 1$, the variance in $w(z)$ increases as expected, given that $\Omega_d/ \Omega_m \ll 1$.\footnote{To avoid unnecessary sampling in the high redshift bins where parameters such as $w^z_9$ are practically unconstrained, we fit with restricted, yet wide, flat prior range $-3<w^z_i<1$.}
\item The constraint on $w_a$, instead, is mainly driven by multiple $w^z_i$ spanning from $z \sim 0.6$ to $z \sim 4$, which consistently lie below $-1$ to various degrees.
Notably, at face value, the data around $z\sim 0.7$ show a departure from $\Lambda$ at almost $2\sigma$, though this should be interpreted with caution, as discussed in the next point.
\item Since the transverse BAO parameter $D_A$ and the luminosity distance $D_L$ are integrated quantities along the line of sight, and supernovae and CMB lensing span multiple redshift bins, the estimated $w^z_i$ exhibit an approximate $\sim 50-70\%$ anticorrelation (see right panel of fig.~\ref{fig:reconst_wz}) between neighbouring bins up to redshift $z \sim 2.2$, the maximal redshift covered by PanPlus supernovae. 
This implies that the apparent deviation in $w(z)$ around redshift $z\sim 0.7$ when inferred from the diagonal of the covariance is not solely driven by data at $z\sim 0.7$ but also influenced by data from lower and higher redshifts. 
\item From the model-independent reconstruction, we find a $1.7\sigma$ deviation from $\Lambda$ (correspondingly $\Delta \chi_{\rm min}^2 = -15.2$), counting $N=9$ degrees of freedom. 
For the first $5$ bins alone (for which deviations from $\Lambda$ might appear significant), we find $2.1\sigma$ instead. 
The data themselves thus provide only mild evidence for departure from $\Lambda$. 
\item The last bin, $w^z_9$, has no impact in the posterior of $(w_0, w_a)$, as deducted by reconstructing $(w_0, w_a)$ using eq.~\eqref{eq:psum} from the posterior of $\{w_i^z\}$, removing $w^z_9$. 
Further removing $w^z_8$, and subsequently $w^z_7$, leads to the $(w_0, w_a)$-posteriors shown in fig.~\ref{fig:w0wa_wzinf}, which we discuss in sec.~\ref{sec:stability}. 
\end{itemize}
Finally, the preference over $\Lambda$ seen with $(w_0, w_a)$ is not solely driven by the multiple deviations from $\Lambda$ visible across cosmic history, but also by the inherent restriction of the $(w_0, w_a)$-parametrisation as discussed in app.~\ref{app:recon}. 
See refs.~\cite{DESI:2024aqx,Raveri:2021dbu,Pogosian:2021mcs,Ye:2024ywg,Yang:2025kgc} for alternative approaches to reconstructing dark energy evolution.

\subsection{Quantifying the phantom evidence}\label{sec:phantom}
For values of $(w_0, w_a)$ satisfying $w_a (a-1) > (w_0 + 1)$, dark energy necessarily becomes phantom ($w(a)<-1$) at some point in the past. 
Given that $(w_0, w_a)$ is merely a parametrisation, enforcing this behaviour appears as a strong theoretical assumption. 
A committed Bayesian could argue that ultimately data serve as the referee: since the data favours $(w_0, w_a)$, this specific `model' is naturally selected, even if it lies in the region where $w<-1$. 
However, to our knowledge, no known UV-complete construction allows for phantom dark energy.\footnote{See however, \textit{e.g.}, refs.~\cite{Csaki:2005vq,Fang:2008sn,Cai:2009zp,Deffayet:2021nnt,Wolf:2024stt}. } 
Given a \emph{zero measure} in the dark energy model space for $w(z) < -1$, this region would then be prohibited in a data analysis. 
When such prior is imposed, and given the inherent constraint of $(w_0, w_a)$ mentioned above, the evidence for evolving dark energy over $\Lambda$ is then immediately eliminated, as illustrated in the left panel of fig.~\ref{fig:stability} (see orange contour). 

Yet, as seen in sec.~\ref{sec:EFTofDE}, stable theories at low energies for $w<-1$ can be build in the EFT, irrespectively of known UV completions. 
We thus quantify the significance for $w<-1$ throughout cosmic history from the constraints obtained in sec.~\ref{sec:results}. 
To do so, we assess how much more likely the data $\mathcal{D}$ support, under a given parametrisation $\pmb \phi$ for $w(t)$ --- \textit{i.e.}, $w(t) \equiv w(t|\pmb{\phi})$ ---, ``\emph{$w<-1$ somewhere}'' than ``\emph{$w\geq-1$ everywhere}'', the latter being our null hypothesis (\textit{i.e.}, no phantom behaviour).
Denoting the events ``\emph{$w<-1$ somewhere}'' (for at least one time $t$) and ``\emph{$w\geq-1$ everywhere}'' (for all times $t$) as $w^{\exists}<-1$ and $w^{\forall}\geq-1$, respectively, this is answered by computing the log-probability ratio
\begin{equation}\label{eq:phantom}
\lambda = -2\ln \left[ \frac{ \sup_{\{ \pmb{ \phi}|w^{\exists} < -1\}} \mathcal{P}(\pmb \phi|\mathcal{D)}}{\sup_{\{ \pmb{\phi}|w^{\forall} \geq -1\}} \mathcal{P}(\pmb \phi|\mathcal{D})} \right] \ ,
\end{equation}
from which the significance level is determined as usual.\footnote{Here $\sup_{\{\pmb{\phi}\in S\}} \mathcal{P}(\pmb{\phi})$ denotes the supremum of $\mathcal{P}(\pmb{\phi})$ over all $\pmb \phi \in S$. }
Eq.~\eqref{eq:phantom} represents the $\chi_{\rm min}^2$-difference found maximising $\mathcal{P}(\pmb \phi|\mathcal{D)}$ with or without imposing $w\geq-1$ throughout cosmic history.\footnote{In practice, we do not impose the condition $w^{\exists} < -1$ to determine $\sup_{\{ \pmb{\phi}|w^{\exists} < -1\}} \mathcal{P}(\pmb \phi|\mathcal{D)}$, since the best fit already lies in a region where $w^{\exists} < -1$. }
Under the $(w_0, w_a)$ parametrisation, we find a preference for $w<-1$ at $2.3 \sigma$ for \textit{Planck} + PanPlus + DESI-BAO, while we obtain $1.4\sigma$ $(2.1 \sigma)$ for $c_s^2 \to 0$ ($c_s^2 \to 1$)  when we replace DESI-BAO by ext-BAO + EFTBOSS, and $2.7\sigma$ with DESI-BAO + EFTBOSS/$r_s^{\rm marg}$ for $c_s^2 \to 0$.
There is thus a marginal hint for phantom behaviour when assuming $(w_0, w_a)$, albeit weaker than the preference for $(w_0, w_a)$ over $\Lambda$. 

Within our model-independent approach using the piece-wise $\{w_z^i\}$ parametrisation, we find instead $\lambda = 14.3$, yielding a $1.6\sigma$ significance for phantom behaviour, where we count $9$ degrees of freedom.\footnote{One may wonder why we count the number of additional degrees of freedom $N_{\rm dof}$ to be the number of model parameters $N$, given that the same model is assumed in the alternative and null hypotheses in eq.~\eqref{eq:phantom}. 
In the null hypothesis, we consider the model parameters only within the region for which $w^\forall \geq -1$. 
The alternative hypothesis provides $N$ extra freedoms to produce $w^\exists<-1$ somewhere. 
Said in a different way, from a frequentist perspective, we are really letting $N$ extra parameters to vary to cover the range inaccessible by the model parameters constrained to $w^\forall \geq -1$. 
Note that in general $N_{\rm dof} \neq N$ since it is model-specific: in some parametrisations, some parameters will not allow to reach $w^\exists<-1$. 
}
This calculation matches the intuitive picture we get from fig.~\ref{fig:reconst_wz} once accounting for the `look-elsewhere' effect: deviations from $w = -1$ arising from random fluctuations should not be mistaken as physical signals.

\subsection{Stability for $w<-1$}\label{sec:stability}

Building on our previous discussion about where $w(z)$ is constrained across cosmic history, one takeaway is that cosmological data primarily constrain $w$ and its local variation within a limited redshift range. 
Consequently, any extrapolation of the linear parametrisation \((w_0, w_a)\) in the far past should only be considered if there is a strong theoretical motivation to assume that $w(z)$ continues to evolve in this specific parametric form.  
Since no such theoretical prior appears evident to us, we appeal for the possibility that $w(z)$ may evolve differently beyond the redshift range where the data are sensitive (see refs.~\cite{Wolf:2023uno,Cortes:2024lgw,Shlivko:2024llw,Gialamas:2024lyw} for related discussions).
In principle, the sensitivity of the data to $w$ decreases smoothly as $\Omega_d(z)/\Omega_m(z) \to 0$, corresponding to redshifts deep in the matter-dominated era. 
Therefore, to simplify the discussion, we introduce a transition redshift, $z_t = 2.26$, as a representative threshold. 
Below this redshift, we assume that $w(z)$ follows the \((w_0, w_a)\) parametrisation, while above $z_t$, we leave it unconstrained.\footnote{Above $z_t$, we use floating parameters, \textit{e.g.}, $w^z_9$ introduced in sec.~\ref{sec:histories}. }  
This choice is well justified, as we find that allowing $w(z)$ to vary freely above $ z_t$ only marginally relaxes the constraints on \((w_0, w_a)\), as shown in fig.~\ref{fig:w0wa_wzinf}. 

\begin{figure}[h]
    \centering
    \includegraphics[width=0.49\textwidth]{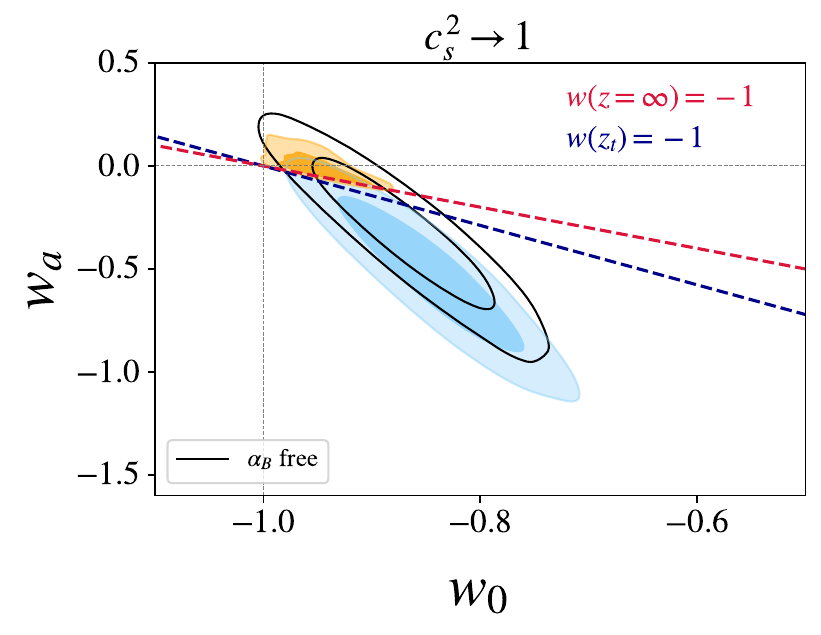} 
    \includegraphics[width=0.49\textwidth]{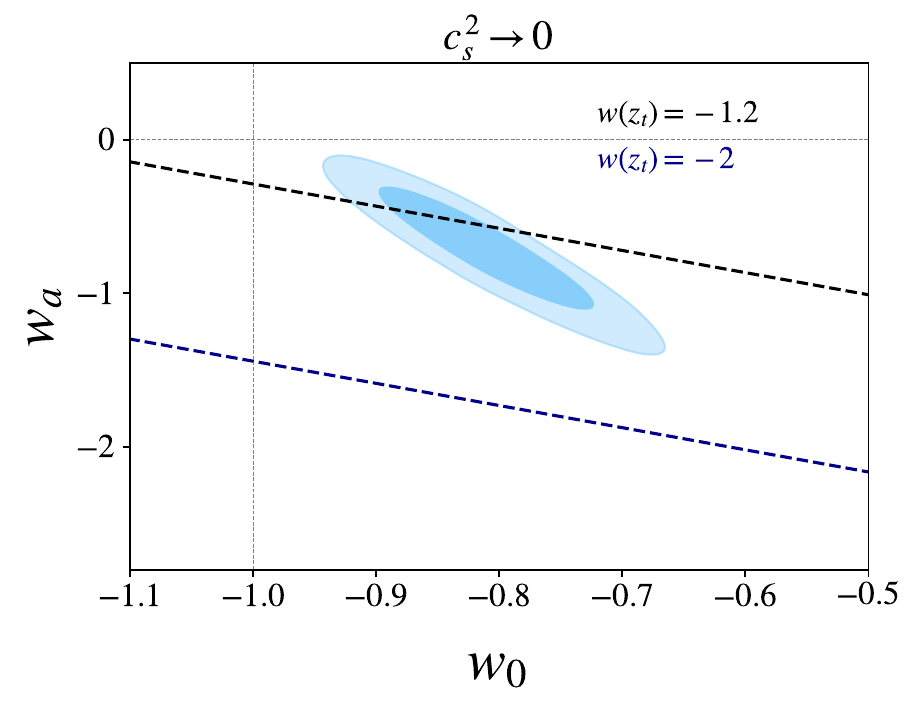}
    \caption{
    \textbf{Stability of dark energy fluctuations} --- 2D posterior distributions in the $w_0-w_a$ plane with representative stability conditions (dashed lines) discussed in the main text for smooth (\textit{left}) and clustering (\textit{right}) quintessence (in cyan). Black and blue lines represent stability conditions on the redshift sensitivity range of the data, extending up to $z_t \simeq 2.26$, while the red line is on the whole cosmic history. 
The resulting posterior restricted to $w > -1$ throughout cosmic history for $c_s^2 \to 1$ (in orange) is shown for illustration. 
We also show in black contour  the posterior obtained in the $c_s^2\rightarrow 1$ limit with the modified Poisson equation~\eqref{eq:modPoisson}, where dark energy fluctuations remain stable in the $w<-1$ limit. }
    \label{fig:stability}
\end{figure}

We now assess the significance of our results in light of the stability conditions governing the propagation of dark energy fluctuations laid in sec.~\ref{sec:EFTofDE}. 
As highlighted there, $w<-1$ is allowed if possible operators in the EFT are present (irrespectively of potential UV completions). 
For each class of theories identified in sec.~\ref{sec:limits}, we show in fig.~\ref{fig:stability} the priors on $w(z)$ that ensure stability,
especially within the redshift range where the data are sensitive to $w$, namely $z \lesssim z_t$. 
For \textit{Planck} + PanPlus + ext-BAO + EFTBOSS, our findings are as follows:
\begin{itemize} 
    \item \emph{$k$-essence models}, for which $w < -1$ is prohibited, remains consistent with $\Lambda$, whether we impose $w<-1$ as a strict prior throughout cosmic history (see the orange contour above the red dashed line in the left panel of fig.~\ref{fig:stability}) or within the data redshift sensitivity window, \textit{i.e.}, up to $z = z_t$ (see the blue dashed line in the left panel of fig.~\ref{fig:stability}). 
    We note that refs.~\cite{Creminelli:2017sry,Creminelli:2018xsv,Creminelli:2019nok} have shown that the EFTofDE reduces to $k$-essence based on constraints imposed by the speed of graviton measured by LIGO/Virgo~\cite{LIGOScientific:2017vwq} and its potential decay into dark energy, assuming the EFTofDE to be valid up to the energy scales observed at LIGO (with caveats discussed in ref.~\cite{deRham:2018red}).
    \item \emph{The $c_s^2 \rightarrow 1$ limit} is stable for $w < -1$ when additional EFTofDE parameters stabilising the gradient are present (see black contour in the left panel of fig.~\ref{fig:stability}), as highlighted in sec.~\ref{sec:EFTofDE}. 
    The case with free $\alpha_B$ was presented in sec.~\ref{sec:mg}, for which a $2.9\sigma$ preference over $\Lambda$ was found. 
    \item \emph{The $c_s^2 \rightarrow 0$ limit} permits stable dark energy fluctuations for $w < -1$, but only if $w \gtrsim -1.2$ when stabilised via $\alpha_B$ or if $w \gtrsim -2$ when stabilised using higher-order operators~\cite{DAmico:2020tty}. 
    As discussed in sec.~\ref{sec:EFTofDE}, assuming the presence of such additional operators has no incident on the cosmological analysis.   
    These stability conditions are displayed in the right panel of fig.~\ref{fig:stability}. 
    While the significance over $\Lambda$ would also reduce, by roughly a half, in the sole presence of $\alpha_B$, it stays intact if the gradient instability is safely confined at large scales by higher-derivative corrections. 
\end{itemize}
In summary, consistently enforcing stability conditions on the propagation of dark energy fluctuations can influence the results. 
Ultimately, a significant preference for evolving dark energy arises in the $c_s^2 \rightarrow 0$ limit assuming the presence of higher-derivative operators or in the $c_s^2 \rightarrow 1$ limit stabilised by $\alpha_B$.


\section{Conclusions}\label{sec:conclusion}
In this paper, we have looked for deviations from $\Lambda$ by probing the time evolution of the dark energy equation of state $w(t)$ in the cosmological data, primarily using the $(w_0, w_a)$ parametrisation. 
One key novelty is the inclusion of the one-loop bispectrum, which is found to be essential for constraining dark energy evolution.
We summarise our main findings as follows:  
\begin{itemize}

    \item Considering the combination of \textit{Planck}, Pantheon+ supernovae, ext-BAO measurements, with BOSS power spectrum and bispectrum, we find a preference for evolving dark energy (using the $w_0w_a$CDM model) at $2.6 \sigma$ for $c_s^2 \to 1$ (smooth quintessence), and at $2.8 \sigma$ for $c_s^2 \to 0$ (clustering quintessence). 
    In particular, the combination of PanPlus and EFTBOSS efficiently breaks degeneracies in the $h - w_0/w_a$ and $\Omega_m- w_0/w_a$ planes (see fig.~\ref{fig:combination}).
    The preference increases at $3.4 \sigma $ ($3.7 \sigma$) when we replace Pantheon+ by Union3 (DESY5) for $c_s^2 \to 1$, while it increases at  $3.6 \sigma $ ($3.9 \sigma$) for $c_s^2 \to 0$.
    
    \item The preference for evolving dark energy {under $w_0w_a$CDM} is observed consistently across two independent galaxy clustering datasets, namely SDSS/BOSS DR12 and DESI Y1, when they are combined with CMB and supernova data. This suggests that it is unlikely that the detected signal is due to an instrumental systematic effect in either DESI or BOSS, as such a bias would have to affect both experiments in the same way. 
    In addition, since the preference vanishes when supernova data are excluded from the analysis, further investigation is required to clarify the role played by supernovae in driving this result (see \textit{e.g.},~refs.~\cite{Efstathiou:2024xcq,Huang:2025som} for discussions regarding DESY5 supernovae).\footnote{{While this paper was under review, a re-analysis of DESY5 supernovae has appeared~\cite{DES:2025sig}, bringing the three supernova datasets considered in this work into better agreement. 
    This development highlights that systematics at the $\sim 1\sigma$ level remain under active investigation. 
    Accordingly, the main results presented in this work, particularly those based on Pantheon+ or Union3, should reflect more accurately the current situation.
    }}

    \item The preference from pre-DESI data shows up upon inclusion of the BOSS bispectrum, which is analysed on an extended $k$-range thanks to the one-loop predictions from the EFTofLSS. Since our analysis follows a different methodology than that used in DESI Y1 BAO, and given the consistency checks outlined in sec.~\ref{sec:checks}, our findings suggest that this evidence is not an artifact of systematic biases in the analysis methods. 
    Moreover, our work motivates an analysis of DESI data including the bispectrum at one-loop. 

    {In fact, the origin of the preference for evolving dark energy is, perhaps, better understood as a tension between CMB, LSS, and SN data along the $\sim \Omega_m h^2$ direction. Within $\Lambda$CDM, and relative to the value inferred from the CMB, DESI prefers a higher one at $\sim 2\sigma$, while SN prefers a lower one at $1\sigma$. 
    Since the inclusion of BOSS one-loop bispectrum shifts $\Omega_m h^2$ upwards by $\sim 1\sigma$ relative to the BOSS power-spectrum-only analysis~\cite{DAmico:2022osl}, the results on $w_0w_a$CDM obtained in the present work are therefore not entirely a surprise, and could have been, to some extent, anticipated. 
    }

    \item Combining DESI Y1 BAO with BOSS power spectrum and bispectrum, together with \textit{Planck} and Pantheon+ supernovae, we find a $3.7\sigma$ preference over $\Lambda$ in the $c_s^2 \to 0$ limit within $w_0w_a$CDM. 
    This preference increases at $3.8 \sigma$ ($4.4 \sigma$) when replacing Pantheon+ with Union3 (DESY5).
    These results are obtained marginalising over the sound horizon in BOSS full-shape analysis {to partially mitigate} the correlation with DESI BAO measurements (see sec.~\ref{sec:dataset} {and app.~\ref{app:bossxdesi}}). 
    Since a significant fraction of objects observed by DESI Y1 were already observed by BOSS~\cite{DESI:2024aax} --- a fraction that will increase with future DESI data releases --- further work is needed to properly account for the correlations between these datasets (at the covariance level) to fully exploit the constraining power of LSS {(see~ref.~\cite{Forero-Sanchez:2026bff} for a recent work in this direction)}.  
    These results also underscore that the bispectrum carries substantial information beyond the BAO, further motivating sound horizon-free analyses~\cite{Lu:2026uph}.  

    \item The preference over $\Lambda$ depends on assumptions regarding the propagation of dark energy fluctuations and their stability conditions. 
    The standard $c_s^2 \to 1$ limit is either disfavoured compared to the $c_s^2 \to 0$ limit if restricting to the $k$-essence regime (practically indistinguishable from $\Lambda$, as shown in fig.~\ref{fig:stability}), or comparably favoured when additional degrees of freedom preventing gradient instabilities for $w<-1$ are introduced (see sec.~\ref{sec:mg}), {although in this latter case, a large fraction of the departure from $\Lambda$ can be attributed to a non-zero $\alpha_B$ stemming from the CMB lensing and ISW information rather than a shift in the $(w_0, w_a)$ plane.}
    In this work, we have considered stabilisation via a single EFTofDE operator, the braiding $\alpha_B$. A more systematic study, following the approach of ref.~\cite{Taule:2024bot}, would be valuable. 
    In that context, a joint fit with galaxy lensing data --- known to efficiently break degeneracies introduced by additional EFTofDE parameters (see \textit{e.g.}, ref.~\cite{DESI:2024hhd}) --- would be useful for strenghtening the constraints in the $c_s^2 \rightarrow 1$ limit.
    Additionally, it would be interesting to investigate dark energy with intermediate values of $c_s^2$, introducing scale dependence in the growth of fluctuations, or in beyond Horndeski theories~\cite{Gleyzes:2014dya}. 

    \item Our results suggest that exploring dark energy evolution would benefit from more general parametrisations of $w(t)$ as touched upon in sec.~\ref{sec:histories}, to allow a consistent link between theoretical models and what the cosmological data can tell us on $w(t)$.\footnote{Departure from $\Lambda$ seen in $(w_0, w_a)$ can still prove useful if dark energy models are systematically mapped onto the $(w_0, w_a)$-space, properly accounting for the data sensitivity in that mapping. 
If the preference aligns with regions where viable models exist, this could serve as a powerful discriminant (see \textit{e.g.}, refs.~\cite{Linder:2015zxa,Shlivko:2024llw,Wolf:2024eph}). } 
{Indeed, when using a more flexible (model-independent) reconstruction of $w(t)$ than the one implied by $(w_0, w_a)$, the deviation from $\Lambda$ falls to $1.7\sigma$ for the representative dataset considered there. 
This reflects that the preferences over $\Lambda$ reported above stem in part by the inherent restriction imposed by the $(w_0, w_a)$ parametrisation, as discussed in length in app.~\ref{app:recon}. 
}
    Intriguingly, as outlined in {sec.~\ref{sec:phantom}}, we find, assuming the $(w_0, w_a)$ parametrisation, a significance for $w<-1$ throughout cosmic history at $1.4-2.7\sigma$ depending on the galaxy clustering dataset considered (combined with \textit{Planck} and Pantheon+) in the $c_s^2 \rightarrow 0$ limit.\footnote{For $c_s^2 \rightarrow 1$, the significance for $w<-1$ is $\lesssim 1\sigma$, while in our model-independent approach, we find $1.6\sigma$. }
    If such hint persists, it would motivate the development of UV-complete descriptions that can accommodate phantom dark energy.
\end{itemize}
{In app.~\ref{app:sys}, we have tested the accuracy of our methods for handling observational systematics. In light of upcoming high-precision surveys, it will be crucial to revisit the full-shape analysis with improved modelling and mitigation of these systematics, in particular the survey mask.} We hope to explore some of these promising directions in future work.

\section*{Aknowledgements}
We thank Davide Bianchi, Guido D'Amico, Matthew Lewandowski, Vivian Poulin, I\~nigo S\'aez-Casares, Leonardo Senatore, and Filippo Vernizzi for useful discussions or comments on the draft. 
We gratefully acknowledge Yi-Fu Cai for support. 
TS and PZ thank the Theory Group at CERN for hospitality during completion of this work. 
We acknowledge the use of computational resources from the computer clusters \textit{Linda \& Judy} at USTC, LUPM’s cloud computing infrastructure founded by Ocevu labex and France-Grilles, as well as the Euler cluster at ETH Z\"urich. 
PZ acknowledges support from Fondazione Cariplo under the grant No 2023-1205.

\appendix

\section{ Details on the EFTofLSS}\label{app:eftoflss}

\subsection{Time dependence in presence of quintessence}
\label{app:time_quint}
The growth factor $D$ is defined as the solution of
\be \label{eq:ode}
\frac{d^2 D}{d\ln a^2}+\left(2+\frac{d\ln H}{d\ln a}-\frac{d\ln C}{d\ln a}\right)\frac{d D}{d\ln a}-\frac{3}{2}\Omega_{m}CD=0 \, ,
\ee
which results from linearising the equations of motion of the smoothed fields, namely eqs.~\eqref{eq:EFTcontinuity} and~\eqref{eq:EFTEuler}. 
Here the Hubble parameter is given by eq.~\eqref{eq:H}. In the smooth ($c_s^2 \rightarrow 1$) limit, $C=1$, while it is given by eq.~\eqref{eq:C} in the clustering ($c_s^2 \rightarrow 0$) limit.  
For a constant $w$, eq.~\eqref{eq:ode} has analytical solutions given in terms of hypergeometric functions~\cite{Lee:2009gb}. 
For a general $w(a)$, while there exists a closed form solution in the clustering limit~\cite{Sefusatti:2011cm}, it is not the case in the $c_s^2 \rightarrow 1$ limit. 
In this work, we thus numerically solve eq.~\eqref{eq:ode}, starting the initial time deep inside matter domination that selects initial conditions from an EdS Universe,
\be \label{eq:inicond}
D_{+}(a)\propto a \quad \text{and}\quad D_{-}(a)\propto a ^{-3/2}\, ,
\ee
where $D_+$ and $D_-$ are the growing and decaying solutions of eq.~\eqref{eq:ode}. 
We have checked that our numerical solutions agree to closed form solutions in the limits where the latter exist. 
Accordingly, we have two solutions for the growth rate $f = \frac{d\ln D}{d \ln a}$, yielding $f_+$ and $f_-$ from the definition of $f$ using $D_+$ or $D_-$. 

Defining the Green's functions from the equations of motion as
\begin{align}
&a \frac{d G^{\delta}_{\sigma}(a,\ta)}{da}-f_{+}(a)G^{\theta}_{\sigma}(a,\ta)=\lambda_{\sigma}\delta_D(a-\ta), \label{Green} \\
&a \frac{d G^{\theta}_{\sigma}(a,\ta)}{da}-f_{+}(a)G^{\theta}_{\sigma}(a,\ta)- \frac{f_{-}(a)}{f_{+}}\left(G^{\theta}_{\sigma}(a,\ta)-G^{\delta}_{\sigma}(a,\ta)\right)=(1-\lambda_{\sigma})\delta_D(a-\ta),
\end{align}
where $\sigma \in \{1,2\}$, while $\lambda_1 = 1$ and $\lambda_2 = 0$. The solutions for the nonlinear time functions with exact time dependence can be written as
\begin{align}
&G^{\delta}_1(a,\ta)=\frac{1}{\ta W(\ta)}\bigg(\frac{d D_{-}(\ta)}{d\ta}D_{+}(a)-\frac{d D_{+}(\ta)}{d\ta}D_{-}(a)\bigg) {\Theta}(a-\ta) \label{gdelta} \ ,\\
&G^{\delta}_2(a,\ta)=\frac{f_{+}(\ta)/\ta^2}{W(\ta)}\bigg(D_{+}(\ta)D_{-}(a)-D_{-}(\ta)D_{+}(a)\bigg){\Theta}(a-\ta) \ , \\
&G^{\theta}_1(a,\ta)=\frac{a/\ta}{f_{+}(a)W(\ta)}\bigg(\frac{d D_{-}(\ta)}{d\ta}\frac{d D_{+}(a)}{d a}-\frac{d D_{+}(\ta)}{d\ta}\frac{d D_{-}(a)}{d a}\bigg) {\Theta}(a-\ta) \ ,\\
&G^{\theta}_2(a,\ta)=\frac{f_{+}(\ta)a/\ta^2}{f_{+}(a)W(\ta)}\bigg(D_{+}(\ta)\frac{d D_{-}(a)}{d a}-D_{-}(\ta)\frac{d D_{+}(a)}{d a}\bigg) {\Theta}(a-\ta) \ .\label{gtheta}
\end{align}
Here $W(\ta)$ is the Wronskian of $D_+$ and $D_-$ and
$\Theta (a-\tilde a)$ is the Heaviside step function. 
We impose the boundary conditions
\begin{align} \label{bound}
& G^{\delta}_\sigma(a,\tilde a) = 0 \quad \quad \text{and} \quad\quad G^{\theta}_\sigma(a, \tilde a)=0 \quad \quad \text{for} \quad \quad \tilde a > a \ , \\
&G^\delta_\sigma ( \tilde a , \tilde a ) = \frac{\lambda_\sigma}{\tilde a} \quad \hspace{.06in} \text{and} \hspace{.2in} \quad G^{\theta}_{\sigma} ( \tilde a , \tilde a ) = \frac{(1 - \lambda_\sigma)}{\tilde a} . \label{bound2}
\end{align}
The generalised nonlinear time functions relevant to our predictions are given by
\begin{align} 
&\mG^{\lambda}_{\sigma}(a)=\int^{1}_0 G^{\lambda}_{\sigma}(a,\ta)\frac{f_{+}(\ta) D_{+}^2(\ta) }{C(\ta) D_+^2 (a) }d\ta \ , \\ \nonumber
&\mU^{\lambda}_{\sigma}(a)=\int^{1}_0 G^{\lambda}_{1}(a,\ta)\frac{f_{+}(\ta)D^3_{+}(\ta)}{ C(\ta)D^3_+ (a) }\mG^{\delta}_{\sigma}(\ta)d\ta, \quad \mV^{\lambda}_{\sigma\tilde\sigma}(a)=\int^{1}_0 G^{\lambda}_{\tilde\sigma}(a,\ta)\frac{f_{+}(\ta)D^3_{+}(\ta)}{ C(\ta)D^3_+ (a) }\mG^{\theta}_{\sigma}(\ta)d\ta, \label{eq:timefunc}
\end{align}
where $\lambda \in \{\delta,\theta\}$ and $\sigma,\tilde{\sigma} \in \{1,2\}$.

The additional function entering the galaxy field expansions, eqs.~\eqref{eq:delta_g_expand}~and~\eqref{eq:theta_g_expand}, is given by
\bea
\tilde Y(a) = -\frac{3}{14}\mG(a)^2+\mV^{\delta}_{11}(a)+\mV^{\delta}_{12}(a) \, ,
\eea
where $\mG = \mG_1^\delta  + \mG_2^\delta$.

\subsection{Time dependence in modify gravity}\label{app:time_MG}
In this subsection, we provide the growth function and time dependence in the EFTofLSS for general dark energy and modified gravity models described by the EFTofDE in the quasi-static limit (assuming $c_s^2 \rightarrow 1$), following refs.~\cite{Piga:2022mge}. 
As discussed in sec.~\ref{sec:EFTofDE}, we need to solve the equations of motion for a modified Poisson equation, that up to third order in fluctuations, reads~\cite{Cusin:2017mzw}:  
\begin{equation}
\begin{aligned}\label{eq:poisson2}
\frac{\partial^2 \Phi}{\mathcal{H}^2} = &\frac{3 \Omega_m}{2} \mu_\Phi(t) \delta + \left( \frac{3 \Omega_m}{2} \right)^2 \mu_{\Phi,2}(t) \left[ \delta^2 - \left(\frac{\partial_i \partial_j \delta}{\partial^2}\right)^2 \right]\\
&+ \left( \frac{3 \Omega_m}{2} \right)^3 \mu_{\Phi,22}(t) \left[ \delta - \frac{\partial_i\partial_j\delta}{\partial^2}\frac{\partial_i\partial_j}{\partial^2}\right]\left[\delta^2-\left(
\frac{\partial_k\partial_l\delta}{\partial^2}
\right)^2\right]\\
&+ \left( \frac{3 \Omega_m}{2} \right)^3 \mu_{\Phi,3}(t) \left[ \delta^3 - 3 \delta \left( \frac{\partial_i \partial_j \delta}{\partial^2} \right)^2 + 2 
\frac{\partial_i\partial_j\delta}{\partial^2}\frac{\partial_k\partial_j\delta}{\partial^2}\frac{\partial_i\partial_k\delta}{\partial^2}
\right] + O(\delta^4)\, .
\end{aligned}
\end{equation}
The time-dependent functions $\mu_\Phi$ in terms of EFTofDE parameters are given in ref.~\cite{Cusin:2017mzw}. 
When only considering the braiding parameter $\alpha_B$, the equation reduces to eq.~\eqref{eq:modPoisson} at linear order. 

The linear growth function $D$ satisfies a modified second order differential equation, 
\begin{equation}
    \frac{d^2D(a)}{d\ln a^2}+\left(2+\frac{d\ln H}{d\ln a}\right)\frac{dD(a)}{d\ln a}-\frac{3}{2}\mu_\Phi(a)\Omega_m(a)D(a)=0\,.
\end{equation}
The initial conditions are set according to the strategy of ref.~\cite{Piga:2022mge} for numerical stability. The growing mode initial condition is still set deep in matter domination as in eq.~\eqref{eq:inicond}, while the decaying mode is solved future to past using the boundary condition in the far future, yielding
\begin{align}
    D_+(a\mid a\ll a_0) \propto a\,,\quad 
    D_-(a\mid a\gg a_0) \propto \left(\frac{a_0}{a}\right)^2.
\end{align}

The derivation for the nonlinear time functions then follows closely the one in sec.~\ref{app:time_quint} but with additional source terms from the modified Poisson equation~\eqref{eq:poisson2}, yielding
\begin{align}
&\mG_1^\lambda(a) = \int_0^1 \left[ G_1^\lambda(a, \tilde{a}) f_+(\tilde{a}) + G_2^\lambda(a, \tilde{a}) \mu_{\Phi,2}(\tilde{a}) M_1(\tilde{a}) \right] \frac{D_+^2(\tilde{a})}{D_+^2(a)} d\tilde{a} \ , \\ \nonumber
&\mG_2^\lambda(a) = \int_0^1 G_2^\lambda(a, \tilde{a}) \left[ f_+(\tilde{a}) - \mu_{\phi,2}(\tilde{a}) M_1(\tilde{a}) \right] \frac{D_+^2(\tilde{a})}{D_+^2(a)} d\tilde{a} \,,\\ 
&\mU_1^\lambda(a) = \int_0^1 \left\{ G_1^\lambda(a, \tilde{a}) f_+(\tilde{a})\mG_1^\delta(\tilde a) + G_2^\lambda(a, \tilde{a}) M_1(\tilde{a}) \left[\mu_{\Phi,2}(\tilde{a}) \mG_1^\delta(\tilde{a}) + M_2(\tilde{a})\right] \right\} \frac{D_+^3(\tilde{a})}{D_+^3(a)} d\tilde{a} \,,\\
&\mU_2^\lambda(a) = \int_0^1 \left\{ G_1^\lambda(a, \tilde{a}) f_+(\tilde{a}) \mG_2^\delta(\tilde{a}) + G_2^\lambda(a, \tilde{a}) M_1(\tilde{a})\left[\mu_{\Phi,2}(\tilde{a}) \mG_2^\delta(\tilde{a}) - M_2(\tilde{a}) \right]\right\} \frac{D_+^3(\tilde{a})}{D_+^3(a)} d\tilde{a}\,,\\
&\mV_{11}^\lambda(a) = \int_0^1 \left\{ G_1^\lambda(a, \tilde{a}) f_+(\tilde{a}) \mG_1^\theta(\tilde{a}) + G_2^\lambda(a, \tilde{a}) M_1(\tilde{a})\left[\mu_{\Phi,2}(\tilde{a}) \mG_1^\delta(\tilde{a}) + M_2(\tilde{a}) \right]\right\} \frac{D_+^3(\tilde{a})}{D_+^3(a)} d\tilde{a}\,,\\
&\mV_{21}^\lambda(a) = \int_0^1 \left\{ G_1^\lambda(a, \tilde{a}) f_+(\tilde{a}) \mG_2^\theta(\tilde{a}) + G_2^\lambda(a, \tilde{a}) M_1(\tilde{a})\left[\mu_{\Phi,2}(\tilde{a}) \mG_2^\delta(\tilde{a}) - M_2(\tilde{a}) \right]\right\} \frac{D_+^3(\tilde{a})}{D_+^3(a)} d\tilde{a}\,,\\
&\mV_{12}^\lambda(a) = \int_0^1 G_2^\lambda(a,\tilde a)\left\{ f_+(\tilde{a}) \mG_1^\theta(\tilde{a}) -M_1(\tilde{a})\left[\mu_{\Phi,2}(\tilde{a}) \mG_1^\delta(\tilde{a}) + M_2(\tilde{a}) \right]\right\} \frac{D_+^3(\tilde{a})}{D_+^3(a)} d\tilde{a}\,,\\
&\mV_{22}^\lambda(a) = \int_0^1 G_2^\lambda(a,\tilde a)\left\{ f_+(\tilde{a}) \mG_2^\theta(\tilde{a}) -M_1(\tilde{a})\left[\mu_{\Phi,2}(\tilde{a}) \mG_2^\delta(\tilde{a}) - M_2(\tilde{a}) \right]\right\} \frac{D_+^3(\tilde{a})}{D_+^3(a)} d\tilde{a}\,,\\
\label{eq:mg_timefunc}
\end{align}
where 
\begin{equation}
    M_1(a)\equiv \frac{1}{f_+(a)}\left(\frac{3\Omega_m(a)}{2}\right)^2\,,\quad
    M_2(a)\equiv \frac{\mu_{\Phi,22}}{2}\frac{3\Omega_m(a)}{2} \ .
\end{equation}

\subsection{Bias expansion}
In real space, the expansion of the galaxy density and velocity divergence up to third order in the presence of dark energy are given by
\begin{eqnarray} \label{eq:delta_g_expand}
		\delta_g(\vec k,t) & = & \tilde{c}_{\delta,1}(t) \; \Big( \mathbb{C}^{(1)}_{\delta, 1}(\vec k, t )+ \mG (t)\mathbb{C}^{(2)}_{\delta, 1}(\vec k, t) + \mG (t)^2\mathbb{C}^{(3)}_{\delta, 1}(\vec k, t) +\tilde Y(a)\mathbb{C}^{(3)}_{Y}(\vec k,a)\Big) 
		\\
		& + & \tilde{c}_{\delta, 2}(t) \; \Big( \mathbb{C}^{(2)}_{\delta, 2}(\vec k, t) + \mG(t) \mathbb{C}^{(3)}_{\delta, 2}(\vec k, t) \Big) \nonumber
		\\
		& + & \tilde{c}_{\delta^2, 1}(t) \; \Big( \mathbb{C}^{(2)}_{\delta^2, 1}(\vec k, t) + \mG(t)\mathbb{C}^{(3)}_{\delta^2, 1}(\vec k, t) \Big) \nonumber
		\\
		& + & \tilde{c}_{\delta, 3}(t) \; \mathbb{C}^{(3)}_{\delta, 3}(\vec k, t)+\tilde{c}_{\delta^2, 2}(t) \; \mathbb{C}^{(3)}_{\delta^2 ,2}(\vec k, t)
		 \nonumber
		\\
		& + &   \tilde{c}_{s^2, 2}(t) \; \mathbb{C}^{(3)}_{s^2, 2}(\vec k, t)+\tilde{c}_{\delta^3}(t) \; \mathbb{C}^{(3)}_{\delta^3}(\vec k, t) \ , \nonumber
\end{eqnarray}

\begin{eqnarray} \label{eq:theta_g_expand}
\theta_g(\vec k,t) & = & \; \Big( \mathbb{C}^{(1)}_{\delta, 1}(\vec k, t )+ \mG(t)\mathbb{C}^{(2)}_{\delta, 1}(\vec k, t) + \mG(t)^2\mathbb{C}^{(3)}_{\delta, 1}(\vec k, t) +\tilde Y(a)\mathbb{C}^{(3)}_{Y}(\vec k,a)\Big) 
\\
& + & \left(\frac{7}{2} - \frac{7}{2}\mG_1^\theta\right) \; \Big( \mathbb{C}^{(2)}_{\delta, 2}(\vec k, t) + \mG(t)\mathbb{C}^{(3)}_{\delta, 2}(\vec k, t) \Big) \nonumber
\\
& + &\left(-\frac{5}{2} + \frac{7}{2}\mG_1^\theta\right) \; \Big( \mathbb{C}^{(2)}_{\delta^2, 1}(\vec k, t) + \mG(t)\mathbb{C}^{(3)}_{\delta^2, 1}(\vec k, t) \Big) \nonumber
\\
& + & \left(\frac{45}{7} - 9 \mV_{12}^\theta - \frac{45}{2} \mV_{21}^\theta\right) \; \mathbb{C}^{(3)}_{\delta, 3}(\vec k, t)+\left(-\frac{22}{7} - \frac{7}{4} \mG_1^\theta + \frac{25}{6} \mV_{12}^\theta + \frac{151}{8} \mV_{21}^\theta\right) \; \mathbb{C}^{(3)}_{\delta^2 ,2}(\vec k, t)
\nonumber
\\
& + &   \left(-\frac{3}{7} + 2 \mV_{12}^\theta + \frac{3}{2} \mV_{21}^\theta)\right) \; \mathbb{C}^{(3)}_{s^2, 2}(\vec k, t)+\left(\frac{10}{7} + \frac{7}{2}\mG_1^{\theta}- 2 \mV_{12}^\theta - \frac{69}{4} \mV_{21}^\theta)\right) \; \mathbb{C}^{(3)}_{\delta^3}(\vec k, t) \ , \nonumber
\end{eqnarray}
where the expressions for the operators $\mathbb{C}$ can be found in ref.~\cite{Fujita:2016dne}, and $\mathbb{C}^{(3)}_Y$ in ref.~\cite{Donath:2020abv}. 
The full list of forth order operators in the same basis of descendants (within EdS time approximation) can be found in ref.~\cite{DAmico:2022ukl} (see also refs.~\cite{Desjacques:2016bnm,Eggemeier:2018qae} for other, equivalent bases).
For reasons mentioned in the main text, we use the exact-time dependence in the loop power spectrum, while for the bispectrum we only keep exact-time dependence at the tree level. 
The explicit redshift-space kernels up to fourth order can be found in the ancillary \texttt{Mathematica} file distributed with ref.~\cite{DAmico:2022ukl}. 

\section{Observational systematics assessment}\label{app:sys}

On the analysis side, the main novelty of this work consists in the inclusion of the galaxy bispectrum of BOSS DR12 LRG data. 
While clustering signal modelled with the EFTofLSS at one loop has been discussed in length in the main text, in this appendix, we assess the accuracy of our modelling of some critical observational effects, namely, the window functions and fibre collisions. 
We finally test our full pipeline against the Nseries $N$-body mock simulations. 

\subsection{Window functions and data covariance}

\begin{figure}[t!]
  \centering
  \includegraphics[width=0.49\textwidth]{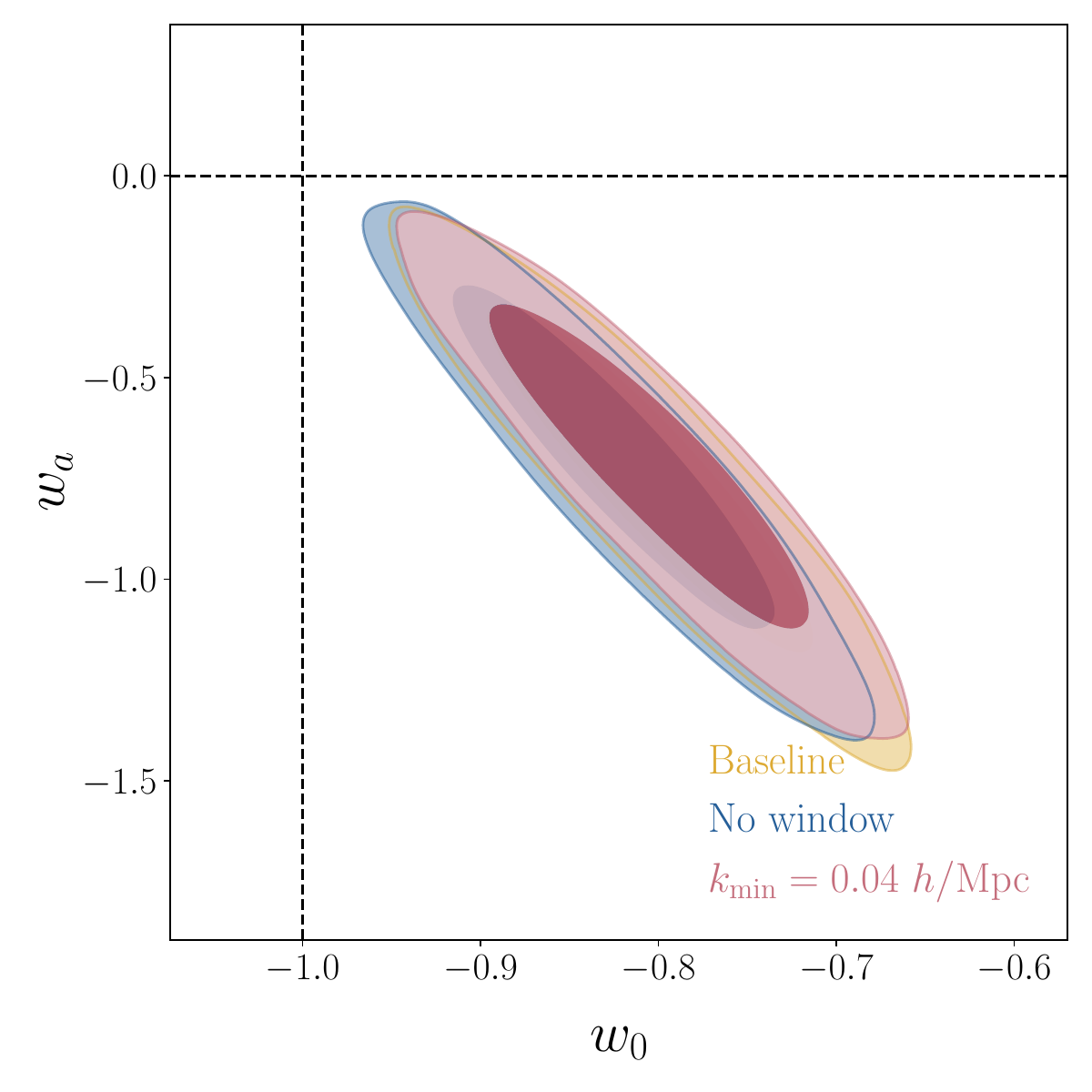}
  \includegraphics[width=0.49\textwidth]{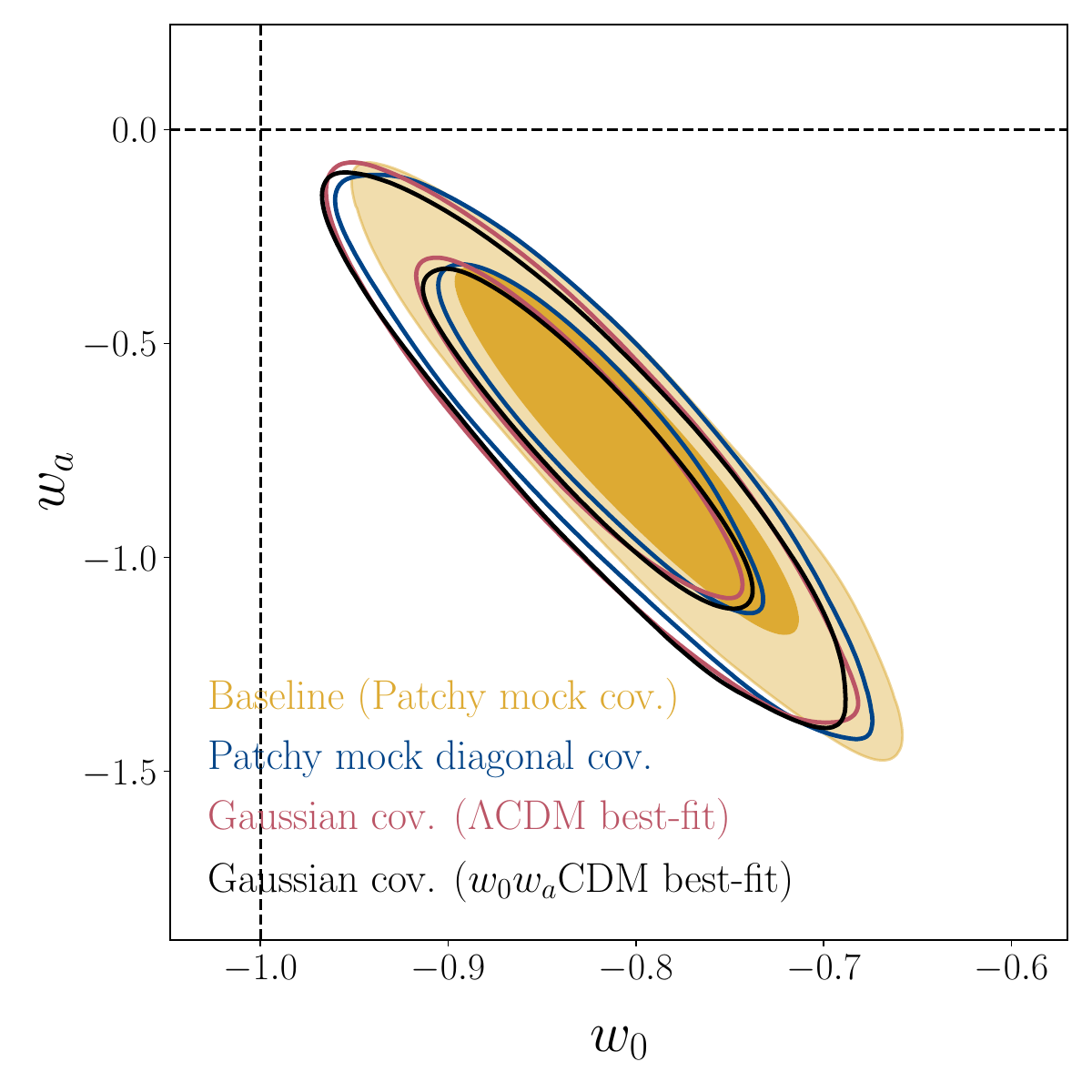}
     \caption{\textbf{Alternative bispectrum analysis setup comparison} --- $(w_0, w_a)$-posteriors from our baseline dataset for various alternative modelling choices for the bispectrum. 
     \textit{Left panel}: baseline mask treatment against no correction. Results using instead $k_{\rm min} = 0.04 \, \hinvMpc$ are also shown. 
     \textit{Right panel}: baseline covariance (patchy) against Gaussian covariance either estimated using $\Lambda$CDM or $w_0w_a$CDM best fits. 
     For all alternative configurations, the shift in the significance over $\Lambda$ are $\lesssim 0.3\sigma$. 
 }
  \label{fig:sys}
  \end{figure}

\paragraph{Window functions}
Our modelling of the window functions relies on the same approximation used in the BOSS analysis of refs.~\cite{Gil-Marin:2014sta,Gil-Marin:2016wya} (also used in a recent DESI analysis~\cite{NovellMasot:2025fju}). 
Instead of applying the window function on the external momenta, the approximation consists in applying it on the input linear power spectrum $P_{11}$. 
Explicitly, applying the correction to the tree-level bispectrum, 
\begin{equation}\label{eq:window}
  B_{211}(\pmb {k_1}, \pmb {k_2}, \pmb {k_3}) = 2  K_1 ( \pmb {k_1} ; \hat z )  K_1 ( \pmb {k_2}  ; \hat z )  K_2 ( \pmb {k_1}, \pmb {k_2} ;  \hat z ) [W \ast P_{11}](\pmb {k_1}) [W \ast P_{11}]( \pmb {k_2}) + \text{ 2 perms.} \ ,
\end{equation}
where $[W \ast P_{11}](\pmb k) = \int \frac{d^3 k'}{(2 \pi)^3} W(\pmb k - \pmb{k'}) P_{11}(\pmb k)$ and $K_1, K_2$ are the galaxy perturbation theory kernels in redshift space defined in refs.~\cite{DAmico:2022ukl,DAmico:2022osl}. 
Ref.~\cite{Pardede:2022udo} checked this approximation on large-volume simulations using a spherical window, yielding a relative difference $<5\%$, except when $k_1 < 0.04 \hinvMpc$ where the errors can reach $7\%$, and decreasing below $2-3\%$ for $k_1 > 0.08 \, \hinvMpc$ with $k_1$ the smallest side of the triangles. 
Compared to BOSS uncertainties (see \textit{e.g.},~ref.~\cite{DAmico:2022osl}) of about $5-7\%$, these represent sizeable, albeit somewhat smaller, errors. 
To assess the impact of our approximate window modelling on the inferred cosmological parameters, we follow the test devised in ref.~\cite{DAmico:2022osl}.
We compare our baseline results (with the bispectrum window functions applied in the approximate way) against results obtained without applying any corrections for the mask. 
{Since the raw effect of the mask is about $10\%$, and since our approximate correction has a relative precision of $\sim 5\%$, the measured shift in the cosmological parameters provides an estimate of the error from the mask modelling \emph{inaccurate} at the $\sim 5\%$ level. }
The results are shown in the left panel of fig.~\ref{fig:sys}. 
{We can see that our approximate treatment of the window function} leads to a tolerably small error on the significance of $(w_0, w_a)$ over $\Lambda$. It is thus unlikely that a fully correct treatment of the mask would decrease the significance. 
To further assess the impact of potential inaccuracies in our modelling of the mask, we remove from the analysis the triangles with sides smaller than $k_{\rm min} = 0.04 \hinvMpc$ (compared to $k_{\rm min} = 0.01 \hinvMpc$ in our baseline). 
As shown in the left panel of fig.~\ref{fig:sys}, the shift along the major axis of the $w_0-w_a$ ellipse is only $0.1\sigma$. 
Based on these tests, we concluded that the approximation in the window function treatment we use introduces a relatively small error in our results, and that our main findings are robust against unaccounted effects from the mask in the bispectrum modelling. {This is also consistent with the treatment adopted in recent works, e.g., refs.~\cite{Cagliari:2025rqe,DAmico:2022osl,NovellMasot:2025fju}. In particular, DESI collaboration~\cite{NovellMasot:2025fju} showed on the Abacus simulations that this window function treatment is robust for recovering cosmological parameters from LRGs, the galaxy sample considered in our BOSS analysis. Although the DESI mask differs from that of BOSS, the comparable data volume for LRGs suggests that these conclusions are unlikely to change significantly.} 
With the increasing data precision, it will be important to {quantify the accuracy of the approximation in eq.~\ref{eq:window} by comparing the measurements from periodic and cutsky mocks and} account for the effect of the mask in the bispectrum in a more accurate way, following either the method developed in ref.~\cite{Pardede:2022udo}, using the window-less estimator developed in refs.~\cite{Philcox:2021ukg,Philcox:2024rqr}, or the estimator developed in refs.~\cite{Sugiyama:2018yzo,Wang:2023zkv} relying on an angular decomposition using tripolar spherical harmonic of the bispectrum for which the window function effects can be corrected for exactly. 
We leave this to future work. 

\paragraph{Covariance matrix estimation}
As described in sec.~\ref{sec:eft_lkl}, our estimation of the data covariance relies on the patchy mock lightcone suites~\cite{Kitaura:2015uqa}. 
Those designed to be rapidly generated have a number of limitations: they are not full $N$-body simulations, nor rely on an actual halo occupation distribution to populate halos with galaxies. 
Moreover, ref.~\cite{Gil-Marin:2016wya} compared the BOSS bispectrum measurements with the mean of the patchy mock measurements, finding a systematic offset in the amplitude of $\sim 2\sigma$. 
To assess the impact of potential inaccuracies in the patchy mock covariance on our results, we fit the bispectrum using instead an analytic Gaussian covariance. 
In fig.~\ref{fig:sys}, we compare the results from our baseline dataset obtained with various covariances. 
The shift in the $(w_0, w_a)$ plane are at most $0.3\sigma$ along the principal axis, suggesting that our results are robust against potential inaccuracies in the covariance modelling. 
{Especially, a fully consistent covariance for our one-loop model and the BOSS sample, including non-Gaussian terms from perturbation theory, survey-window effects, and related contributions, is unlikely to produce a larger shift, since the difference between the Patchy mocks and our Gaussian approximation captures the leading corrections listed above.}
Yet, it would be valuable to re-assess our analysis with a better modelling of the covariance, \textit{e.g.}, based on the Uchuu-GLAM lightcone simulations~\cite{2311.14456}. 
We leave this to future work.

\subsection{Forward modelling fibre collisions}
For SDSS/BOSS, fibre collisions are corrected at the catalog level using completeness weights~\cite{BOSS:2015ewx}. 
In general, the fibre-completeness weight should be proportional to the inverse probability that a target is resolved (assigned a fibre), \textit{i.e.}~$w_{\rm fc} \propto 1/P_{\rm resolved}$. 
In practice, BOSS implements the simplest, ``order-zero'' scheme: nearest-neighbour (NN) upweighting~\cite{SDSS:2001fzq,SDSS:2004oes}. 
In this method, the statistical weight of each unobserved (collided) target is transferred to its nearest neighbour on the sky: the receiving galaxy's fibre-collision weight becomes $w_{\rm fc}=1+n_{\rm miss}$, where $n_{\rm miss}$ is the number of collided neighbours it represents. 
Consequently, when collisions are rare $w_{\rm fc}\to 1$, and where collisions are more frequent $w_{\rm fc}>1$. 

Effectively, the observed density $\delta_o$ at position $\pmb{x}$ is then related to the true one $\delta$ as
\begin{equation}
\delta_o(\pmb{x}) = \delta(\pmb{x}) + f_c \, \delta_c(\pmb{s}) \ , 
\end{equation}
where $0 < f_c < 1$, is the fraction of unresolved galaxies assigned to $\pmb{x}$. 
Here $\pmb{s}$ is the true position of the unresolved galaxies, 
\begin{equation}\label{eq:fc_shift}
\pmb{s} = \pmb{x} + D_A^{\rm fid}(z_{\rm eff}) \delta \theta \, \hat{x}_\perp \ ,
\end{equation}
where $\hat{x}_\perp$ is a unit vector pointing in a direction perpendicular to the line of sight (LOS), and $\delta\theta < \delta \theta_{\rm max}$ is the angular separation between the unresolved galaxy and its NN. 
In BOSS, the minimal angular scale resolution before fibres collide is $\delta \theta_{\rm max} = 62''$. 
Here we assume the usual flat-sky approximation to convert the angle $\delta \theta$ to a distance through the angular diameter distance $D_A^{\rm fid}(z_{\rm eff})$ evaluated at a fiducial cosmology $\Omega_m^{\rm fid}$ and effective redshift $z_{\rm eff}$. 
We further assume the usual effective-redshift approximation (shown to work well for galaxy-clustering modelling~\cite{Zhang:2021uyp}), such that the unresolved galaxy and its NN are taken both at $z_{\rm eff}$, therefore leading us to consider only the shift in eq.~\eqref{eq:fc_shift} to be in the direction of $\hat{x}_\perp$. 
{The rationale is that if the unresolved galaxy sits too far away from its NN, their \emph{unequal-time} correlation is diluted at the relatively small scales where fibre collision effects kick in (see the power suppression at small scales from unequal time in \textit{e.g.},~refs.~\cite{Chisari:2019tig,Zhang:2021uyp}).}
In fact, neglecting the LOS dependence of fibre collisions has been shown to be a good approximation at the two-point level in ref.~\cite{Hahn:2016kiy}, which demonstrated using the realistic simulations of BOSS (the $N$-series presented below) that their effects along the radial direction are negligible.\footnote{Note that our main conclusion below, namely that fibre collision effects are captured in an expansion of terms degenerate with terms already present in the EFT expansion, remains unaffected if the vector between the unresolved galaxy and its NN has a radial component. }

The following proceeds in straight analogy with redshift-space distortions (see~\textit{e.g.},~app.~B of~ref.~\cite{Simon:2022csv} for a pedantic derivation), with the shift being now in the direction perpendicular to the LOS instead of being parallel. 
As in both cases, they are just coordinate re-mapping, mass conservation between infinitesimal volumes $d^3x$ and $d^3s$ implies 
\begin{equation}
(1 + \delta_c(\pmb{x}))d^3 x = (1 + \delta_c(\pmb{s}))d^3 s = (1 + \delta_c(\pmb{s})) \left| \frac{\partial \pmb{s}}{\partial \pmb{x}} \right| d^3 x \ , 
\end{equation}
where $\left| \frac{\partial \pmb{s}}{\partial \pmb{x}} \right|$ is the Jacobian. 
Here the unresolved galaxies density follows the same distribution of the resolved galaxies density, \textit{i.e.}, $\delta_c(\pmb x) \sim \delta(\pmb x)$. 
The Fourier conjugate of the field $\delta_c(\pmb{s})$ is then given by
\begin{equation}
\delta_{c}(\pmb{k}) \equiv \int d^3s \, e^{-i \pmb{k} \cdot \pmb{s}} \delta_{c}(\pmb{s})  = \delta(\pmb{k}) + \int d^3 x \, e^{-i \pmb{k} \cdot \pmb{x}} \left( e^{-i \pmb{k_\perp} \cdot \hat{x}_\perp D_A^{\rm fid} \delta \theta } -1 \right) (1+ \delta(\pmb{x})) \, . \label{eq:delta_r}
\end{equation}
To proceed, we expand the exponential of the position shift, 
\begin{equation}\label{eq:exp_fc}
e^{-i \pmb{k_\perp} \cdot \hat{x}_\perp D_A^{\rm fid} \delta \theta } = 1 - i k \sqrt{1-\mu^2} D_A^{\rm fid} \delta \theta - \frac{1}{2} k^2 (1-\mu^2) (D_A^{\rm fid})^2 \delta \theta \delta \theta + \dots \ , 
\end{equation}
where $\sqrt{1-\mu^2}$ and $\mu$ are respectively the sinus and cosinus of the angle between $\pmb{k}$ and the LOS. 
It is safe to assume that $\delta \theta$ is a variable with mean zero but non-zero variance $\sigma_c^2  \equiv (D_A^{
\rm fid})^2 \braket{\delta \theta \delta \theta}$. 
Taking the expectation value over eq.~\eqref{eq:exp_fc}, we see that the linear term in $k$ vanishes, such that the first correction starts at $k^2$ and is proportional to $(1-\mu^2)\sigma_c^2$. 
It is easy to convinced ourselves that such term lead to degenerate contributions with terms already present in our EFT expansion (for a close, explicit derivation at the power-spectrum level, see again~app.~B of~ref.~\cite{Simon:2022csv}). 
For instance, at leading order in $f_c$, the correction to the power spectrum is degenerate with $k^2$-stochastic contributions. 
For $f_c \sim 5\%$, taking $P_{g,r}(k\rightarrow0) \sim \bar n^{-1}$, with $\bar n$ the mean galaxy number density, and approximating $\sigma_c$ by the comoving distance corresponding to $\delta \theta_{\rm max}$ (as correlation decreases with scales, most of the unresolved galaxies sit far,~\textit{i.e.},~closer to the threshold $\delta \theta_{\rm max}$ than $0$), this correction is of the order of $\mathcal{O}(100)$ at $k \sim 0.2 \, \hinvMpc$, in agreement with the findings of ref.~\cite{Hahn:2016kiy}. 
Therefore, the size of the fibre-collision corrections lies well within our prior on the degenerate EFT contributions, effectively absorbing them. 
We note that a key difference from ref.~\cite{Hahn:2016kiy} is that, here, we expand the effects of fibre collisions at the field level rather than at the two-point level.\footnote{
At leading order in $f_c$, our derivation reproduces the same correction to the power spectrum as reported in ref.~\cite{Hahn:2016kiy} (see $\Delta P_{\rm corr}$ in their eq.~(37)). 
We further note that the additional term $\Delta P_{\rm uncorr}$ from fibre collisions discussed in ref.~\cite{Hahn:2016kiy} enters only at order $f_c^2$, consistent with their conclusion that its impact is small. 
Moreover, neglecting this correction has been shown to induce only negligible shifts in cosmological posteriors~\cite{DAmico:2019fhj}. 

} 
This approach enables us to derive a complete effective model for fibre collisions applicable to all $N$-point statistics. 

\begin{figure}[t!]
  \centering
  \includegraphics[width=0.49\textwidth]{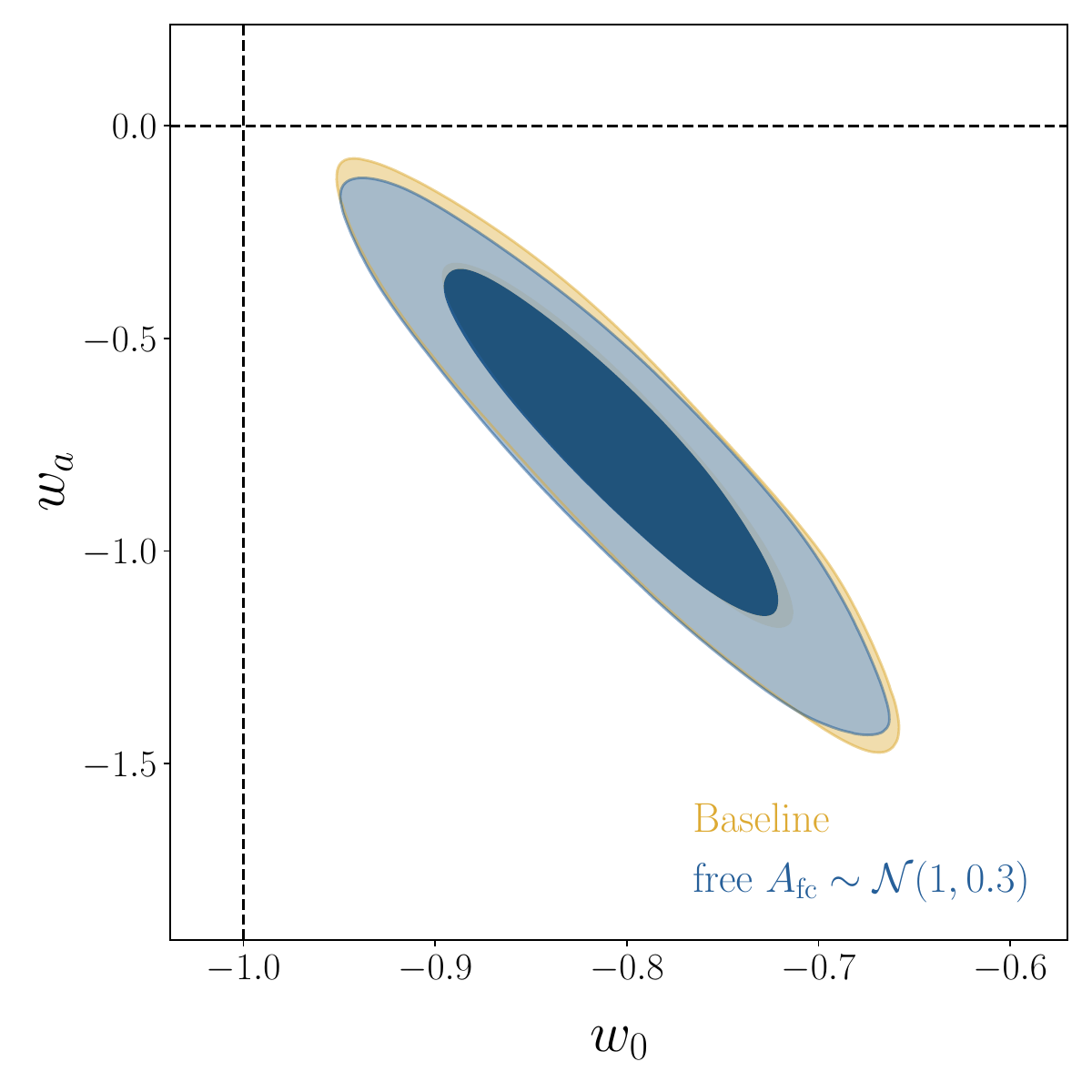}
  \includegraphics[width=0.49\textwidth]{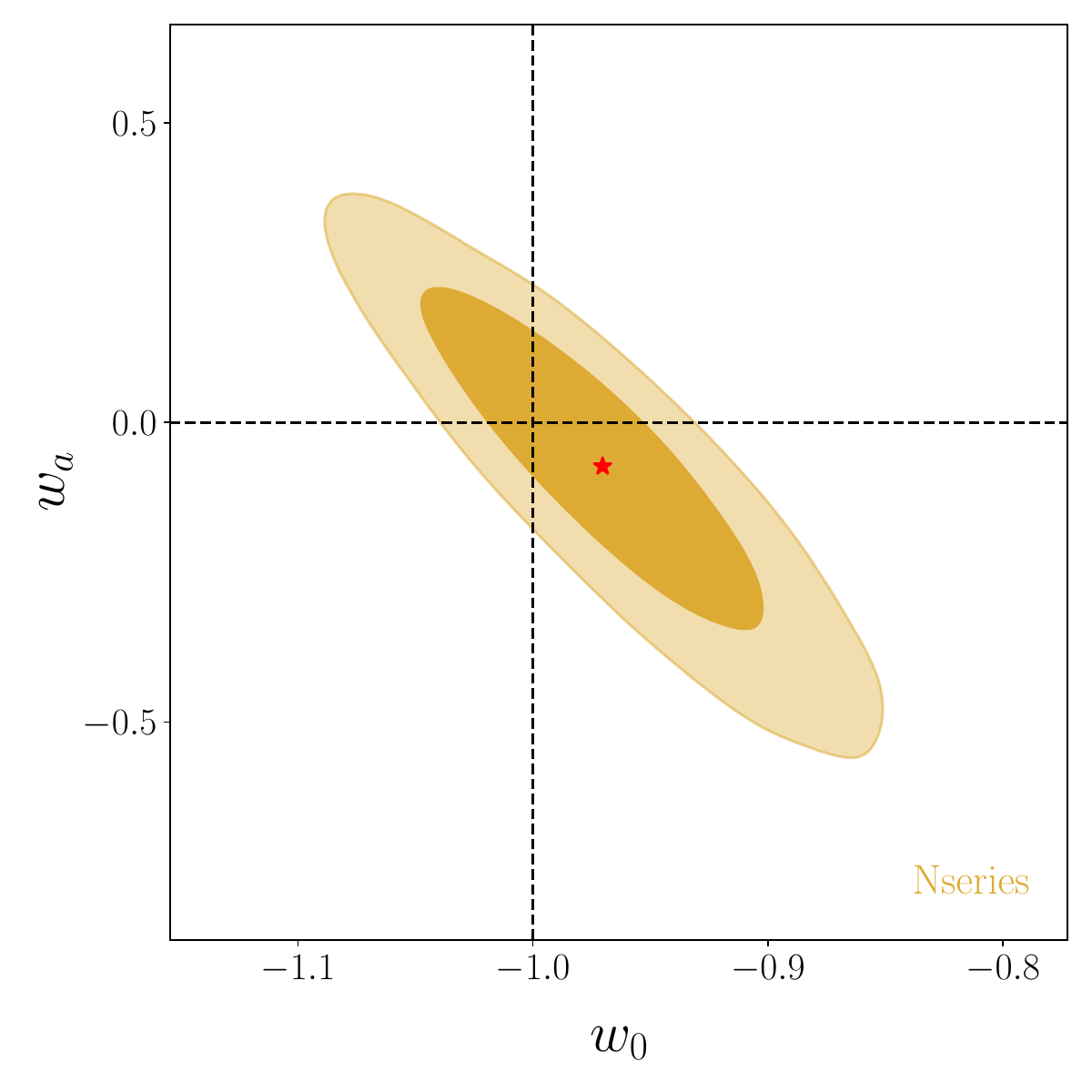}
     \caption{\textit{Left panel}: comparison against baseline results of the case where the whole bispectrum amplitude (controlled through a multiplicative fudge factor $A_{\rm fc}$) is allowed to vary by $\pm 30\%$; 
     \textit{Right panel}: Baseline analysis on $N$-series simulation data, jointly fit with mock CMB, SN, and BAO, in accordence to our baseline dataset. The star denotes the posterior mean, which is $<0.3\sigma$ away from the fiducial value in the $(w_0-w_a)$ plane. 
     Shifts from the truth are even smaller for the other cosmological parameters (not shown).}
  \label{fig:Abisp}
\end{figure}

Finally, there are two other modifications on the bispectrum measurements in the presence of fibre collisions that affect the number counts: shot noise estimation and estimator normalisation. 
Those are in principle corrected by the completeness weights. 
Any residual mismatch in the shot noise estimation is captured by our free $k^0$-stochastic terms, proportional to $\bar{n}^{-1}$. 
Next, to gauge the impact of a potential residual difference in the normalisation, we rescale the bispectrum with amplitude parameter $A_{\rm fc}$ free to vary within a Gaussian prior centred on 1 of width largely incompassing the variation in the observed number count squared from fibre collisions, $\mathcal{N}(1, 0.3)$. 
As shown in fig.~\ref{fig:Abisp}, we find that the shift in the $w_0w_a$CDM plane is negligible (and a value of $A_{\rm fc}$ consistent with no modification at $1\sigma$, $A_{\rm fc} = 1.09 \pm 0.08$). 
We conclude that effects from fibre collisions are captured by our modelling in both the power spectrum and the bispectrum given our large priors accommodating the effects above, while uncertainties in the estimator normalisation do not change significantly our cosmological results. 

We note that our formalism should apply equally well for fibre completeness beyond the NN scheme. 
We thus anticipate that any EFTofLSS analysis \emph{at the one loop} based on fibre-assigned spectroscopic targets, provided preliminary estimates of the size of the corrections derived here to potentially accommodate priors on the degenerate EFT contributions, should be immune to fibre collision systematics. 
We leave such a dedicated study for future work.

\subsection{Tests on simulations}\label{app:sims}
To validate our analysis setup used in this work, we fit the Nseries mocks~\cite{BOSS:2016wmc}, a suites of high-fidelity $N$-body simulations with realistic halo-to-galaxy connection and observational effects mimicking BOSS as the ones considered above. 
We use the mean of the $84$ pseudo-independent realisations, such that the data vector is almost noiseless, allowing for a high-precision test. 
To stay close to our realistic case, we however use the covariance corresponding to BOSS volume. 
The results are shown in the right panel of fig.~\ref{fig:Abisp}. 
For our baseline analysis setup, we see that the shift in the $(w_0-w_a)$ plane with respect to $\Lambda$ is less than $< 0.3 \sigma$, which makes for a negligible systematic error on our report of the significance once summed in quadrature. 
This validates our analysis pipeline of BOSS power spectrum and bispectrum likelihood in our combined baseline probe.


\begin{figure}[!ht]
    \centering
    \includegraphics[width=1\textwidth]{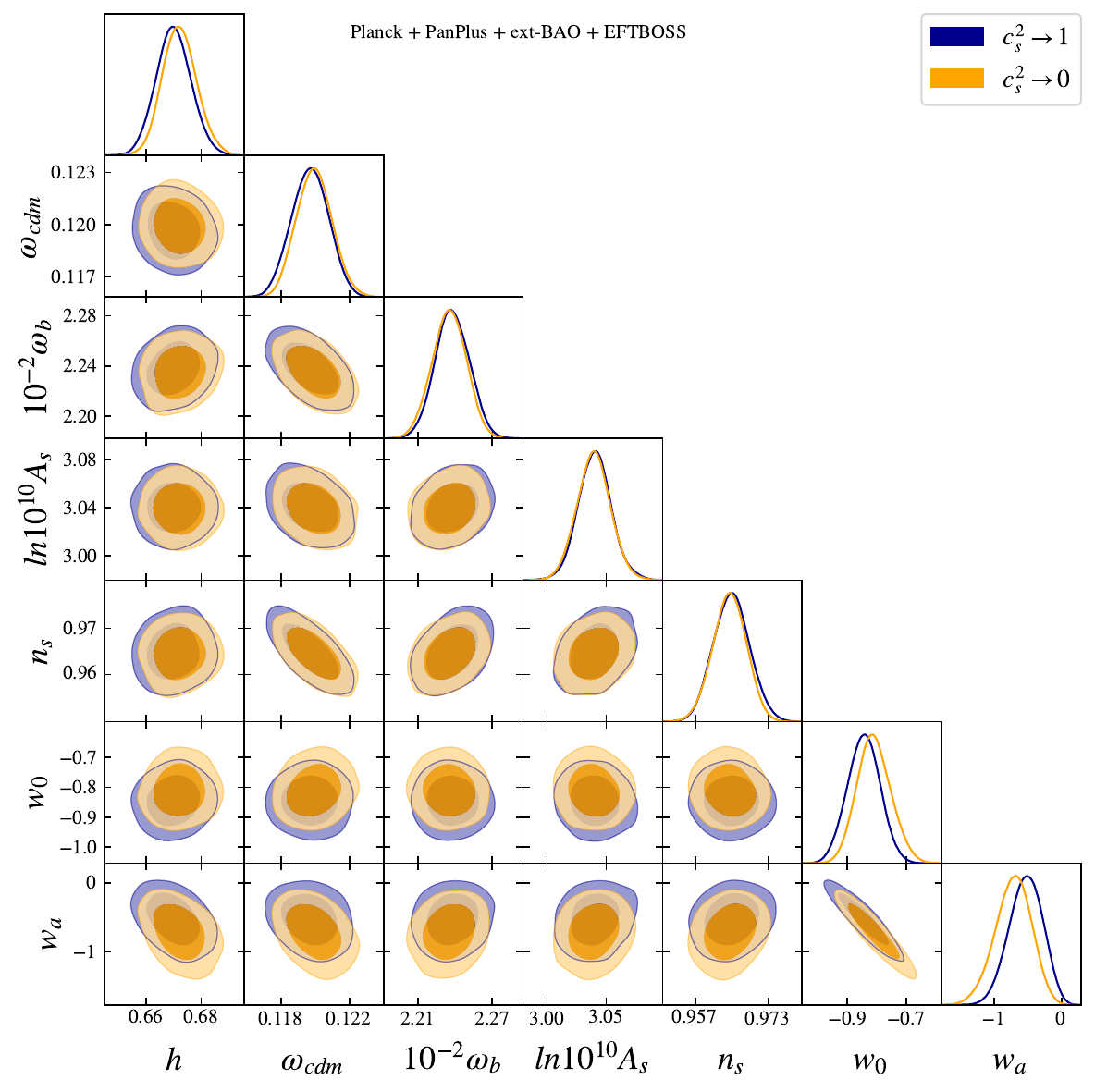}
    \caption{1D and 2D posterior distributions of the cosmological parameters inferred from \textit{Planck} + PanPlus + ext-BAO + EFTBOSS within $w_0w_a$CDM for both smooth and clustering quintessence. }
    \label{fig:w0wa_all}
\end{figure}

\section{Supplementary analysis products}\label{app:supp_material}

This appendix contains supplementary analysis products. 
In fig.~\ref{fig:w0wa_all}, we display the 1D and 2D posterior distributions of the whole cosmological parameter space from \textit{Planck} + PanPlus + ext-BAO + EFTBOSS for both smooth and clustering quintessence, while in tab.~\ref{tab:bestfit_params} we show the corresponding maximum a posteriori estimates as well as the $68\%$ credible intervals. 
We also provide {an assessment of the impact of our prior assumption that the galaxy bias parameters are correlated between the various BOSS skies, followed by} two additional discussions on the relevance of  the one-loop correction to the bispectrum and on the cross-correlation between the sound-horizon-marginalised BOSS full shape and DESI BAO. 

\begin{table}[!ht]
    \centering
    \begin{tabular}{|c|c|c|}
        \hline
        Quintessence & $c_s^2 \to 1$ & $c_s^2 \to 0$  \\
        \hline
        \hline
        $10^{-2}\omega_b$ & 2.237 & 2.237  \\
         & $2.238\pm 0.014$ 
         &$2.235\pm 0.014$ \\
        \hline
        $\omega_{\mathrm{cdm}}$ & 0.1199 & 0.1199  \\
        & $0.1197\pm 0.0011 $
        &$0.1199\pm 0.0010$\\
        \hline
        $h$ & 0.6713 & 0.6714\\
        &$0.6700\pm 0.0063$ 
        & $0.6722^{+0.0058}_{-0.0065}$\\
        \hline
        $\ln(10^{10}A_s)$  & 3.044 & 3.044 \\
        &$3.040\pm 0.014$
        & $3.039\pm 0.014$\\
        \hline
        $n_s$ & 0.9658 & 0.9656\\
        &$0.9650\pm 0.0039$ 
        &$0.9645\pm 0.0037$\\
        \hline
        $\tau_{\mathrm{reio}}$ & 0.0544 & 0.0544\\
        &$0.0531\pm 0.0073$ 
        & $0.0520\pm 0.0072$\\
        \hline
        $w_0$ & -0.831  & -0.820\\
        &$-0.844\pm 0.055$
        & $-0.809^{+0.053}_{-0.061}$\\
        \hline
        $w_a$     & -0.61 & -0.66\\
        &$-0.53^{+0.26}_{-0.23}$ 
        & $-0.72^{+0.28}_{-0.25}$\\
        \hline
        $\Omega_m$  & 0.317 & 0.317 \\
        &$0.3180\pm 0.0067$ 
        & $0.3164\pm 0.0064$\\
        \hline
        $\sigma_8$   & 0.8115  & 0.8119 \\
        &$0.8059\pm 0.0093$ 
        & $0.8104\pm 0.0096$\\
        \hline
        \hline
        $\chi^2_{\rm min}$ & 4835.36 & 4834.22\\
        \hline
        $\Delta\chi^2_{\rm min}$ & -9.2 & -10.3 \\
        \hline
        p-value& $2.6\sigma$ & $2.8\sigma$ \\
        \hline
    \end{tabular}
    \caption{Maximum a posteriori estimates and $68\%$ credible intervals of the cosmological parameters inferred from \textit{Planck} + PanPlus + ext-BAO + EFTBOSS within $w_0w_a$CDM for both smooth quintessence and clustering quintessence, where EFTBOSS includes both the power spectrum and bispectrum.  }
    \label{tab:bestfit_params}
\end{table}

\begin{figure}[h!]
  \centering
  \includegraphics[width=0.8\textwidth]{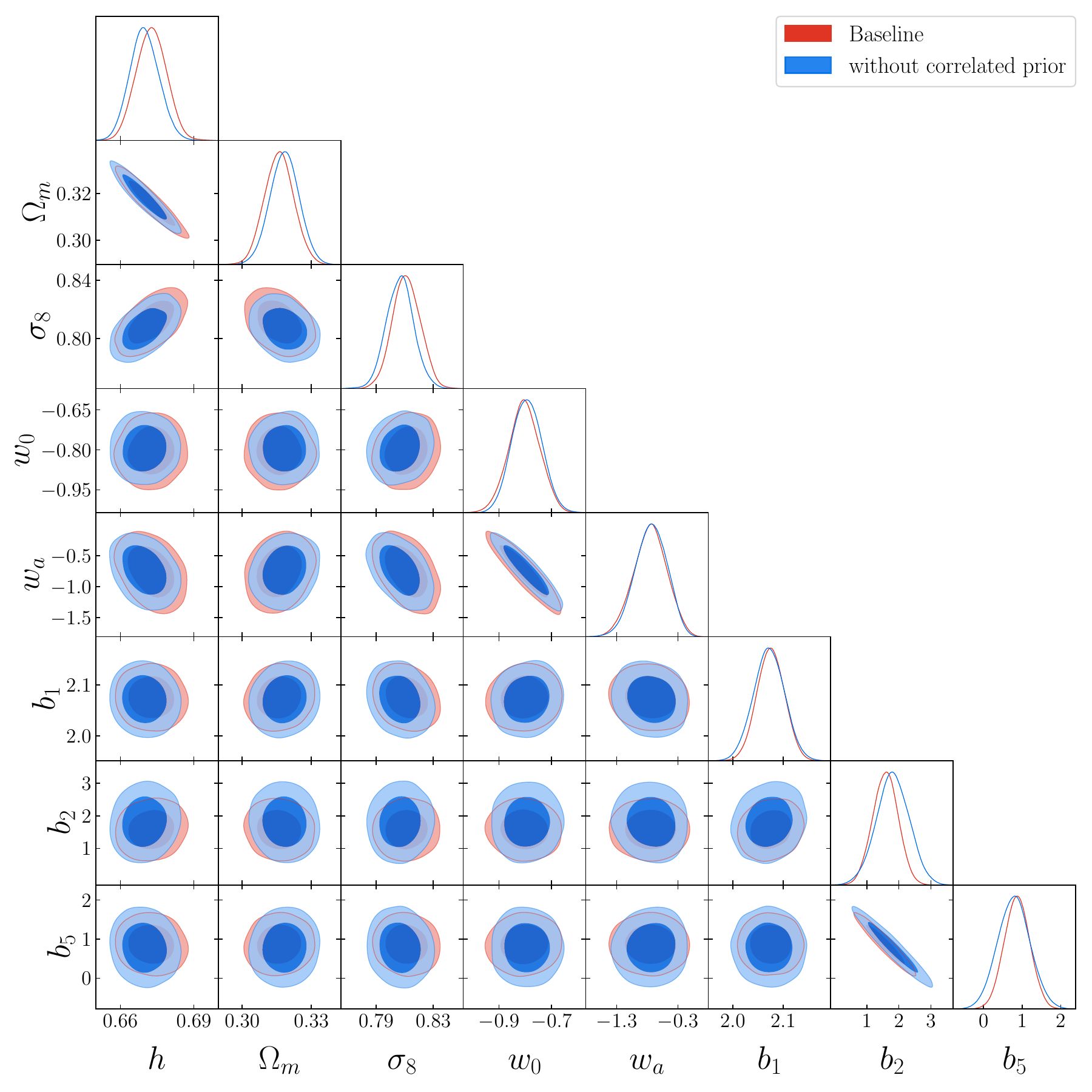}
  \caption{{1D and 2D posterior distributions of representative cosmological and galaxy bias parameters, with and without correlated priors across BOSS skies on the latter. For clarity, only the biases from BOSS CMASS NGC are shown. }}
  \label{fig:test_boss_correlated_prior}
\end{figure}

{
\paragraph{Galaxy bias correlated prior}
We further test the impact of the correlated prior on the bias parameters between different skies introduced in sec.~\ref{sec:eft_lkl}, completing the consistency tests presented in sec.~\ref{sec:checks}.  
To this end, we repeat the baseline analysis while removing the correlated priors. 
The results are shown in fig.~\ref{fig:test_boss_correlated_prior}. 
As expected, the bias parameters (in particular $b_2$ and $b_5$) become noticeably less constrained once the correlated priors are removed. 
In contrast, the impact on the cosmological parameters is mild, especially in the $(w_0, w_a)$ plane. 
Quantitatively, quoting the FoM ratio, the constraints on $(b_1, b_2, b_5)$ are relaxed by a factor $1.74$, while constraints on $(\Omega_m, h, \sigma_8, w_0, w_a)$ remains consistent at $0.98$. 
In particular, the projection onto $(w_0, w_a)$ remains largely consistent in the mean values and the uncertainties (FoM ratio of $0.92$). 
We therefore conclude that our quoted cosmological results remains consistent with our prior assumption on bias correlation from galaxy bias redshift evolution and BOSS selection function. 
}

{
\subsection{Tree-level vs. one-loop}\label{app:loop}

In the main text, we observe an improvement of $\sim 30\%$ on the contours of $(w_0, w_a)$ when including the one-loop bispectrum, while no significant gain is obtained when using only the tree-level bispectrum. 
The origin of this difference can be understood in terms of the number of data points included in the analysis. Within the EFTofLSS framework, the inclusion of loop corrections, counterterms, and stochastic contributions allows the analysis to be extended into the mildly nonlinear regime. 
In practice, based on their respective perturbation-theory truncation error, the one-loop analysis is performed up to $k_{\rm max}^{\rm bisp}=0.23\,h\,\mathrm{Mpc}^{-1}$, while the tree-level prediction is restricted to $k_{\rm max}^{\rm bisp}=0.08\,h\,\mathrm{Mpc}^{-1}$~\cite{DAmico:2019fhj,DAmico:2022osl}. This difference in scale cuts leads to a larger number of modes entering the one-loop analysis, and consequently stronger constraints.

\paragraph{Information content} To quantify the resulting difference in the recovered cosmological information, we use the Fisher matrix information of the two analyses, adopting the same configurations as in the main text. Namely, we compute the Fisher matrix for four setups: 
\textit{(i)} power spectrum only; 
\textit{(ii)} power spectrum combined with the tree-level bispectrum ($k_{\rm max}^{\rm bisp}=0.08\,h\,\mathrm{Mpc}^{-1}$); 
\textit{(iii)} power spectrum combined with the one-loop bispectrum with highest wavenumber $k_{\rm max}^{\rm bisp}=0.08\,h\,\mathrm{Mpc}^{-1}$; 
and \textit{(iv)} power spectrum combined with the one-loop bispectrum with highest wavenumber $k_{\rm max}^{\rm bisp}=0.23\,h\,\mathrm{Mpc}^{-1}$. 
As shown in fig.~\ref{fig:tree_vs_loop}, adding the tree-level bispectrum to the power spectrum leads to only marginal improvement of $23\%$, while the inclusion of the one-loop bispectrum yields instead a significant gain of $260\%$ in constraining power in terms of FOM on the $h-\Omega_m$ plane. Note that the improvement here comes from  BOSS only while in the main text we do combined probe analysis.
Moreover, restricting the one-loop bispectrum to $k_{\rm max}^{\rm bisp}=0.08\,h\,\mathrm{Mpc}^{-1}$ results in the same constraints as those from tree-level. This demonstrates that the primary source of improvement is the inclusion of additional small-scale modes.

\paragraph{Relevance of the loop correction} 
The one-loop contribution to access small scales is important to recover unbiased constraints, as shown in ref.~\cite{DAmico:2022osl} (see also app.~\ref{app:sims}). 
To further illustrate the relevance of the loop corrections to the bispectrum, we plot in the right panel of fig.~\ref{fig:tree_vs_loop} the tree-level and one-loop contributions using the best-fit parameters from CMASS NGC of tab.~\ref{tab:bestfit_params}. 
For clarity, only equilateral configurations $(k_1 = k_2 = k_3)$ are presented. 
We see that around $k \sim 0.1 \, \hinvMpc$, the one-loop contribution starts to dominate the size of the data error bars, and becomes larger and larger at higher $k$s. 
This indicates that loop corrections can not be neglected at these scales. 
In particular, we see that all loop contributions are, roughly, of the same size (\textit{i.e.},~order one of the total loop), showing that all loop diagrams are necessary to reach these scales. 
We also have checked that removing the counterterms and stochastic terms modify the loop contributions by an order one, further supporting that all contributions considered in our modelling play an important role to fit the data up to $k_{\rm max} = 0.23 \, \hinvMpc$.

\begin{figure}[h!]
  \centering
  \includegraphics[width=0.49\textwidth]{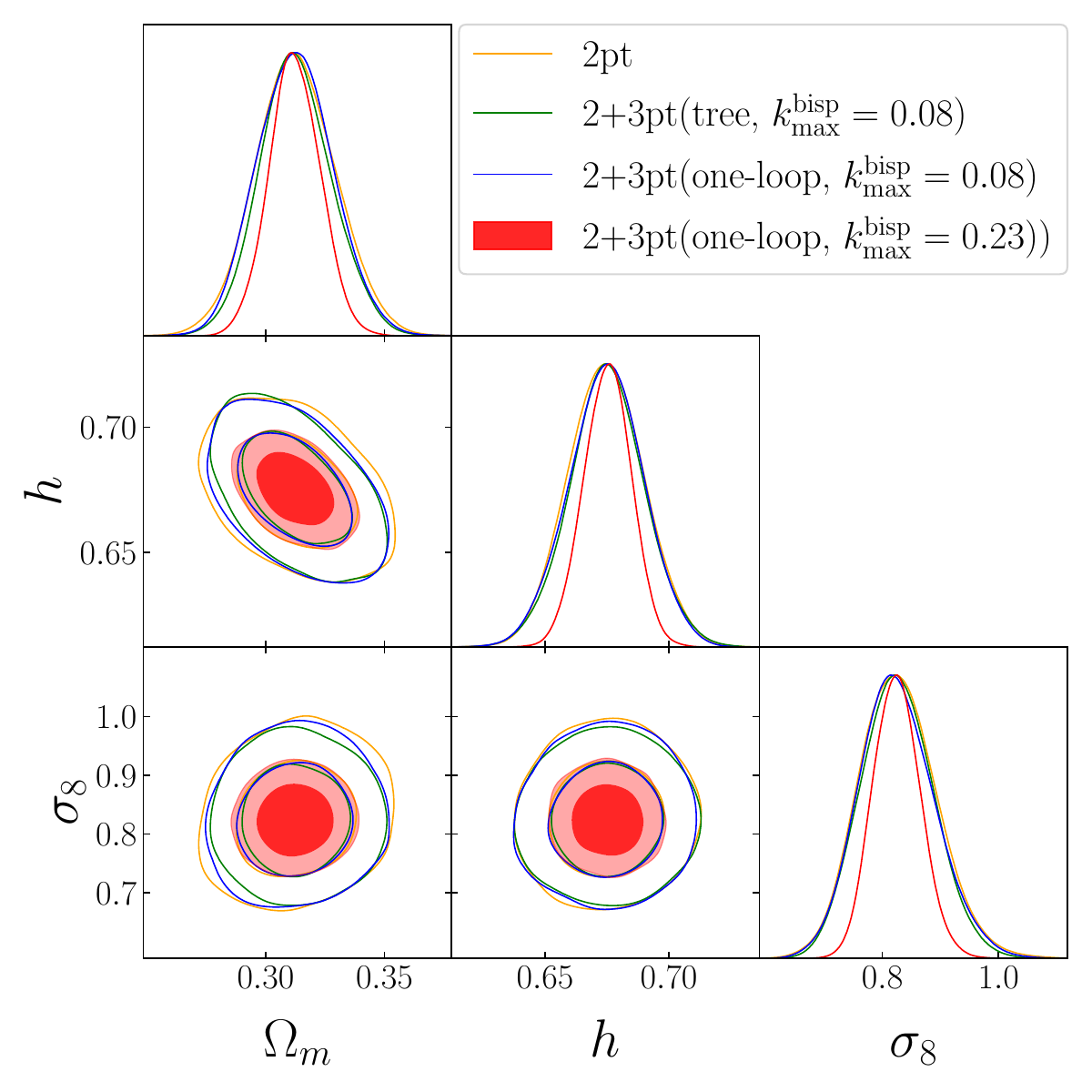}
  \includegraphics[width=0.39\textwidth]{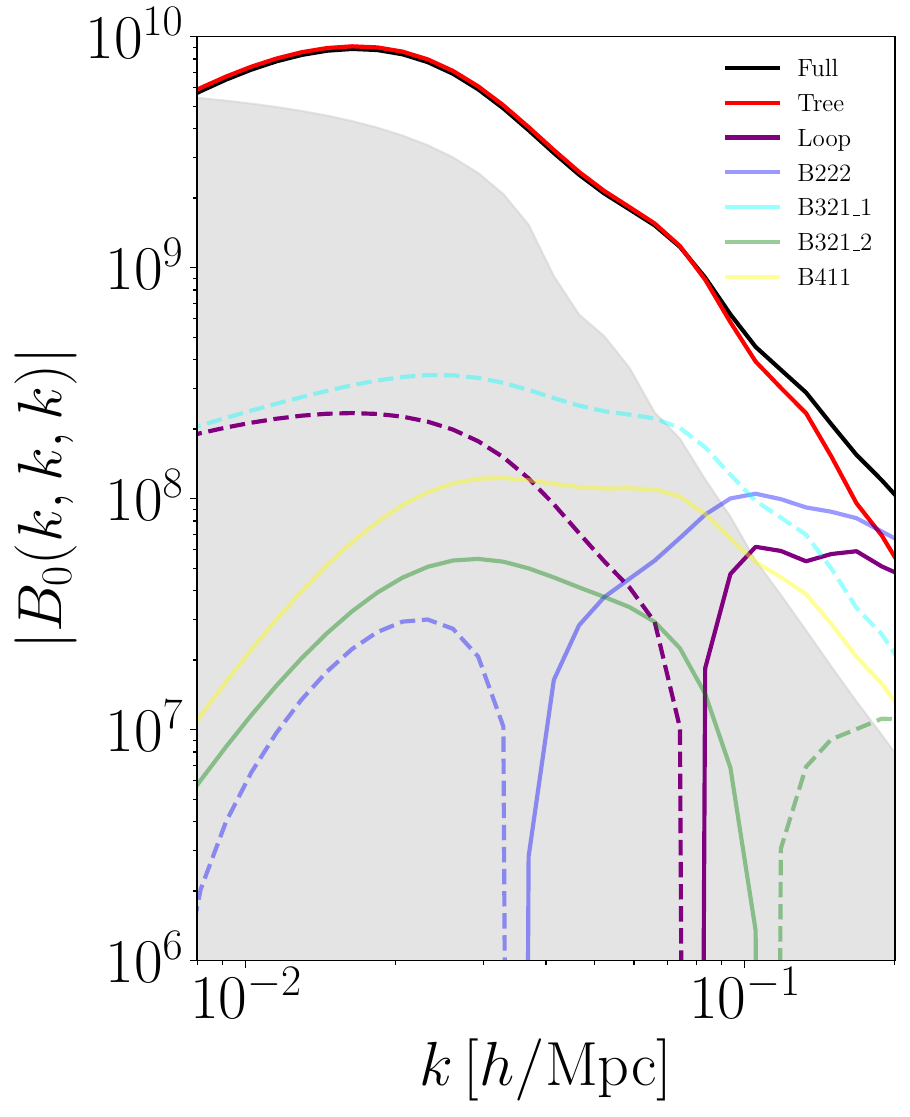}
  \caption{{\textit{Left panel}: Fisher information from the bispectrum (3pt) at tree-level and one-loop order at different scale cuts. The power spectrum (2pt) configuration is kept fixed throughout. \textit{Right panel}: Summary of tree-level and one-loop contributions to the bispectrum, using the best-fit parameters of tab.~\ref{tab:bestfit_params}. The grey region indicates a representative error level from BOSS. For clarity, only equilateral configurations are shown. `Loop' denotes the total sum of all renormalised diagrams, $222,321I,321II,411$, including counterterms and stochastic terms, and `Full' is the total bispectrum including both tree-level and one-loop contributions. }}
  \label{fig:tree_vs_loop}
\end{figure}

}

\subsection{EFTBOSS/$r_{s,\rm marg}$ and DESI BAO cross-correlations}\label{app:bossxdesi}

When combining BOSS full-shape analysis with DESI DR1 BAO, we marginalise over the sound-horizon information following the methodology presented in sec.~\ref{sec:dataset} to mitigate some of the overlap between the two experiments. 
The BAO likelihood consists in two compressed parameters $\lbrace \alpha^{\rm BAO}_{\parallel}, \alpha^{\rm BAO}_{\perp}\rbrace$, which are defined as
    \begin{equation}\label{eq:alpha_bao}
        \alpha^{\rm BAO}_{\parallel} = \frac{H_{\rm fid}(z)\, r_{s, \rm fid}}{H(z)\, r_s} \ , \qquad \alpha^{\rm BAO}_{\perp} = \frac{D_A(z)\, r_{s, \rm fid}}{D_{A, \rm fid}(z)\, r_s} \ , 
    \end{equation}
    where $H(z)$ and $D_A(z)$ are the Hubble and angular diameter distance, respectively. 
    Here, the subscript `$\textrm{fid}$' refers to fiducial cosmology used to convert celestial and redshift coordinates into physical distances, which is also the one used to evaluate the fixed template in the BAO analysis. 
Thus, our procedure removes only partially the cross-correlation between BOSS and DESI, as the BAO information is a combination of late-time distances ($H(z), D_A(z)$) and early-time rulers ($r_s$). 
In the full-shape analysis, these quantities enter through the Alcock-Paczynski (AP) geometrical distortions on the one hand, and through the position of the BAO peak controlled by and built-in the input linear matter power spectrum on the other hand. 

\paragraph{Understanding the correlation}
To understand the situation, let us focus the discussion on a Kaiser power spectrum for definiteness, 
\begin{equation}
P_g(k,\mu) = (b_1+ f\mu^2)^2 P_{\rm lin}(k) \ .
\end{equation}
The argument in the following is generalised straightforwardly to the one-loop modelling and to the bispectrum. 
Distortions are introduced in the forward model of the power spectrum to account for the mismatch between the observed comoving wavenumbers ($k', \mu'$) converted from celestial coordinates and redshifts assuming a fixed chosen fiducial cosmology, with the actual wavenumbers ($k, \mu$) in the probed `true' cosmology. 
They are related as 
\begin{equation}
    k = \frac{k'}{\alpha^{\rm AP}_{\perp}} \left[1 + \mu'^2 \left(\frac{1}{F^2} -1 \right) \right]^{1/2} \ , \qquad \mu = \frac{\mu'}{F} \left[1 + \mu'^2 \left(\frac{1}{F^2} -1 \right) \right]^{-1/2} \ ,
\end{equation}
where $F = \alpha^{\rm AP}_{\parallel} / \alpha^{\rm AP}_{\perp}$, and where the AP parameters are defined as
\begin{equation}\label{eq:alpha_AP}
\alpha^{\rm AP}_{\parallel} = \frac{H_{\rm fid}(z)}{H(z)} \ , \qquad \alpha^{\rm AP}_{\perp} = \frac{D_A(z)}{D_{A, \rm fid}(z)} \ .
\end{equation}
To match the observations, we then forward model of the AP distortions as
\begin{equation}
P_g(k', \mu') = \frac{1}{\alpha^{\rm AP}_{\parallel} (\alpha^{\rm AP}_{\perp})^2} P_g(k(k', \mu'), \mu(\mu')) \ . 
\end{equation}
To make the dependence of the BAO on $r_s$ explicit, we introduce the following parameter that controls the shift in the position of the peak with respect to some fiducial, 
\begin{equation}\label{eq:alpha_rs}
\alpha_{r_s} := \frac{r_{s}}{r_{s,\rm fid}} \ . 
\end{equation}
The linear power spectrum then reads
\begin{equation}
P_{\rm lin}(k) = P_{\rm nw}(k) + P_{\rm w}(\alpha_{r_s}k) \ . 
\end{equation}
Thus, the model for the galaxy power spectrum is given by
\begin{equation}
P_g(k', \mu') = \frac{1}{\alpha^{\rm AP}_{\parallel} (\alpha^{\rm AP}_{\perp})^2} (b_1 + f\mu(\mu')^2)^2 \left[ P_{\rm nw}(k(k',\mu')) + P_{\rm w}(\alpha_{r_s}k(k',\mu')) \right] \ . 
\end{equation}
At this point it becomes clear what we call the BAO information is. 
In the BAO analysis, $P_{\rm lin}$ is a fixed template evaluated in the fiducial cosmology, and the broadband $P_{\rm nw}$ is marginalised over (using a polynomial fit to account for potential differences for example) together with amplitude parameters ($b_1, f$). 
As such, the remaining information is captured by the position of the peak controlled by the combination $\alpha_{r_s} k(k',\mu')$ entering $P_{\rm w}$, together with the AP rescaling acting (mainly) on $P_{\rm w}$. 
It is then easy to see from eqs.~\eqref{eq:alpha_rs}~and~\eqref{eq:alpha_AP} that this information can be recasted into the so-called BAO parameters~\eqref{eq:alpha_bao}. 
At the same time, it is now also clear why marginalising over $\alpha_{r_s}$ in a full-shape analysis does not remove the full BAO information: the AP parameters remain constrained through the overall geometrical distortion effect on the power spectrum shape. 

\begin{figure}[!h]
    \centering
    \includegraphics[width=0.9\textwidth]{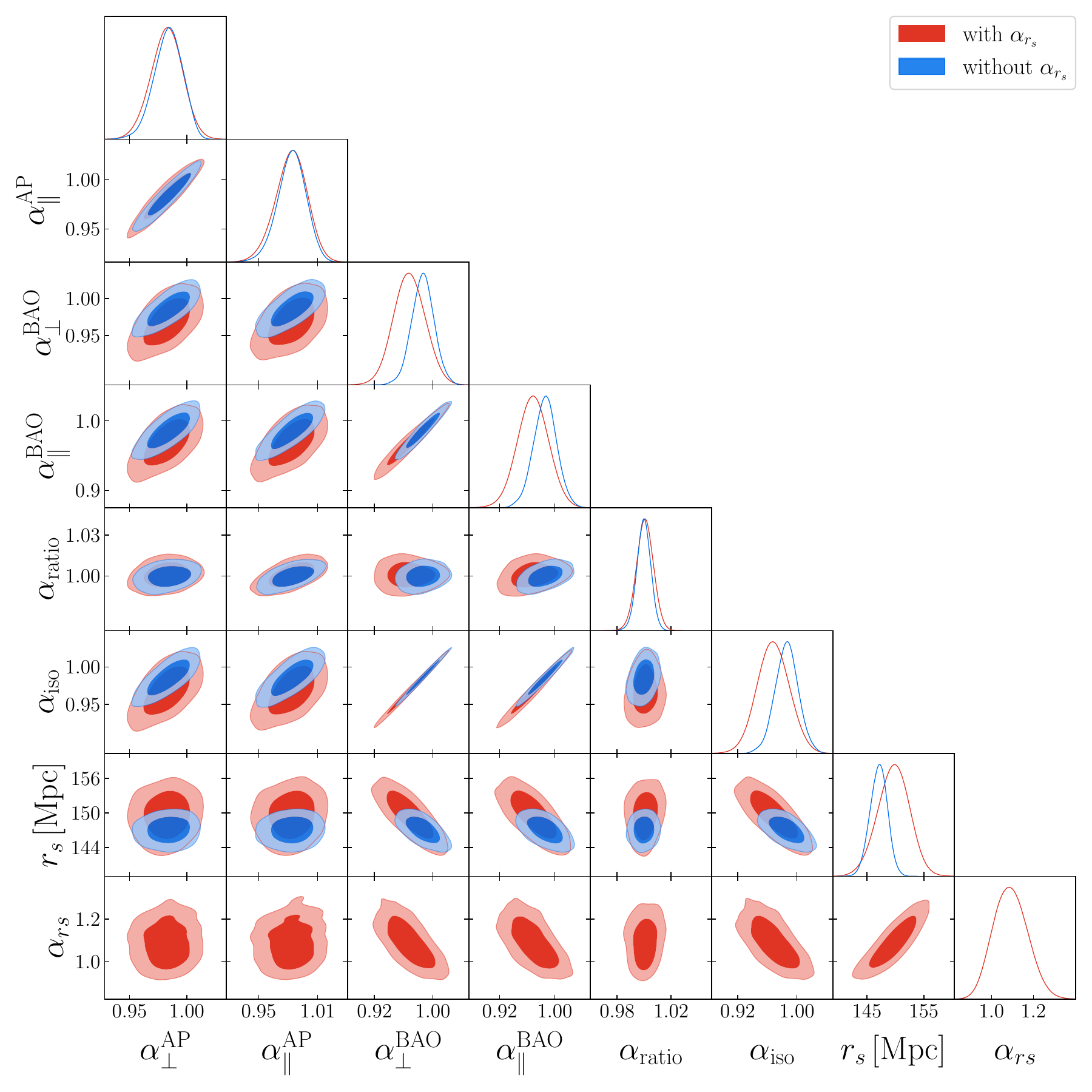}
    \caption{1D and 2D posterior distributions on BAO parameters $\alpha^{\rm BAO}_{\parallel,\perp}$, AP parameters $\alpha^{\rm AP}_{\parallel,\perp}$, and the sound horizon $r_{s}$, obtained assuming $w_0w_a$CDM by fitting synthetic data of EFTBOSS with our without $\alpha_{r_s}$-marginalisation. {For reference, we also show the derived quantities $\alpha_{\rm ratio} \equiv \alpha_{\parallel}^{\rm BAO} / \alpha_{\perp}^{\rm BAO}$ and $\alpha_{\rm iso} \equiv \left[\alpha_{\parallel}^{\rm BAO} (\alpha_{\perp}^{\rm BAO})^2 \right]^{1/3}$}.
    }
    \label{fig:alpha_bao}
\end{figure}

\paragraph{{Quantifying the amount of common information}}
First, we quantify how much BAO information is removed by marginalising over $\alpha_{r_s}$ in EFTBOSS. 
The results are shown in fig.~\ref{fig:alpha_bao}. 
The constraint in the $\alpha^{\rm BAO}$-space is relaxed by $\sim 40\%$ (computed by taking the ratio of the FoM).\footnote{{For reference, the bare FoM before and after $\alpha_{r_s}$-marginalisation are respectively $4970$ and $3143$ for the BAO parameters $(\alpha_{\parallel}^{\rm BAO}, \alpha_{\perp}^{\rm BAO})$, while for the AP parameters $(\alpha_{\parallel}^{\rm AP}, \alpha_{\perp}^{\rm AP})$, we have respectively $4880$ and $3033$. }} 
In comparison, the constraint in the $\alpha^{\rm AP}$-space is relaxed only by $\sim 20\%$. 
Those results are in accordance with the intuition built earlier: marginalising over $\alpha_{r_s}$ marginally relaxes the AP parameters, while the BAO information is more washed out since the information on $r_s$ contained in the BAO is marginalised. 
Note that the constraint on $r_s$ is not fully gone as in our full-shape analysis, we assume $w_0w_a$CDM, such that $r_s \equiv r_s(\omega_m, w_0, w_a)$ (for fixed $\omega_b$) is still determined indirectly through the constraints on $\Omega_m, h, w_0, w_a$ coming from the shape ($\sim k_{\rm eq}$) and amplitude ($f$) information, as well as the AP parameters.
As a consistency check, we find that the BAO information is reduced similarly by $\sim 50\%$ in the simple Kaiser model presented earlier. 

{
To quantify the overlap (or redundancy) in information between two experiments $A$ and $B$, we define the Fisher information correlation as
\begin{equation}
    \rho_{AB} = \frac{|\mathcal{F}_{A \times B}|}{\sqrt{|\mathcal{F}_A| |\mathcal{F}_B|}} \ , 
\end{equation}
where $\mathcal{F}_{A \times B} = \mathcal{F}_A + \mathcal{F}_B - \mathcal{F}_{A,B}$, with $\mathcal{F}_{A,B}$ the joint Fisher information. 
If there is no overlap, the two experiments are independent, such that $\mathcal{F}_{A,B} = \mathcal{F}_A + \mathcal{F}_B$, and so $\rho_{AB} = 0$. 
Conversely, if $A=B$ (perfect correlation), then $\rho_{AB}=1$. 
For now, let us assume that $A$ and $B$ correspond to the same type of experiment (\textit{e.g.}, BAO) but are performed on two, possibly overlapping, datasets. 
Let $V_A$ and $V_B$ be survey volumes of the two datasets, and $V_{A \times B}$ the overlapping volume. 
Since the Fisher information is proportional to $V$, we have 
\begin{equation}
    \rho_{AB} = \frac{V_{A \times B}}{\sqrt{V_A V_B}} \ . 
\end{equation}
Now suppose we reduce the information in $B$ (\textit{e.g.}, via $r_s$-marginalisation), such that $V_B' = r V_B$, with $0< r < 1$. 
Accordingly, the shared information is reduced, $V_{A \times B}' = r V_{A \times B}$. 
The correlation then decreases as
\begin{equation}
\rho_{AB}' = \frac{V_{A \times B}'}{\sqrt{V_A V_B'}} = \sqrt{r} \, \rho_{AB}. 
\end{equation}
Assuming that all information in BOSS comes from the BAO, we have that $r = V_{\textrm{BOSS}}^{r_s-\textrm{marg}}/V_{\rm BOSS} = (\textrm{FoM}_{\textrm{BAO}}^{r_s-\textrm{marg}} / \textrm{FoM}_{\textrm{BAO}})^2 = 0.6^2$, 
implying that the information overlap is reduced by $40\%$. 
In reality, only a fraction $g$ of the BOSS information originates from BAO, with the remainder $(1-g)$ coming from other sources.
In this case, the effective reduction is
\begin{equation}
\rho_{AB}' = \frac{V_{A \times B}'}{\sqrt{V_A V_B'}} = \frac{r}{\sqrt{1+g(r-1)}} \, \rho_{AB} \ .
\end{equation}
In the limit $g \rightarrow 1$ (pure BAO), this expression reduces to the previous $\sqrt{r}$ scaling.
For $g < 1$, the reduction is actually even stronger. 
As a conservative lower bound for $g$, \textit{i.e.}, how much the BAO contribute to the overall information in our EFTBOSS analysis, we can take $g \approx 1-r$. 
We then find $\rho_{AB}' / \rho_{AB} \sim 45\% $. 
}
To summarise, our $r_{s, \rm marg}$-procedure removes {at least} $55\%$ of the correlation between EFTBOSS and DESI BAO.

\begin{figure}[!ht]
    \centering
    \includegraphics[width=0.7\textwidth]{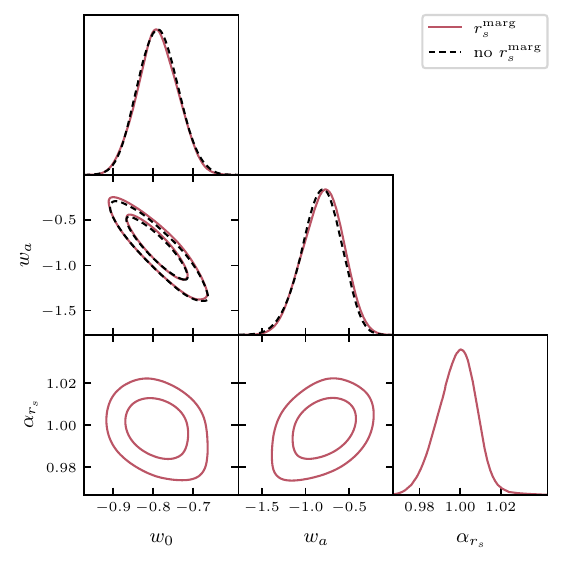}
    \caption{1D and 2D posterior distributions of $(w_0, w_a, \alpha_{r_s)}$ inferred from \textit{Planck} + PanPlus + DESI-BAO + EFTBOSS/$r_s^{\rm marg}$ within $w_0w_a$CDM for clustering quintessence. 
    Results combining DESI-BAO + EFTBOSS without marginalising over the sound horizon $r_s$ are shown for reference. 
    }
    \label{fig:rsmarg}
\end{figure}

What is then left? 
We remind that only $27\%$ of DESI BGS and LRG1 galaxies, along with $9\%$ of DESI LRG2 galaxies, were already observed by BOSS. 
As a rough estimate, assuming that the DESI BGS, LRG1, and LRG2 galaxies make for half of the full DESI BAO samples, and that BOSS volume is about half the one of DESI, we are left with a $\lesssim 5\%$ residual correlation after removing $55\%$ of the BAO information in BOSS.
{This suggests a small residual correlation within our order-of-magnitude
estimate,} especially considering that the BAO is only a fraction of the full information from EFTBOSS. 
Had we removed none of it, the information overlap would have been $\sim 10\%$, which still makes for a small correlation. 

To make a quantitive statement on the impact on cosmological parameters from the double counting, we compare the constraints on $(w_0, w_a)$ in fig.~\ref{fig:rsmarg} obtained by fitting \textit{Planck} + PanPlus + DESI-BAO + EFTBOSS whether or not marginalising over $r_s^{\rm marg}$. 
We see that the constraints are barely modified when removing $55\%$ of the double counting. 
Therefore, we do not expect that the residual $45\%$ of the redundant information to change our results significantly. 
{In conclusion, while we were able to derive order-of-magnitude estimate on the size and the impact of the residual correlation after our $r_s$-marginalisation procedure, we caution that our combined DESI BAO + EFTBOSS/$r_s^{\rm marg}$ analysis does not replace a full likelihood analysis with a joint covariance matrix that properly accounts for their correlations, which we leave for future work. }


\section{Reconstructing $w(z)$ histories}\label{app:recon}

In this appendix, we first lay down the methodology to reconstruct the credible region of $w(z)$ from a given posterior distribution of its underlying parameters. 
For the exercise, we do so by comparing two parametrisations: $(w_0, w_a)$, serving as a representative candidate model of dark energy, and an (arbitrarily) general one, providing a model-agnostic way to probe the evolution of dark energy. 
We explain the difference in the reconstructed $w(z)$, highlighting that when no model is assumed, we obtain an estimate of the data sensitivity on $w(z)$ across cosmic history. 
In passing, we derive approximate analytical formulae for assessing the preference over $\Lambda$ or for phantom behaviour. 

Let $\pmb{\phi}$ and $\pmb{\psi}$ be two sets of parameters of \emph{non-equivalent} models for the observational data $\mathcal{D}$. 
Let $\pmb \theta$ be the $\Lambda$CDM parameters that they share, such that the remaining ones offer two parametrisations of $w(z)$. 
We consider $\pmb{\phi} \setminus \pmb{\theta} = (w_0, w_a)^T$ and $\pmb{\psi} \setminus \pmb{\theta} = (w^z_{1}, w^z_{2}, \dots, w^z_{N})^T$, where $w^z_i$ is the value of $w$ in the redshift bin $z_i$ and $N$ is the number of bins. 
To avoid clutter, in the following we will simply denote $\pmb{\phi}$ for $\pmb{\phi} \setminus \pmb{\theta}$ and similarly for $\pmb{\psi}$, implicitly implying that $\pmb{\theta}$ are marginalised when relevant. 
With $N$ sufficiently large, we assume that $\pmb{\psi}$ forms a partition of $\mathcal{D}$. 
Since $\pmb{\phi}$ is just a parametric form for $w(z)$ guiding our search, we do not know a priori if $\pmb{\phi}$ forms a partition of $\mathcal{D}$. 
We thus do not assume so. 
We define the map $M: \Omega_\phi \subset \mathbb{R}^{N_\phi} \to \mathcal{T} \subset \mathbb{R}^{N}$ as $\phi_\alpha \mapsto M_{i\alpha} \phi_\alpha$,\footnote{In general, $M$ does not need to be a linear map, but for the case we consider, $M$ consists in applying $w(a) = w_0 + (1-a)w_a$, i.e., $M_{i\alpha} = (1, 1-a_i)^T$ (with eventually a binning matrix). } where $\Omega_\phi \sim \mathcal{P}(\pmb{\phi})$, and where $\mathcal{T}$ is the target space of the map $M$. As such, $\mathcal{T}$ is a subspace of $\Omega_\psi \sim \mathcal{P}(\pmb{\psi})$, the domain of $\pmb{\psi}$.

Reconstructing $w(z)$ amounts to estimate $\mathcal{P}(\pmb{\psi}|\mathcal{D})$, that can be sampled from the likelihood $\mathcal{L}(\mathcal{D}|\pmb{\psi})$, defined in sec.~\ref{sec:setup} with $w(z)$ parametrised following $\pmb{\psi}$. 
The reconstructed $w(z)$ is shown in fig.~\ref{fig:reconst_wz}. 

If $\pmb{\phi}$ is not a partition of $\mathcal{D}$, we can not infer $\mathcal{P}(\pmb{\psi}|\mathcal{D})$ from $\mathcal{P}(\pmb{\phi}|\mathcal{D})$. 
In contrast, if we believe that the data $\mathcal{D}$ is well-specified by the model $M$ parametrised by $\pmb{\phi}$, we may write 
\begin{equation}\label{eq:pwrong}
\mathcal{P}_\mathcal{T}(\pmb{\psi}|\mathcal{D}) = \int \diff^{N_\phi} \pmb{\phi} \ \mathcal{P}(\pmb{\psi}|\pmb{\phi}) \mathcal{P}(\pmb{\phi} | \mathcal{D}) \ ,
\end{equation}
where $\mathcal{P}_\mathcal{T}$ means that the distribution lives in the target space $\mathcal{T}$ of $M$. 
Since $\mathcal{P}(\pmb{\psi}|\pmb{\phi}) = \prod_{i=1}^N \delta_D(\psi_i - M_{i\alpha}\phi_\alpha)$, eq.~\eqref{eq:pwrong} is equivalent to sampling $\mathcal{P}_\mathcal{T}(\pmb{\psi}|\mathcal{D})$ by repeatedly drawing $\pmb{\phi}$ according to $\mathcal{P}(\pmb{\phi} | \mathcal{D})$ and applying the map $M$. 
This reconstruction appears clearly distinct from $\mathcal{P}(\pmb{\psi}|\mathcal{D})$ in fig.~\ref{fig:reconst_wz}. 
Since the two models specified by $\pmb \phi$ and $\pmb \psi$ are non-equivalent, we have that $\mathcal{P}_\mathcal{T}(\pmb{\psi}|\mathcal{D}) \neq \mathcal{P}(\pmb{\psi}|\mathcal{D})$.

To develop a more intuitive understanding of the situation, we can approach the problem from the opposite perspective.
Since $\pmb{\psi}$ is a partition of $\mathcal{D}$, for a given $\mathcal{D}$ there exists a $\pmb{\psi}$ that generates it. 
This implies that $\mathcal{P}(\pmb{\phi}|\pmb{\psi},\mathcal{D}) \equiv \mathcal{P}(\pmb{\phi}|\pmb{\psi})$, and using the law of total probability, we can write
\begin{equation}\label{eq:psum}
\mathcal{P}(\pmb{\phi}|\mathcal{D}) = \int \diff^{N} \pmb{\psi} \ \mathcal{P}(\pmb{\phi}|\pmb{\psi},\mathcal{D}) \mathcal{P}(\pmb{\psi} | \mathcal{D})  \ .  
\end{equation}
Given a mapping $\mathcal{P}(\pmb{\phi}|\pmb{\psi}) = \prod_{i=1}^N \delta_D(\psi_i - M_{i \alpha} \phi_{\alpha})$, eq.~\eqref{eq:psum} translates the possibility to compress the data $\mathcal{D}$ into the posterior distribution $\mathcal{P}(\pmb{\psi}|\mathcal{D})$ and use it as a likelihood to sample the probability distribution of the model parameters $\pmb{\phi}$, similarly to other situations in cosmology (see e.g.,~\cite{Wandelt:2003uk,BOSS:2016wmc}). 
In our case, eq.~\eqref{eq:psum} means we can reconstruct $\mathcal{P}(\pmb{\phi}|\mathcal{D})$, the posterior of $(w_0, w_a)$, from $\mathcal{P}(\pmb{\psi}|\mathcal{D})$, the posterior of $(w^z_1, \dots, w^z_N)$. 
Since $M$ is not surjective, the converse is not true. 

To close the loop, we can finally insert eq.~\eqref{eq:psum} into eq.~\eqref{eq:pwrong} to relate the two reconstructions of $w(z)$, yielding
\begin{equation}
\mathcal{P}_\mathcal{T}(\pmb{\psi}|\mathcal{D}) = \int_{\pmb{\psi'} \in \mathcal{T}} \diff^{N} \pmb{\psi'} \ \delta_D(\pmb{\psi}-\pmb{\psi'}) \mathcal{P}(\pmb{\psi'}|\mathcal{D}) \ . 
\end{equation}
This translates the fact that $\mathcal{P}_\mathcal{T}$ is the projection of $\mathcal{P}(\pmb{\psi}|\mathcal{D})$ onto the target space $\mathcal{T}$ of $M$. 
We conclude that the reconstructed $w(z)$ being constrained by $(w_0, w_a)$ is in general not representative of the data sensitivity on $w(z)$. 

\paragraph{Analytical posterior for Gaussian distributions}
For near Gaussian distributions (as we deal with in this paper), we can derive analytical solutions for the formulae above to get posterior distributions and significance (likelihood-ratio) at low cost. 
In the following, we assume that $\mathcal{P}(\pmb{\psi}|\mathcal{D})$ is given to us and derive the evidence of the model $\pmb{\phi}$ over $\Lambda$, the null hypothesis. 

Assuming $\mathcal{P}(\pmb{\psi} | \mathcal{D})$ to be a multivariate normal distribution $\mathcal{N}(\pmb{\psi}|\pmb{\bar \psi}, \mathcal{C})$, we can solve eq.~\eqref{eq:psum} exactly by projecting the $N$-D Gaussian onto the domain $\Omega_\phi$, yielding $\mathcal{P}(\pmb{\phi} | \mathcal{D}) = \mathcal{N}(\pmb{\phi}|\pmb{\bar \phi}, \Sigma)$, where 
\begin{equation}\label{eq:CI_phi}
\pmb{\bar \phi} = \Sigma M^T \mathcal{C}^{-1} \pmb{\bar \psi} \ , \quad \Sigma = (M^T \mathcal{C}^{-1} M)^{-1} \ . 
\end{equation}
For completeness, we also provide the expressions to get $\mathcal{P}_\mathcal{T}(\pmb{\psi}|\mathcal{D})$. 
We can solve eq.~\eqref{eq:pwrong} exactly by projecting the 2D Gaussian onto the 1D constraints $\psi_i = M_{i\alpha}\phi_\alpha$, yielding $\mathcal{P}_\mathcal{T}(\pmb{\psi}|\mathcal{D}) = \mathcal{N}(\pmb{\psi}|\pmb{\bar\psi}_\mathcal{T}, \mathcal{C}_\mathcal{T})$, where 
\begin{equation}\label{eq:CI_psiT}
\pmb{\bar \psi}_\mathcal{T} = M \pmb{\bar\phi} \ , \quad \mathcal{C}_\mathcal{T} = M \Sigma M^T \ .
\end{equation}
Notice that $\mathcal{C}_\mathcal{T}$ is nothing but eq.~\eqref{eq:fish} for $z_i = z_*$. 
Using eqs.~\eqref{eq:CI_phi}~and~\eqref{eq:CI_psiT}, we find that
\begin{equation}
\pmb{\bar \psi}_\mathcal{T} = M (M^T \mathcal{C}^{-1} M)^{-1} M^T \mathcal{C}^{-1} \pmb{\bar \psi} \ , \quad \mathcal{C}_\mathcal{T} = M (M^T \mathcal{C}^{-1} M)^{-1} M^T \ .
\end{equation}
Note that $\mathcal{C}_\mathcal{T} \neq \mathcal{C}$ given that $\pmb\psi$ and $\pmb\phi$ are not equivalent parametrisations of the same model (\textit{i.e.}, $M$ is not a bijective map --- a square invertible matrix). 

We now show how we quantify the evidence for evolving dark energy over $\Lambda$ as well as the evidence for phantom dark energy.
For a normal distribution $\mathcal{N}(\pmb{\phi}|\pmb{\bar \phi}, \Sigma)$, the log-likelihood ratio with respect to the null hypothesis $\Lambda$ can be found as the square distance to a point $\pmb{\phi_\Lambda} \equiv (-1, 0)^T$, reading
\begin{equation}\label{eq:distance_point}
\lambda(\pmb{\phi_\Lambda}) = (\pmb{\phi_\Lambda}-\pmb{\bar \phi})^T \Sigma^{-1} (\pmb{\phi_\Lambda}-\pmb{\bar \phi}) \ .
\end{equation}
Given that the posterior of $(w_0, w_a)$ is near Gaussian, we find that~\eqref{eq:distance_point} agrees well with the $\Delta \chi_{\rm min}^2$ from full numerical minimisation, allowing us to quantify the evidence for evolving dark energy over $\Lambda$. 
Regarding the evidence for $w < -1$, we evaluate the difference between the maximum of $\mathcal{N}(\pmb{\phi}|\pmb{\bar \phi}, \Sigma)$ (where $\pmb{\phi}$ is taken as $(w_0, w_a)$) with the one found profiling the line $w(a)=-1$.\footnote{This is true only because $\sup_{\{ \pmb{\phi}|w^{\forall} \geq -1\}} \mathcal{P}(\pmb \phi|\mathcal{D})$ lies on the $w(t) = -1$ line.}
The square distance to a line $\pmb{\phi}(\tau) = \pmb{\phi_\Lambda} + \tau \pmb{v}$ is given by
\begin{equation}\label{eq:distance_line}
\lambda(\pmb{\phi}(\tau)) = \lambda(\pmb{\phi_\Lambda}) - \frac{(\pmb{v}^T \Sigma^{-1} (\pmb{\phi_\Lambda}-\pmb{\bar \phi}))^2}{\pmb{v}^T \Sigma^{-1} \pmb{v}}\ .
\end{equation} 

\bibliography{references}
\bibliographystyle{JHEP}

\end{document}